\documentclass[preprint,12pt]{elsarticle}
\usepackage{amsmath}
\usepackage{amssymb}
\usepackage{booktabs}
\usepackage{multirow}
\usepackage{float}
\usepackage{subfigure}
\usepackage{lineno,hyperref}
\usepackage{epstopdf}
\usepackage[noend]{algpseudocode}
\usepackage{algorithmicx,algorithm}

\modulolinenumbers[5]
\begin{document}
\begin{frontmatter}
\title{An Adaptive Characteristic-wise Reconstruction WENO-Z scheme for Gas Dynamic Euler Equations}

\author[IAPCM,IMECH]{Jun Peng\corref{cor}}
\ead{pengjun62@163.com}
\author[IAPCM]{Chuanlei Zhai}
\ead{zhai\_chuanlei@iapcm.ac.cn}
\author[IAPCM]{Guoxi Ni}
\ead{gxni@iapcm.ac.cn}
\author[IAPCM]{Heng Yong}
\ead{yong\_heng@iapcm.ac.cn}
\author[IMECH]{Yiqing Shen}
\ead{yqshen@imech.ac.cn}

\cortext[cor]{Corresponding author}
\address[IAPCM]{Institute of Applied Physics and Computational Mathematics, Beijing 100094, China}
\address[IMECH]{State Key Laboratory of High Temperature Gas Dynamics, Institute of Mechanics, Chinese Academy of Sciences, Beijing 100190, China}

\begin{abstract}
Due to its excellent shock-capturing capability and high resolution, the WENO scheme family has been widely used in varieties of compressive flow simulation. However, for problems containing strong shocks and contact discontinuities, such as the Lax shock tube problem, the WENO scheme still produces numerical oscillations. To avoid such numerical oscillations, the characteristic-wise construction method should be applied. Compared to component-wise reconstruction, characteristic-wise reconstruction leads to much higher computational cost and thus is not suitable for large scale simulation such as direct numerical simulation of turbulence. In this paper, an adaptive characteristic-wise reconstruction WENO-Z scheme, i.e. the AdaWENO-Z scheme, is proposed to improve the computational efficiency of the characteristic-wise reconstruction method. By defining shared smoothness functions, shared smoothness indicators as well as shared WENO weights are firstly introduced to reduce the computational cost of the component-wise reconstruction procedure and to define a global switch function capable of detecting discontinuity. According to the given switch function, the new scheme performs characteristic-wise reconstruction near discontinuities and switches to component-wise reconstruction for smooth regions. Several one dimensional and two dimensional numerical tests are performed to validate and evaluate the AdaWENO-Z scheme. Numerical results show that AdaWENO-Z maintains essentially non-oscillatory flow field near discontinuities as with the characteristic-wise reconstruction method. Besieds, compared to the component-wise reconstruction method, AdaWENO-Z is about 20\% to 40\% faster which indicates its excellent efficiency.
\end{abstract}


\end{frontmatter}

\section{Introduction}\label{sec1}

Numerical simulation of compressible flow levers engineering and scientific researches by providing detailed and high-quality flow field information. For high-resolution and accurate simulation of compressible flow, numerical methods being applied shall be able to capture all important features, e.g. turbulence and shock waves, in the flow field.

The family of weighted essentially non-oscillatory (WENO) finite difference schemes \cite{liu1994weighted,jiang1996efficient} has been widely used in compressible flow simulations due to its high resolution of small structures and good shock-capturing capability. Within the general framework of smoothness indicators and non-linear weights proposed by Jiang and Shu \cite{jiang1996efficient}, many efforts have been made to improve the accuracy and efficiency of the WENO scheme. Henrick et al. \cite{henrick2005mapped} improved the accuracy of the WENO scheme at critical points by suggesting a mapping function. Borges et al. \cite{borges2008improved} proposed the WENO-Z scheme which calculates the non-linear weights with a high order smoothness indicator. The WENO-Z scheme achieves lower dissipation and higher resolution than the classical WENO scheme of Jiang and Shu and has lower computational cost than the mapping function method of Henrick et al.. The accuracy of the WENO-Z scheme was further improved in \cite{Castro:2011cz,DON2013347} as well as by Yamaleev and Carpenter \cite{Yamaleev:2009hs,Yamaleev:2009gp} and Fan et al. \cite{Fan:2014kj}. Fu et al.\cite{fu2016family} proposed a family of high-order targeted ENO schemes which combines the idea of both the ENO scheme and the WENO scheme. While most methods mainly focus on the improvement of the accuracy of WENO schemes at smooth regions. Shen and Zha \cite{Shen:2014gf} showed that at transitional points, which connect smooth region and discontinuity, the accuracy of fifth order WENO schemes is second order and a multi-step weighting method \cite{shen:2014gfa,Peng:2015hn} was developed to improve the accuracy. Except for these improvements for the fifth order WENO scheme, higher order WENO schemes (higher than fifth order) were also developed \cite{balsara2000monotonicity,Gerolymos:2009dn,martin2006bandwidth}. For the reconstruction of point values, the Central WENO (CWENO) schemes are proposed in \cite{levy_puppo_russo_1999} and extended in \cite{cravero2018cweno,CRAVERO2017}. Different from WENO, CWENO reconstructs polynomials defined in the whole cell instead of values at given points.

In spite of their excellent performances for scalar problems, WENO schemes still produce numerical oscillations for problems like the Lax shock tube problem of the gas dynamic Euler equations. To get rid of such oscillations, the characteristic-wise reconstruction method \cite{harten1987uniformly,jiang1996efficient,ren2003characteristic} should be applied. Compared to the component-wise reconstruction method, the characteristic-wise reconstruction method results in much higher computational cost. Therefore, for efficiency consideration, in practical large scale simulations, the component-wise reconstruction method is always preferred, e.g. \cite{pirozzoli2004direct}, that some numerical oscillations are tolerable. However, such compromise may reduce the reliability of the simulation result that numerical oscillations disturb the flow field and may change the whole flow structure. To prevent numerical oscillations and avoid the use of characteristic-wise reconstruction, He et al. \cite{he2015preventing} analyzed the WENO weights and proposed a new method to calculate the final smoothness indicators. This method reduces but is not free of numerical oscillations.  Hu et al.\cite{hu2015efficient} proposed a discontinuity detector to combine characteristic-wise WENO with low dissipation linear scheme that several free parameters were introduced. Puppo \cite{puppo2003adaptive} proposed an adaptive method to combine the component-wise reconstruction method and the characteristic-wise reconstruction method and showed good performance and efficiency.

In this paper, we focus on developing a characteric-wise reconstruction WENO scheme which is more efficient for large scale simulation of compressible flow. Shared smoothness indicators and WENO weights are firstly introduced to reduce the computational cost of the component-wise reconstruction procedure. Based on the WENO-Z non-linear weights calculated from the shared smoothness indicators, a global switch function is then introduced to detect discontinuity. Utilizing the given switch function, a new scheme is proposed that it performs the characteristic-wise reconstruction near discontinuities and switches to the component-wise reconstruction for smooth regions. This paper is organized as following. In Section \ref{sec2}, the WENO scheme and its component-wise and characteristic-wise reconstruction implementations are introduced. In section \ref{sec3}, a new adaptive method is developed based on the analysis of the two implementations. In section \ref{sec4}, several numerical tests are presented to illustrate the performance and efficiency of the new method. Concluding remarks are given in Section \ref{sec5}.

\section{Solving the gas dynamic Euler equations with WENO}\label{sec2}
\subsection{The gas dynamic Euler equations}\label{subsec2.1}
The one dimensional Euler equations of inviscid ideal gas is given by:
\begin{equation}\label{eq2.1}
\frac{\partial \vec{U}}{\partial t}+\frac{\partial \vec{F}}{\partial x} = 0
\end{equation}
in which $\vec{U}$ and $\vec{F}$ are the conserved variable and the convective flux vectors:
\begin{equation}\label{eq2.2}
\vec{U}=\left[ \begin{matrix}
   \rho  \\
   \rho u  \\
   E  \\
\end{matrix} \right],
\vec{F}=\left[ \begin{matrix}
   \rho u  \\
   \rho u^2+p  \\
   u(E+p)  \\
\end{matrix} \right],
\end{equation}
where $\rho$ is the density, $u$ is the velocity, and $E=\frac{p}{\gamma-1}+\frac{1}{2}\rho u^2$ is the total energy with $\gamma=1.4$.

For the convective flux vector $\vec{F}$, its Jacobian matrix $\bf{A}$ is defined as:
\begin{equation}\label{eq2.3}
\frac{\partial \vec{F}(\vec{U})}{\partial x}=A\frac{\partial \vec{U}}{\partial x},
\end{equation}
where $\mathbf{A}$ is written as:
\begin{equation}\label{eq2.4}
\bf{A}= \bf{R}\bf{\Lambda}\bf{L}=\left[ \begin{matrix}
   0 & 1 & 0  \\
   -\frac{3-\gamma}{2}u^2 & (3-\gamma)u & \gamma-1 \\
   \frac{\gamma-2}{2}u^3-\frac{uc^2}{\gamma-1} & \frac{c^2}{\gamma-1}+\frac{3-2\gamma}{2}u^2 & \gamma u \\
\end{matrix} \right]
\end{equation}
 where $c = \sqrt{\gamma p/\rho}$ is the sound speed. Here, $\mathbf{\Lambda}$ is the eigen matrix of $\bf{A}$:
\begin{equation}\label{eq2.5}
\mathbf{\Lambda}=\left[ \begin{matrix}
    u-c &  &   \\
    & u &  \\
    &  & u+c \\

\end{matrix} \right],
\end{equation}
and $\mathbf{L}$ and $\mathbf{R}$ are the left and right eigen vectors:
\begin{equation}\label{eq2.6}
\mathbf{L}=\left[
\begin{matrix}
   \mathbf{l}_{0}  \\
   \mathbf{l}_{1}  \\
   \mathbf{l}_{2}  \\
\end{matrix} \right]=\left[ \begin{matrix}
   \frac{\gamma -1}{4}\frac{u^2}{c^2}+\frac{1}{2}\frac{u}{c} & -\frac{\gamma -1}{2}\frac{u}{c^2}-\frac{1}{2}\frac{1}{c} & \frac{\gamma -1}{2}\frac{1}{c^2}  \\
   1-\frac{\gamma -1}{2}\frac{u^2}{c^2} & \frac{\gamma -1}{2}\frac{u}{c^2} & -\frac{\gamma -1}{2}\frac{1}{c^2}  \\
   \frac{\gamma -1}{4}\frac{u^2}{c^2}-\frac{1}{2}\frac{u}{c} & -\frac{\gamma -1}{2}\frac{u}{c^2}+\frac{1}{2}\frac{1}{c} & \frac{\gamma -1}{2}\frac{1}{c^2}  \\
\end{matrix} \right]
\end{equation}
\begin{equation}\label{eq2.7}
\mathbf{R}=\left[\mathbf{r}_{0},\mathbf{r}_{1},\mathbf{r}_{2}\right]=\left[\begin{matrix}
   1 & 1 & 1  \\
   u-c & u & u+c  \\
   \frac{u^2}{2}+\frac{c^2}{\gamma -1}-uc & \frac{u^2}{2} & \frac{u^2}{2}+\frac{c^2}{\gamma -1}+uc  \\
\end{matrix} \right],
\end{equation}

To introduce correct upwinding, the flux vector $\vec{F}$ is generally splitted into two parts:
\begin{equation}\label{eq2.8}
\vec{F}=\vec{F}^{+}+\vec{F}^{-}
\end{equation}
where
\begin{equation}\label{eq2.9}
\nonumber \frac{d\vec{F}^{+}}{d\vec{U}}\ge 0, \quad \frac{d\vec{F}^{-}}{d\vec{U}}< 0.
\end{equation}
In this paper, the Lax-Friedrichs splitting method \cite{jiang1996efficient} is used:
\begin{equation}\label{eq2.10}
\vec{F}_{i}^{\pm }=\frac{1}{2}(\vec{F}_{i}\pm \alpha _i \vec{U}_{i}).
\end{equation}
For the Euler equations, $\alpha_i$ is taken as:
\begin{equation}\label{eq2.11}
	\alpha_i=\alpha=\underset{i}\max(|u_i|+c_i)
\end{equation}
for simplicity and robustness as being discussed in \cite{jiang1996efficient,he2015preventing}.

\subsection{The WENO scheme}\label{subsec2.2}
To introduce the finite difference WENO scheme, let us consider the semi-discrete form of eq.\eqref{eq2.1} on equally spaced grid, i.e. $\Delta x =x_{i+1}-x_{i}$:
\begin{equation}\label{eq2.12}
	\frac{\partial \vec{U}_{i}}{\partial t}=-\frac{\partial \vec{F}_{i}}{\partial x}\approx -\frac{1}{\Delta x}({{\vec{F}}_{i+1/2}}-{{\vec{F}}_{i-1/2}})
\end{equation}
where $\vec{F}_{i+1/2}=\vec{F}_{i+1/2}^{+}+\vec{F}_{i+1/2}^{-}$ is the numerical flux at cell interface. Each component of the numerical fluxes $\vec{F}_{i+1/2}^{+}$ and $\vec{F}_{i+1/2}^{-}$,  i.e. ${}^{k}\hat{f}_{i+\frac{1}{2}}^{\pm}$, is then reconstructed by the WENO scheme. For simplicity, $k$ and $\pm$ in the superscript are dropped in the following parts of this paper.

The numerical flux component $\hat{f}_{i+\frac{1}{2}}$ can be obtained by high order schemes. The fifth order upstream-biased scheme is written as:
\begin{equation}\label{eq2.13}
	\hat{f}_{i+\frac{1}{2}}=\frac{2}{60}{f}_{i-2}-\frac{13}{60}{f}_{i-1}+\frac{47}{60}{f}_{i}+\frac{27}{60}{f}_{i+1}-\frac{3}{60}{f}_{i+2},
\end{equation}
where $f_i=f(\vec{U_i})$ is the point value of the flux component. Eq.\eqref{eq2.13} is a convex combination of three third order schemes on three substencils $S_0=(x_{i-2},x_{i-1},x_{i})$, $S_1=(x_{i-1},x_{i},x_{i+1})$, and $S_2=(x_i,x_{i+1},x_{i+2})$:
\begin{flalign}
\hat{f}_{0,i+1/2} &=\frac{1}{3}{{f}_{i-2}}-\frac{7}{6}{{f}_{i-1}}+\frac{11}{6}{{f}_{i}},\label{eq2.14}\\
\hat{f}_{1,i+1/2} &=-\frac{1}{6}{{f}_{i-1}}+\frac{5}{6}{{f}_{i}}+\frac{1}{3}{{f}_{i+1}},\label{eq2.15}\\
\hat{f}_{2,i+1/2} &=\frac{1}{3}{{f}_{i}}+\frac{5}{6}{{f}_{i+1}}-\frac{1}{6}{{f}_{i+2}}.\label{eq2.16}
\end{flalign}
with linear weights $c_0=0.1$, $c_1=0.6$ , and $c_2=0.3$ respectively. By substituting the linear weights with the WENO weights, we have the fifth order WENO scheme:
\begin{equation}\label{eq2.17}
	WENO5:\quad \hat{f}_{i+1/2}={\omega }_{0}\hat{f}_{0,i+1/2}+\omega_{1}\hat{f}_{1,i+1/2}+\omega_{2}\hat{f}_{2,i+1/2}.
\end{equation}
The WENO weights $\omega_k$ are given by:
\begin{equation}
\omega_{k} =\frac{\alpha_k}{\sum\limits_{i}{\alpha_i}},\label{eq2.18a}
\end{equation}
where ${\alpha_k}$ are the non-linear weights:
\begin{equation}
{\alpha_k} =\frac{c_k}{(\varepsilon +\beta_k)^p},k=\{0,1,2\},p=1,2,..,\label{eq2.18b}
\end{equation}
in which $\epsilon$ is a small number to avoid dividing by zero. In this paper, $\epsilon$ is taken to be $1\times 10^{-6}$. The smoothness indicators $\beta_i$ are:
\begin{flalign}
{{\beta }_{0}} &=\frac{13}{12}{{({{f}_{i-2}}-2{{f}_{i-1}}+{{f}_{i}})}^{2}}+\frac{1}{4}{{({{f}_{i-2}}-4{{f}_{i-1}}+3{{f}_{i}})}^{2}},\label{eq2.19}\\
{{\beta }_{1}} &=\frac{13}{12}{{({{f}_{i-1}}-2{{f}_{i}}+{{f}_{i+1}})}^{2}}+\frac{1}{4}{{({{f}_{i-1}}-{{f}_{i+1}})}^{2}},\label{eq2.20}\\
{{\beta }_{2}} &=\frac{13}{12}{{({{f}_{i}}-2{{f}_{i+1}}+{{f}_{i+2}})}^{2}}+\frac{1}{4}{{(3{{f}_{i}}-4{{f}_{i+1}}+{{f}_{i+2}})}^{2}}.\label{eq2.21}
\end{flalign}

As shown by Henrick et al. \cite{henrick2005mapped}, for the fifth order WENO scheme of Jiang and Shu \cite{jiang1996efficient}, the non-linear weights \eqref{eq2.18b} do not satisfy the necessary and sufficient conditions for fifth order convergence. A mapping function was introduced to improve the accuracy of the final weights (the WENO-M scheme). In \cite{borges2008improved}, Borges et al. introduced a parameter $\tau_5=|\beta_0-\beta_2|$ to calculate the non-linear weights as:
\begin{equation}\label{eq2.22}
\alpha_k=c_k\left(1+(\frac{\tau_5}{\beta_k+\varepsilon})^q\right), k=\{0,1,2\}, q=1,2,...
\end{equation}
This new scheme (the WENO-Z scheme) is less computational expensive than the WENO-M scheme. Optimal definitions of different orders of accuracy for $\tau$ can be found in \cite{DON2013347}. In this paper, the fifth order WENO-Z scheme is used as the base WENO scheme.

There are two ways to calculate the numerical fluxes of the Euler equations with WENO schemes, namely, the component-wise reconstruction method and the characteristic-wise reconstruction method. The former method reconstructs the numerical flux vector component-by-component while the latter method performs the reconstruction in the characteristic space. In the following two subsections(\ref{subsec2.3} and \ref{subsec2.4}), details of these two methods will be presented.

\subsection{Component-wise reconstruction}\label{subsec2.3}
By implementing the WENO reconstruction for the numerical flux vector $\vec{F}$ component-by-component, the resulted numerical flux at cell interface can be written as:
\begin{equation}\label{eq2.23}
	\vec{F}^{CP}_{i+1/2}=\left[\begin{matrix}
   {}^{0}\hat{f}^{CP}_{i+1/2}  \\
   {}^{1}\hat{f}^{CP}_{i+1/2}  \\
   {}^{2}\hat{f}^{CP}_{i+1/2}  \\
\end{matrix} \right]=\sum\limits_{k=0}^{2}{\mathbf{\omega }_{k}^{CP}\mathbf{\hat{f}}_{k,i+1/2}^{CP}}
\end{equation}
in which
\begin{equation}\label{eq2.24}
	\mathbf{\omega }_{k}^{CP}\text{=}\left[ \begin{matrix}
   {}^{0}\omega _{k}^{CP} & {} & {}  \\
   {} & {}^{1}\omega _{k}^{CP} & {}  \\
   {} & {} & {}^{2}\omega _{k}^{CP}  \\
\end{matrix} \right]
\end{equation}
and
\begin{equation}\label{eq2.25}
	\mathbf{\hat{f}}_{k,i+1/2}^{CP}=\left[ \begin{matrix}
   {}^{0}\hat{f}_{k,i+1/2}  \\
   {}^{1}\hat{f}_{k,i+1/2}  \\
   {}^{2}\hat{f}_{k,i+1/2}  \\
\end{matrix} \right], \quad k=0,1,2.
\end{equation}
The WENO weights in Eq.\eqref{eq2.24} are calculated according to the corresponding flux component at each stencil:
\begin{equation}\label{eq2.26}
	{}^{s}\omega _{k}^{CP}={}^{s}\omega _{k}^{CP}({}^{s}{{f}_{i+k-2}},\cdots ,{}^{s}{{f}_{i+k}}),\quad k=0,1,2, \quad s=0,1,2,
\end{equation}
where $s$ is the component index and $k$ is the stencil index. The right superscript 'CP' stands for component-wise reconstruction.

It can be observed that the component-wise reconstruction method is easy to be implemented that only one single WENO reconstruction subroutine is needed in one's code. However, numerical oscillations may present in solutions obtained by the component-wise reconstruction method. In the following parts, the component-wise method will be referred to as the CP method.

\subsection{Characteristic-wise reconstruction}\label{subsec2.4}
Compared to the CP method, the characteristic-wise method produces less numerical oscillations. To perform reconstruction in the characteristic space, the flux vector $\vec{F}$ should firstly be projected onto the left eigenvector of its Jacobian \eqref{eq2.4} on cell interface $x_{i+1/2}$. The left eigenvectors on cell interface are obtained from Roe-averaged \cite{roe1981approximate} primitive variables:
\begin{equation}\label{eq2.27}
	\bar{u}=\frac{\sqrt{{{\rho }_{i}}}{{u}_{i}}+\sqrt{{{\rho }_{i+1}}}{{u}_{i+1}}}{\sqrt{{{\rho }_{i}}}+\sqrt{{{\rho }_{i+1}}}}
\end{equation}
\begin{equation}\label{eq2.28}
	\bar{h}=\frac{\sqrt{{{\rho }_{i}}}{{h}_{i}}+\sqrt{{{\rho }_{i+1}}}{{h}_{i+1}}}{\sqrt{{{\rho }_{i}}}+\sqrt{{{\rho }_{i+1}}}}
\end{equation}
\begin{equation}\label{eq2.29}
	\bar{c}=\sqrt{(\gamma -1)(\bar{h}-\tfrac{1}{2}{{{\bar{u}}}^{2}})}
\end{equation}
\begin{equation}\label{eq2.30}
	h=\frac{p}{(\gamma -1)\rho }+\tfrac{1}{2}{{u}^{2}}+\frac{p}{\rho }
\end{equation}
The averaged left eigenvector matrix is  therefore written as:
\begin{equation}\label{eq2.31}
	{{\mathbf{\bar{L}}}_{\text{i+1/2}}}\text{=}\left[ \begin{matrix}
   {{{\mathbf{\bar{l}}}}_{0}}  \\
   {{{\mathbf{\bar{l}}}}_{1}}  \\
   {{{\mathbf{\bar{l}}}}_{2}}  \\
\end{matrix} \right]=\left[ \begin{matrix}
   \frac{\gamma -1}{4}\frac{{{{\bar{u}}}^{2}}}{{{{\bar{c}}}^{2}}}+\frac{1}{2}\frac{{\bar{u}}}{{\bar{c}}} & -\frac{\gamma -1}{2}\frac{{\bar{u}}}{{{{\bar{c}}}^{2}}}-\frac{1}{2}\frac{1}{{\bar{c}}} & \frac{\gamma -1}{2}\frac{1}{{{{\bar{c}}}^{2}}}  \\
   1-\frac{\gamma -1}{2}\frac{{{{\bar{u}}}^{2}}}{{{{\bar{c}}}^{2}}} & \frac{(\gamma -1){\bar{u}}}{{{{\bar{c}}}^{2}}} & -\frac{\gamma -1}{{{{\bar{c}}}^{2}}}  \\
   \frac{\gamma -1}{4}\frac{{{{\bar{u}}}^{2}}}{{{{\bar{c}}}^{2}}}-\frac{1}{2}\frac{{\bar{u}}}{{\bar{c}}} & -\frac{\gamma -1}{2}\frac{{\bar{u}}}{{{{\bar{c}}}^{2}}}+\frac{1}{2}\frac{1}{{\bar{c}}} & \frac{\gamma -1}{2}\frac{1}{{{{\bar{c}}}^{2}}}  \\
\end{matrix} \right]
\end{equation}
The WENO reconstruction is performed component-by-component to the projected variable $\hat{w}$:
\begin{equation}\label{eq2.32}
	\mathbf{\hat{w}}_{k,i+1/2}^{C\text{H}}=\left[ \begin{matrix}
   {}^{0}\hat{w}_{k,i+1/2}  \\
   {}^{1}\hat{w}_{k,i+1/2}   \\
   {}^{2}\hat{w}_{k,i+1/2}   \\
\end{matrix} \right]={{\mathbf{\bar{L}}}_{i+1/2}}\mathbf{\hat{f}}_{k,i+1/2}^{CP},k=0,1,2
\end{equation}
\begin{equation}\label{eq2.33}
	\mathbf{\vec{W}}_{i+1/2}^{CH}=\sum\limits_{k=0}^{2}{\mathbf{\omega }}_{k}^{CH}\mathbf{\hat{w}}_{k,i+1/2}^{C\text{H}}
\end{equation}
Different from the CP method, the WENO weights are computed according to the projected variables on each stencil:
\begin{equation}\label{eq2.34}
	\mathbf{\omega}_{k}^{CH}=\left[ \begin{matrix}
   {}^{0}\omega _{k}^{CH} & {} & {}  \\
   {} & {}^{1}\omega _{k}^{CH} & {}  \\
   {} & {} & {}^{2}\omega _{k}^{CH}  \\
\end{matrix} \right]
\end{equation}
\begin{equation}\label{eq2.35}
	{}^{s}\omega_{k}^{CH}={}^{s}\omega_{k}^{CH}(\mathbf{\bar{l}}_{s}\vec{F}_{i+k-2},\cdots ,\mathbf{\bar{l}}_{s}\vec{F}_{i+k}),\quad k=0,1,2,\quad s=0,1,2
\end{equation}
After the WENO reconstruction of the projected variables, the obtained values need to be transformed back to the physical space by projecting onto the averaged right eigenvectors on cell interface:
\begin{equation}\label{eq2.36}
	{{\vec{F}}^{CH}}_{i+1/2}=\left[ \begin{matrix}
   {}^{0}{{{\hat{f}}}_{i+1/2}^{CH}}  \\
   {}^{1}{{{\hat{f}}}_{i+1/2}^{CH}}  \\
   {}^{2}{{{\hat{f}}}_{i+1/2}^{CH}}  \\
\end{matrix} \right]={{\mathbf{\bar{R}}}_{i\text{+}1/2}}\vec{W}_{i+1/2}^{C\text{H}}
\end{equation}
The averaged right eigenvectors are given by:
\begin{equation}\label{eq2.37}
	{{\mathbf{\bar{R}}}_{i+1/2}}\text{=}[{{\mathbf{\bar{r}}}_{0}},{{\mathbf{\bar{r}}}_{1}},{{\mathbf{\bar{r}}}_{2}}],=\left[ \begin{matrix}
   1 & 1 & 1  \\
   \bar{u}-\bar{c} & {\bar{u}} & \bar{u}+\bar{c}  \\
   \frac{{{{\bar{u}}}^{2}}}{2}+\frac{{{{\bar{c}}}^{2}}}{\gamma -1}-\bar{u}\bar{c} & \frac{{{{\bar{u}}}^{2}}}{2} & \frac{{{{\bar{u}}}^{2}}}{2}+\frac{{{{\bar{c}}}^{2}}}{\gamma -1}+\bar{u}\bar{c}  \\
\end{matrix} \right]
\end{equation}
The whole process of the characteristic-wise WENO reconstruction can be summarized into one formula:
\begin{equation}\label{eq2.38}
\vec{F}^{CH}_{i+1/2}=\left[\begin{matrix}
{}^{0}\hat{f}^{CH}_{i+1/2}  \\
{}^{1}\hat{f}^{CH}_{i+1/2}  \\
{}^{2}\hat{f}^{CH}_{i+1/2}  \\
\end{matrix} \right]=\sum\limits_{k=0}^{2}\mathbf{\bar{R}}_{i+1/2}\mathbf{\omega}_{k}^{CH}\mathbf{\bar{L}}_{i+1/2}\mathbf{\hat{f}}_{k,i+1/2}^{CP},
\end{equation}
where the superscript 'CH' stands for characteristic-wise reconstruction. In the following parts, the characteristic-wise method will be referred to as the CH method.

Compared to the CP method, the CH method requires several matrix constructions and projections resulting in much higher computational cost. Although the computational cost of the CH method is high, it produces less numerical oscillations.

\section{The new scheme}\label{sec3}
On one hand, in spite of its high computational cost, the merit of the CH method is that it leads to less spurious oscillations. On the other hand, the advantage of the CP method is its computational efficiency regardless of the numerical oscillations it brings. To get rid of their drawbacks and utilize their advantages, the two methods can be combined in a dynamic way. In this section, the two methods will be analyzed to show where their differences and similarities lie. A new adaptive method that combines these two method will be introduced according to our analysis.

\subsection{Comparison of the two reconstruction methods}\label{subsec3.1}
Eq.\eqref{eq2.38} can be written in a more compact form by considering:
\begin{equation}\label{eq3.1}
	\mathbf{\tilde{\omega }}_{k}^{CH}={{\mathbf{\bar{R}}}_{i\text{+}1/2}}\mathbf{\omega }_{k}^{C\text{H}}{{\mathbf{\bar{L}}}_{i+1/2}}.
\end{equation}
Substituting Eq.\eqref{eq3.1} into Eq.\eqref{eq2.38}, we have:
\begin{equation}\label{eq3.2}
	{{\vec{F}}^{CH}}_{i+1/2}=\left[ \begin{matrix}
   {}^{0}{{{\hat{f}}}^{CH}}_{i+1/2}  \\
   {}^{1}{{{\hat{f}}}^{CH}}_{i+1/2}  \\
   {}^{2}{{{\hat{f}}}^{CH}}_{i+1/2}  \\
\end{matrix} \right]=\sum\limits_{k=0}^{2}{\mathbf{\tilde{\omega }}_{k}^{CH}\mathbf{\hat{f}}_{k,i+1/2}^{CP}}
\end{equation}
Comparing Eq.\eqref{eq2.23} and Eq.\eqref{eq3.2}, one immediately finds that the difference between the CP method and the CH method lies in their WENO weight matrices:
$\mathbf{\tilde{\omega }}_{k}^{CH}$ and $\mathbf{\omega }_{k}^{CP}$.

For smooth region, the WENO weights approximate the linear weights. Therefore, the WENO weight matrix $\mathbf{\omega }_{k}^{CP}$ of the CP method becomes:
\begin{equation}\label{eq3.3}
\mathbf{\omega}_{k}^{CP} \approx c_k\mathbf{I},
\end{equation}
where $\mathbf{I}$ is an identity matrix. For the CH method, its weight matrix $\mathbf{\tilde{\omega }}_{k}^{CH}$ reads:
\begin{equation}\label{eq3.4}
\mathbf{\tilde{\omega }}_{k}^{CH} \approx \mathbf{\bar{R}_{i+1/2}}c_k\mathbf{I}\mathbf{\bar{L}_{i+1/2}}=c_k\mathbf{\bar{R}_{i+1/2}}\mathbf{I}\mathbf{\bar{L}_{i+1/2}},
\end{equation}
Considering that
\begin{equation}\label{eq3.5}
	{{\mathbf{\bar{R}}}_{i+1/2}}={{\mathbf{\bar{L}}}_{i+1/2}}^{-1},
\end{equation}
we have:
\begin{equation}\label{eq3.6}
\mathbf{\omega}_{k}^{CP} \approx  c_k \mathbf{I}\approx \tilde{\mathbf{\omega}}_{k}^{CH}.
\end{equation}
Eq.\eqref{eq3.6} reveals that for smooth flow the numerical fluxes obtained by the two methods are approximately equal:
\begin{flalign}\label{eq3.7}
\begin{cases}
\mathbf{\omega}_{k}^{CP} \approx \mathbf{\bar{R}}_{i+1/2}\mathbf{\omega}_{k}^{CH}\mathbf{\bar{L}}_{i+1/2} &\\
 \vec{F}^{CP}_{i+1/2} \approx \vec{F}^{CH}_{i+1/2} &\\
\end{cases}&
\end{flalign}

\subsection{Puppo's method}\label{subsec3.2}
According to the analysis above, the CP method and the CH method give approximately equal results in smooth region. As the CH method results in less numerical oscillations, it is preferred in discontinuous region. Although the CP method produces oscillations in some discontinuous regions, it is more computational efficient than the CH method. A straightforward strategy to combine the advantages of these two methods is to use the CP method in smooth region and to use the CH method in discontinuous region.

In \cite{puppo2003adaptive}, an adaptive projection method was proposed to select between component-wise reconstructed and characteristic-wise reconstructed variables. The switch being applied is a smoothness indicator that measures the smoothness of the whole stencil:
\begin{equation}
	\beta^{TOT}_{i} = \sum_{l=0}^{2}\beta_{l,i}
\end{equation}
in which $\beta_{l,i}$ is the global smoothness indicator of the substencil $S_i^l=(x_{i+l-2},x_{i+l-1},x_{i+l})$ of the five-point stencil $S_i=(x_{i-2},x_{i-1},x_{i},x_{i+1},x_{i+2})$:
\begin{equation}
	\beta_{l,i} = \frac{1}{m}\sum_{r=1}^{m}\frac{1}{||{}^{r}f||_2} {}^{r}\beta_{l,i}.
\end{equation}
Here, $m$ is the number of components, $||{}^{r}f||_2$ is the $L^2$ norm of the $r$th component of the reconstruction candidate $\vec{F}$(the splitted fluxes $\vec{F}^{\pm}$ in our implementation) :
\begin{equation}
	||{}^{r}f||_2 = \left(\sum_{all\ i}|{}^{r}f_{i}|^2h\right)^{1/2}
\end{equation}
where $h$ is the grid size, ${}^{r}f_{i}$ is the r-th component of $\vec{F}_{i}$, and ${}^{r}\beta_{k,i}$ is the smoothness indicator given by \eqref{eq2.19}-\eqref{eq2.21} of ${}^{r}f_{i}$ at stencil $S_i^l$. To avoid division by zero, $||{}^{r}f||_2$ is set to be 1 when it is smaller than $1\times 10^{-10}$.

The switch criterion is then given by:
\begin{equation}\label{Puppo}
   \vec{F}_{i+1/2}=\left\{ \begin{matrix}
   \vec{F}^{CP}_{i+1/2}, \quad \beta^{TOT}_{i}<1  \\
   \vec{F}^{CH}_{i+1/2}, \quad \text{otherwise}  \\
\end{matrix} \right.
\end{equation}
In the following part of this paper, this method will be referred to as the TOT method.

\subsection{Adaptive WENO-Z method}\label{subsec3.2}
The TOT method is faster than the CH method. However, TOT requires the computation of the smoothness indicators for each flux component that it still needs more computational time than the CP method\cite{puppo2003adaptive}.

The calculation of the smoothness indicators occupies most of the computational time of the WENO scheme according to \cite{jiang1996efficient} as well as our experiences. To reduce the computational cost of the component-wise reconstruction, instead of computing the smoothness indicators for each vector component, we propose to use the following variable to compute a set of smoothness indicators $\beta_{k,G}^{\pm}$ for all components:
\begin{equation}\label{eqG}
	G^{\pm}=\rho + (\rho u^2+p\pm \alpha \rho u),
\end{equation}
where the $\pm$ sign denotes $G$ and smoothness indicators for the positive and the negative fluxes and $\alpha$ is given by Eq.\eqref{eq2.11}. We refer to $G^{\pm}$ as the shared smoothness functions and $\beta_{k,G}^{\pm}$ as the shared smoothness indicators.

The non-linear weights $\alpha_{k,G}^{\pm}$ and WENO weights $\omega_{k,G}^{\pm}$ are then obtained from $\beta_{k,G}^{\pm}$ according to \eqref{eq2.22} and \eqref{eq2.18a} respectively. We call $\alpha_{k,G}^{\pm}$ the shared non-linear weights and $\omega_{k,G}^{\pm}$ the shared WENO weights. By computing the shared smoothness indicators $\beta_{k,G}^{\pm}$ based on $G^{\pm}$, the number of smoothness indicators needed to be calculated is reduced to 6 ($3\ \text{smoothness indicators}\times2\ \text{splitted fluxes}$) from 18 ($3\ \text{smoothness indicators}\times 3\ \text{components}\times2\ \text{splitted fluxes}$) for the one dimensional Euler equations. The number of non-linear weights and WENO weights needed to be computed is also reduced accordingly.

\begin{figure}[H]
\begin{center}
\includegraphics[width=0.8\textwidth]{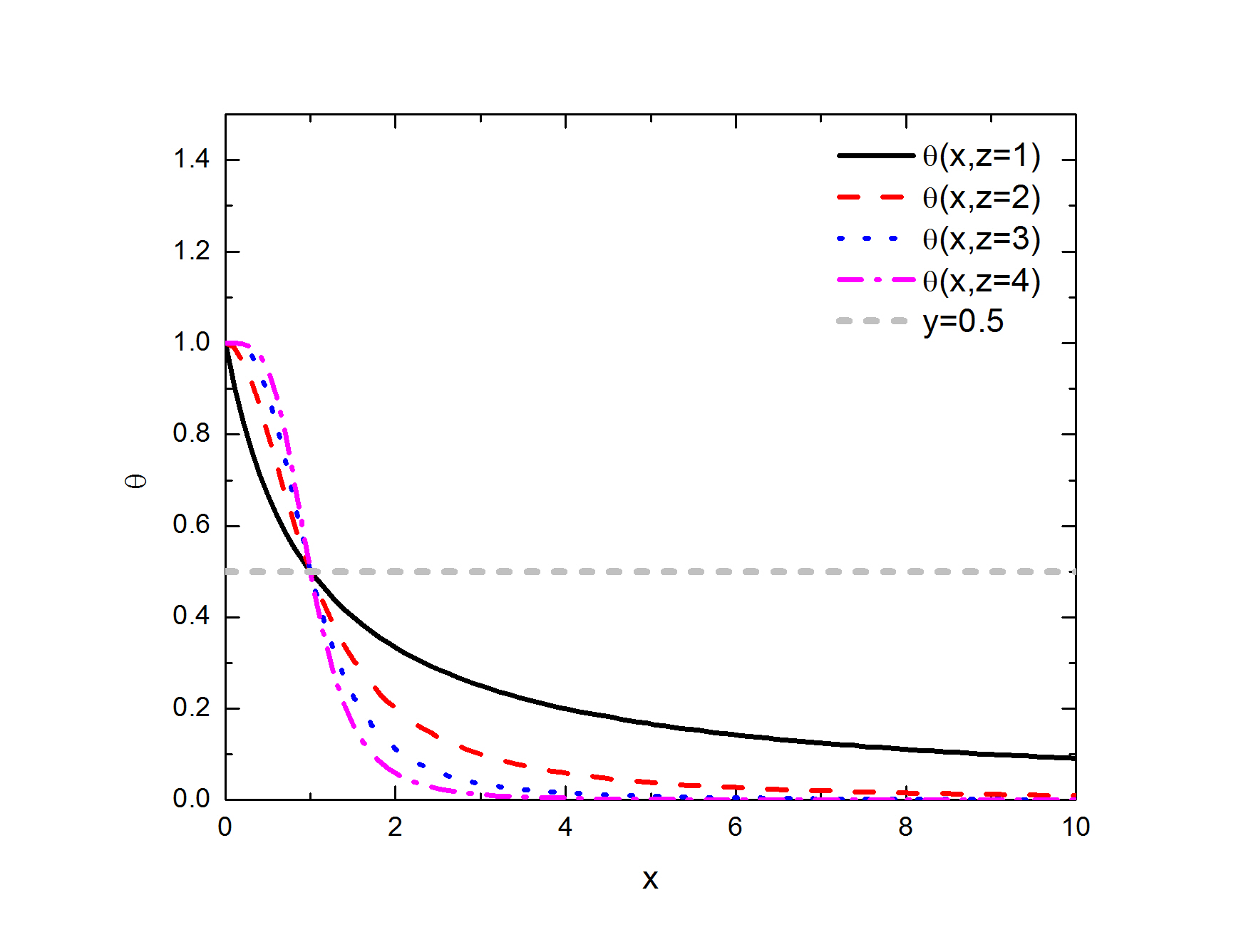}
\caption{$\theta(x)$ with different $z$ values}
\label{fig1}
\end{center}
\end{figure}

To shift from the component-wise reconstruction to the characteristic-wise reconstruction, a switch function capable of detecting discontinuities is required. The switch function employed in this paper is:
\begin{equation}
	\theta (x,z)=\frac{1}{1+|x{{|}^{z}}}, \quad z\ge 1.
\end{equation}
It varies from $1$ to $0$ rapidly and smoothly with increasing $x$. Fig.\ref{fig1} illustrates $\theta$ with different values of parameters. Detailed analysis of this function can be found in \cite{peng2017novel}.
Here, we take:
$$
x=(\sum\limits_{k=0}^{2}\alpha_{k,G}^{\pm}-1)
$$
and get
\begin{equation}\label{eqtheta0}
	\theta^{\pm}=\frac{1}{1+(\sum\limits_{k=0}^{2}\alpha_{k,G}^{\pm}-1)^z},
\end{equation}
where $\alpha_{k,G}^{\pm}$ are the shared non-linear weights for the positive and the negative splitted fluxes respectively given by Eq.\eqref{eq2.22}. Taking the properties of the non-linear weights $\alpha_{k,G}^{\pm}$ into consideration (see \cite{borges2008improved,DON2013347,peng2017novel} for more details), we have:
\begin{equation}
\sum\limits_k\alpha_{k,G}^{\pm} \approx 1
\end{equation}
for smooth region and
\begin{equation}
\sum\limits_k\alpha_{k,G}^{\pm} \gg 1
\end{equation}
for discontinuous region.
This property leads to
\begin{equation}\label{eqtheta}
\theta \approx
\begin{cases}
1, & smooth,\\
0, & discontinuous.
\end{cases}	
\end{equation}
Taking advantage of Eq.\eqref{eqtheta}, we propose the following adaptive approach:
\begin{equation}\label{eqAdaWENO}
	{{\vec{F}}^{{}}}_{i+1/2}\text{=}\left\{ \begin{matrix}
   {{{\vec{F}}}^{C\text{P}}}_{i+1/2},\theta >{{\theta }_{0}}  \\
   {{{\vec{F}}}^{CH}}_{i+1/2},\theta \le {{\theta }_{0}}  \\
\end{matrix} \right.
\end{equation}
where $\theta_0$ is a threshold which can be simply taken to be:
\begin{equation}
	\theta_0 = \theta (1,z)=0.5.
\end{equation}
Bearing in mind that the $\theta$ function can be interpreted as how much we can believe the function being measured is smooth, it is not hard to understand the choice of $\theta_0$ made above.

The final adaptive characteristic-wise WENO-Z scheme (referred to as AdaWENO-Z) algorithm is given in Algorithm \ref{algorithm}. We refer to the component-wise part of AdaWENO-Z as SWENO-Z(Shared-weight WENO-Z) for convenience. It is worth of noting that the shared WENO weights $\omega_{k,G}^{\pm}$ are only used by the SWENO-Z part. For the characteristic-wise part, the calculation method of the WENO weights is not changed.

\begin{algorithm}[t]
\caption{The AdaWENO-Z algorithm}\label{algorithm}
\begin{algorithmic}[1]
\State Calculate $G^{\pm}_{i}$ for the positive and the negative splited fluxes $\vec{F}^{\pm}_{i}$ respectively according to Eq.\eqref{eqG};

\For{Each stencil $S_i=(i-2,i-1,i,i+1,i+2)$, $i=1$ to $n$}
\State Calculate $\beta_{k,G,i}^{\pm}$ according to Eq.\eqref{eq2.19}-\eqref{eq2.22} for the positive and the negative splitted fluxes $\vec{F}^{\pm}_{i}$ based on the calculated $G^{\pm}_{i}$ ;

\State Calculate $\omega_{k,G,i}^{\pm}$ and $\theta_{i}^{\pm}$ according to Eq.\eqref{eq2.18a} and Eq.\eqref{eqtheta0}.
\EndFor

\For{Each stencil $S_i=(i-2,i-1,i,i+1,i+2)$, $i=1$ to $n$}
	\If{$\theta_{i}^{\pm}\geq 0.5$}
        \State ${}^{s}\hat{f}_{i+1/2}^{\pm}=\sum_{k=0}^{2}\omega_{k,G,i}^{\pm}{}^{s}\hat{f}_{k,i+1/2}^{\pm,CP}$, $s=0,1,2$
	\Else
        \State ${}^{s}\hat{f}_{i+1/2}^{\pm}={}^{s}\hat{f}_{i+1/2}^{\pm,CH}$, $s=0,1,2$
	\EndIf
\EndFor
\end{algorithmic}
\end{algorithm}
\section{Numerical tests}\label{sec4}
In this section, several numerical tests including one dimensional and two dimensional problems are considered to validate and evaluate the performance of the new method. Numerical results are compared between CP, CH, TOT, and AdaWENO-Z. The WENO-Z non-linear weights are used for all of the methods to make the results comparable.

The third order TVD Runge-Kutta method \cite{shu1988total} is used for time advancing:
\begin{eqnarray}
&{{u}^{(1)}}  &={{u}^{n}}+\Delta tL({{u}^{n}}),\label{eq4.1}\\
&{{u}^{(2)}}  &=\frac{3}{4}{{u}^{n}}+\frac{1}{4}{{u}^{(1)}}+\frac{1}{4}\Delta tL({{u}^{(1)}}),\label{eq4.2}\\
&{{u}^{n+1}}  &=\frac{1}{3}{{u}^{n}}+\frac{2}{3}{{u}^{(2)}}+\frac{2}{3}\Delta tL({{u}^{(2)}}).\label{eq4.3}
\end{eqnarray}
Unless specified, the time step $\Delta t $ is given by:
\begin{equation}\label{eq.dt1}
\Delta t = \sigma \frac{\Delta x}{\max\limits_{i}(|u_i|+\alpha_{i})}
\end{equation}
for one dimensional cases and
\begin{equation}\label{eq.dt2}
\Delta t = \sigma\frac{\Delta t_x \Delta t_y}{\Delta t_x+ \Delta t_y},\ \Delta t_x=\frac{\Delta x}{\max\limits_{i,j}(|u_{i,j}|+\alpha_{i,j})},\ \Delta t_y=\frac{\Delta y}{\max\limits_{i,j}(|v_{i,j}|+\alpha_{i,j})}
\end{equation}
for two dimensional cases, where $\sigma$ is the Courant-Friedrichs-Lewy number.

For the TOT method, considering that the CWENO framework is different from that of the finite difference WENO, we implement the criterion \eqref{Puppo} as the switch between the characteristic-wise and the component-wise WENO-Z schemes instead of the CWENO counterparts. Hence, the computational results of TOT in the following part of the paper may be different from those in \cite{puppo2003adaptive}. Except the switch criterion calculation processes, the remaining code for TOT and AdaWENO-Z are the same in our implementation.

To better demonstrate the performance of AdaWENO-Z, we also tested SWENO-Z, i.e. the component-wise part of AdaWENO-Z, for the Sod problem, the Lax problem and the Shu-Osher problem.

The data in the following sections were obtained from running our code compiled by the GCC-gfortran compiler with the '-O2' flag enabled on an Intel Core i7 8700K 3.7GHz CPU.
\subsection{One dimensional cases}\label{subsec4.1}
\subsubsection{1D Density perturbation advection}\label{subsec4.1.0}
To measure the orders of accuracy of different schemes, the smooth density perturbation advection problem is considered. The initial condition of this problem is given by:
\begin{equation}\label{eqDensity}
\rho(x,0) = 1+0.2sin(\pi x), \quad u(x,0)=1, \quad p(x,0)=1.
\end{equation}
The exact solution is:
\begin{equation}
\rho(x,t) =1+0.2sin(\pi (x-t)),\ u(x,t) =1,\ p(x,t)  =1.
\end{equation}
The computational domain is $[0,2]$. Solutions are integrated to $t=2.0$. To rule out the effect of the time integration method on the order of accuracy, the time step is set to be $\Delta t = 0.05 \Delta x^{5/3}$.

The $L_2$ errors:
\begin{equation}
L_2 = \sqrt{\frac{\sum_{i}((\rho_i-\rho_{i,exact})^2+(u_i-u_{i,exact})^2+(p_i-p_{i,exact})^2)}{N}}
\end{equation}
 and orders of accuracy of different methods at $t=2.0$ are shown in Tab.\ref{tab:l2.1}. It can be observed that the $L_2$ errors and orders of accuracy of the tested methods are all the same for this smooth problem as essentially these methods are the same method for smooth solution. The CPU time of each method is shown in Tab.\ref{tab:cpu.1}. The CH method requires about 1.5 times more CPU time than the CP method. TOT is faster than the CH method, however, as it requires extra effort to compute the global weights along with computing all the smoothness indicators of each flux component, it is slower than the CP method. AdaWENO-Z is the fastest method as only one third of the smoothness indicators are needed to be computed.
\begin{table}[H]
   \caption{The $L_2$ errors and orders of accuracy of different methods at $t=2.0$ for problem \eqref{eqDensity}.}\label{tab:l2.1}
  \begin{center}\footnotesize
  \begin{tabular}{ccccccccc}
  \toprule
  \multirow{2}{*}{N} &
  \multicolumn{2}{c}{CP} &
  \multicolumn{2}{c}{CH} &
  \multicolumn{2}{c}{TOT} &
  \multicolumn{2}{c}{AdaWENO-Z} \\
  & $L_2$ & order & $L_2$ & order & $L_2$ & order & $L_2$ & order\\
  \midrule
  8   & 9.17E-03 & -  	& 9.17E-03 & -     & 9.17E-03 & -    & 9.17E-03  & -    \\
  16  & 3.07E-04 & 4.90 & 3.07E-04 & 4.90  & 3.07E-04 & 4.90 & 3.07E-04 & 4.90 \\
  32  & 9.81E-06 & 4.97 & 9.81E-06 & 4.97  & 9.81E-06 & 4.97 & 9.81E-06 & 4.97 \\
  64  & 3.11E-07 & 4.98 & 3.11E-07 & 4.98  & 3.11E-07 & 4.98 & 3.11E-07 & 4.98 \\
  128 & 9.76E-09 & 4.99 & 9.76E-09 & 4.99  & 9.76E-09 & 4.99 & 9.76E-09 & 4.99 \\
  256 & 3.04E-10 & 5.00 & 3.04E-10 & 5.00  & 3.04E-10 & 5.00 & 3.04E-10 & 5.00 \\
  \bottomrule
  \end{tabular}
  \end{center}
\end{table}
\begin{table}[H]
   \caption{The total CPU time (s) of different methods for problem \eqref{eqDensity}.}\label{tab:cpu.1}
  \begin{center}\footnotesize
  \begin{tabular}{ccccccccc}
  \toprule
   N  &  CP  &  CH  &  TOT  &  AdaWENO-Z \\
  \midrule
  8   & 9.385E-03 & 1.256E-02 & 1.104E-02 & 8.259E-03  \\
  16  & 4.804E-02 & 7.083E-02 & 5.607E-02 & 3.779E-02  \\
  32  & 0.324     & 0.488     & 0.376     & 0.241      \\
  64  & 2.405     & 3.638     & 2.790     & 1.722      \\
  128 & 18.471    & 28.088    & 21.355    & 13.120     \\
  256 & 144.67    & 219.973   & 166.724   & 101.069    \\
  \bottomrule
  \end{tabular}
  \end{center}
\end{table}

\subsubsection{The Sod problem}\label{subsec4.1.1}
The initial condition of the Sod problem is given by:
\begin{equation}\label{eq4.4}
	(\rho ,u,p)=\left\{ \begin{array}{*{35}{l}}
   (1,0,1) & x\le0  \\
   (0.125,0,0.1) & x>0  \\
\end{array} \right.
\end{equation}
The final solution time is $t=0.14$. The CFL number is set to be 0.1.
\begin{figure}[H]
    \begin{center}
    \subfigure[N=200]{
    \includegraphics[width=0.45\textwidth]{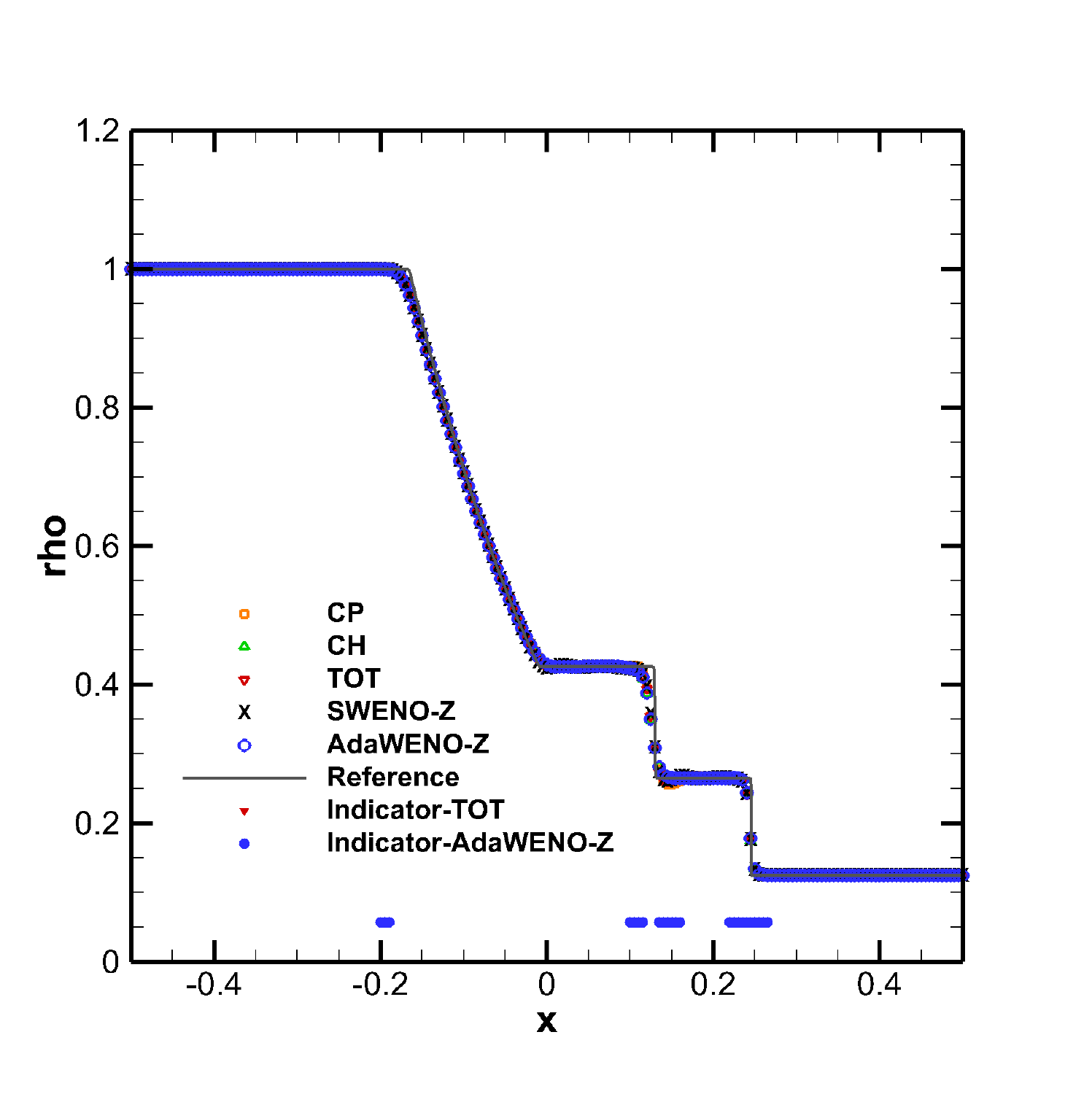}}
    \subfigure[Zoom-in view of (a) near the contact wave]{
    \includegraphics[width=0.45\textwidth]{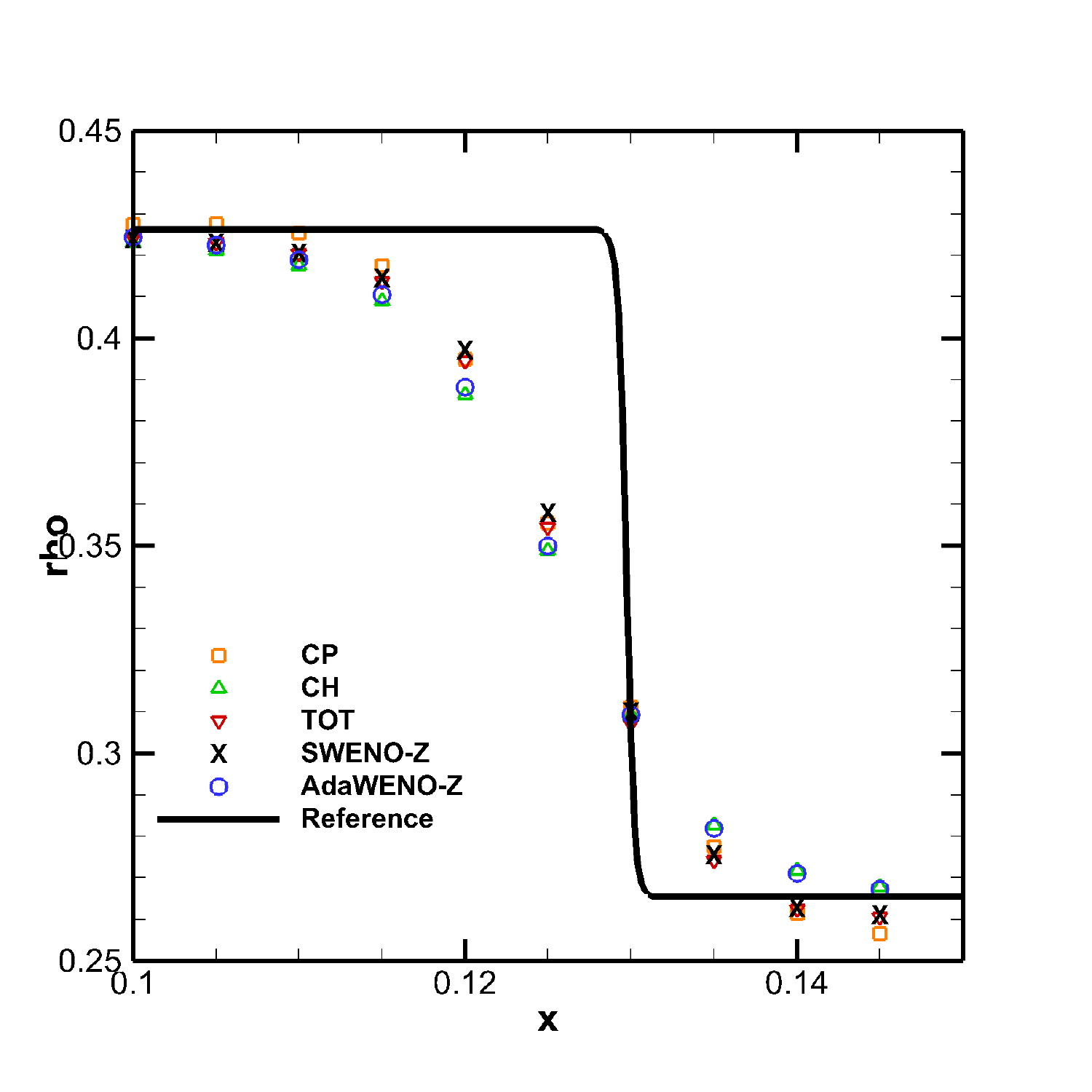}}
    \end{center}
\end{figure}
\begin{figure}[H]
    \begin{center}
    \subfigure[N=400]{
    \includegraphics[width=0.45\textwidth]{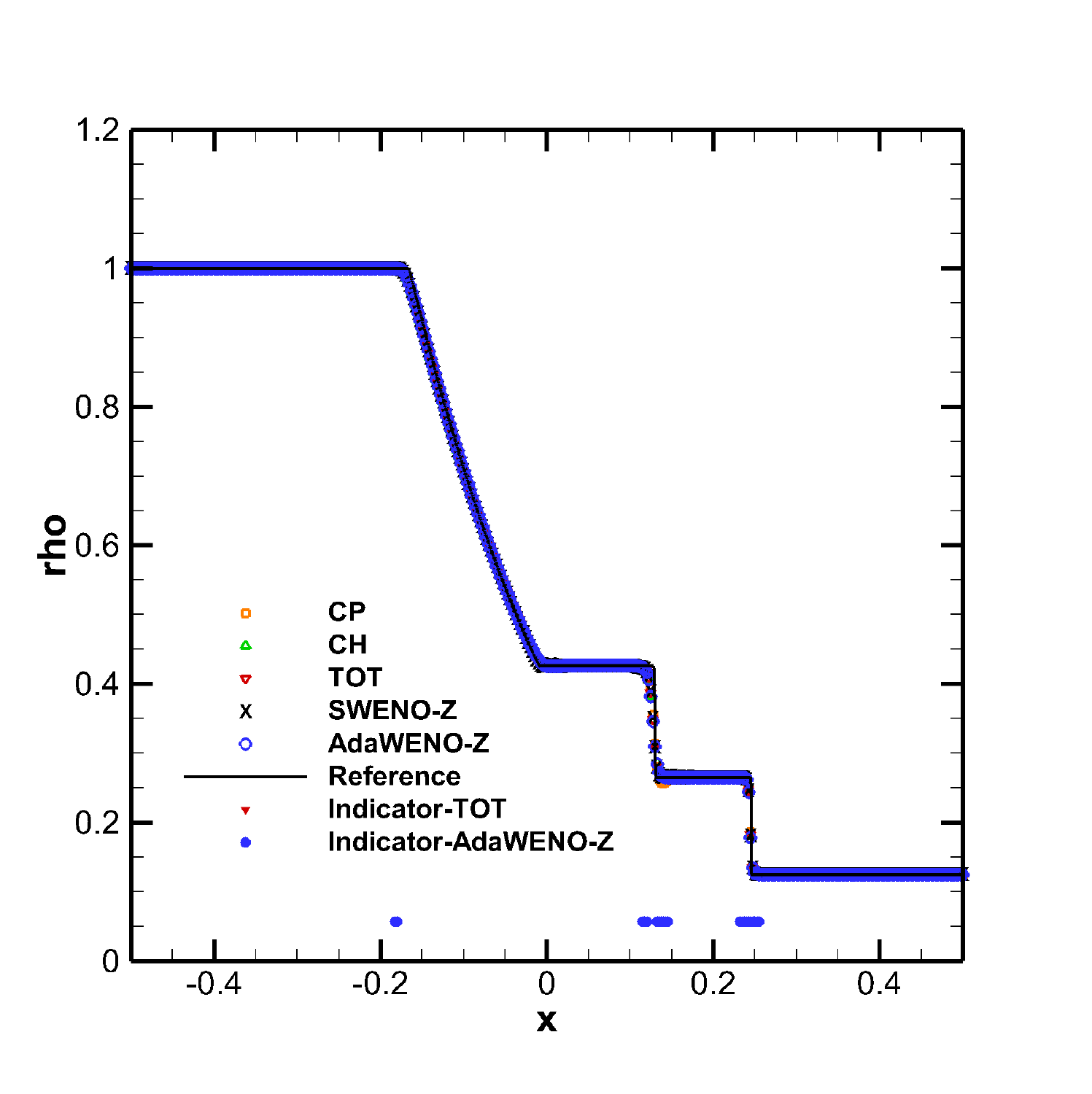}}
    \subfigure[Zoom-in view of (c) near the contact wave]{
    \includegraphics[width=0.45\textwidth]{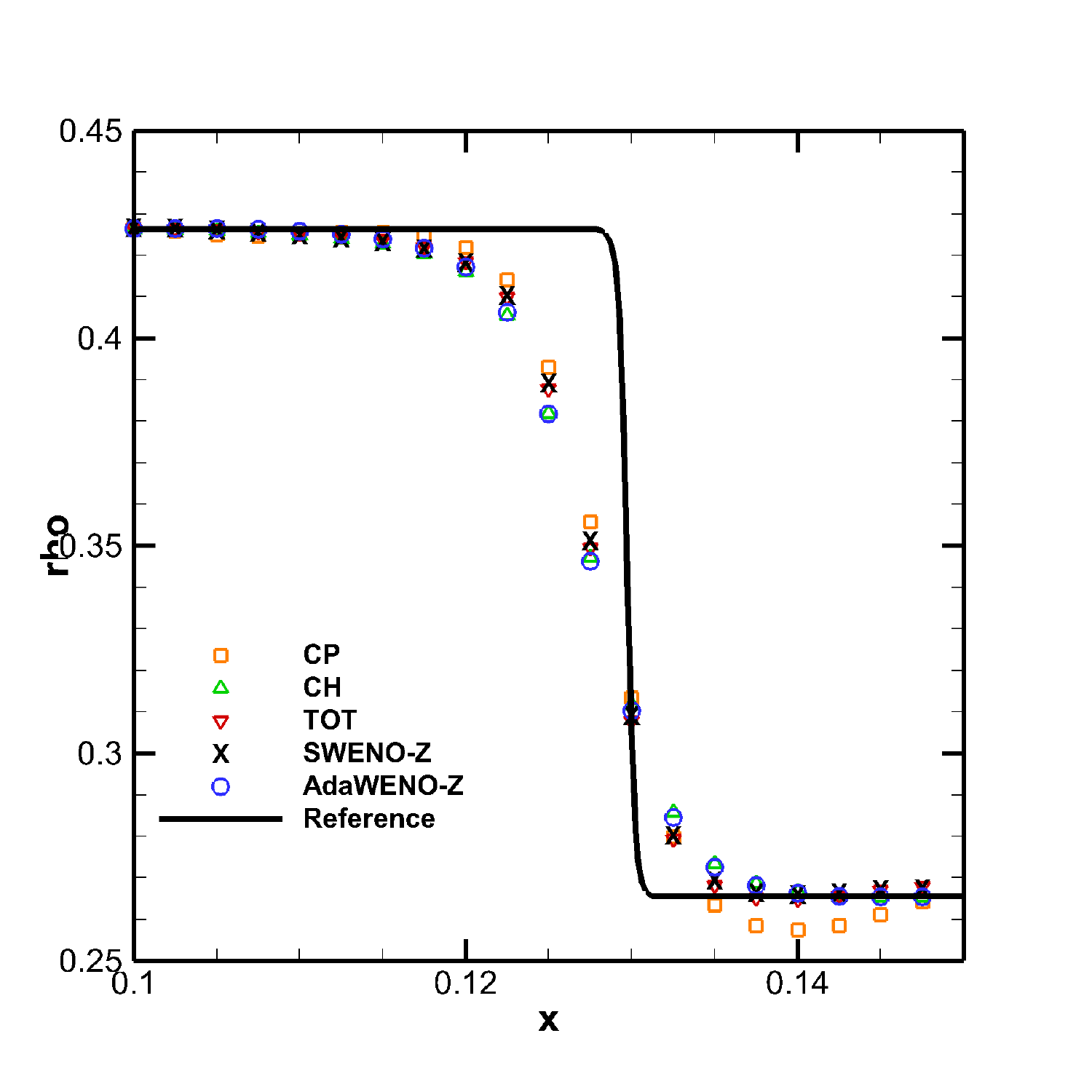}}
    \end{center}
\end{figure}
\begin{figure}[H]
    \begin{center}
    \subfigure[N=600]{
    \includegraphics[width=0.45\textwidth]{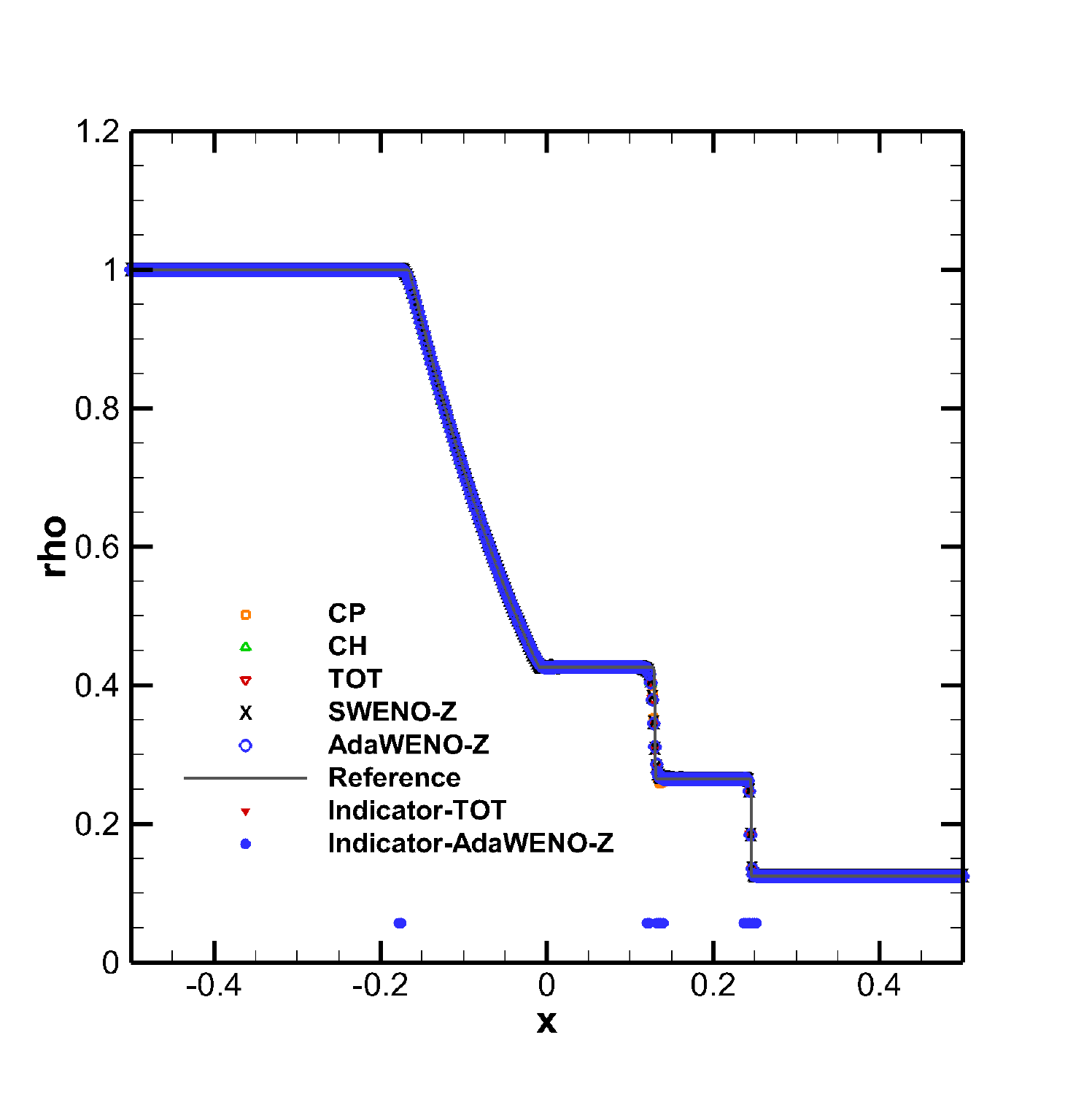}}
    \subfigure[Zoom-in view of (e) near the contact wave]{
    \includegraphics[width=0.45\textwidth]{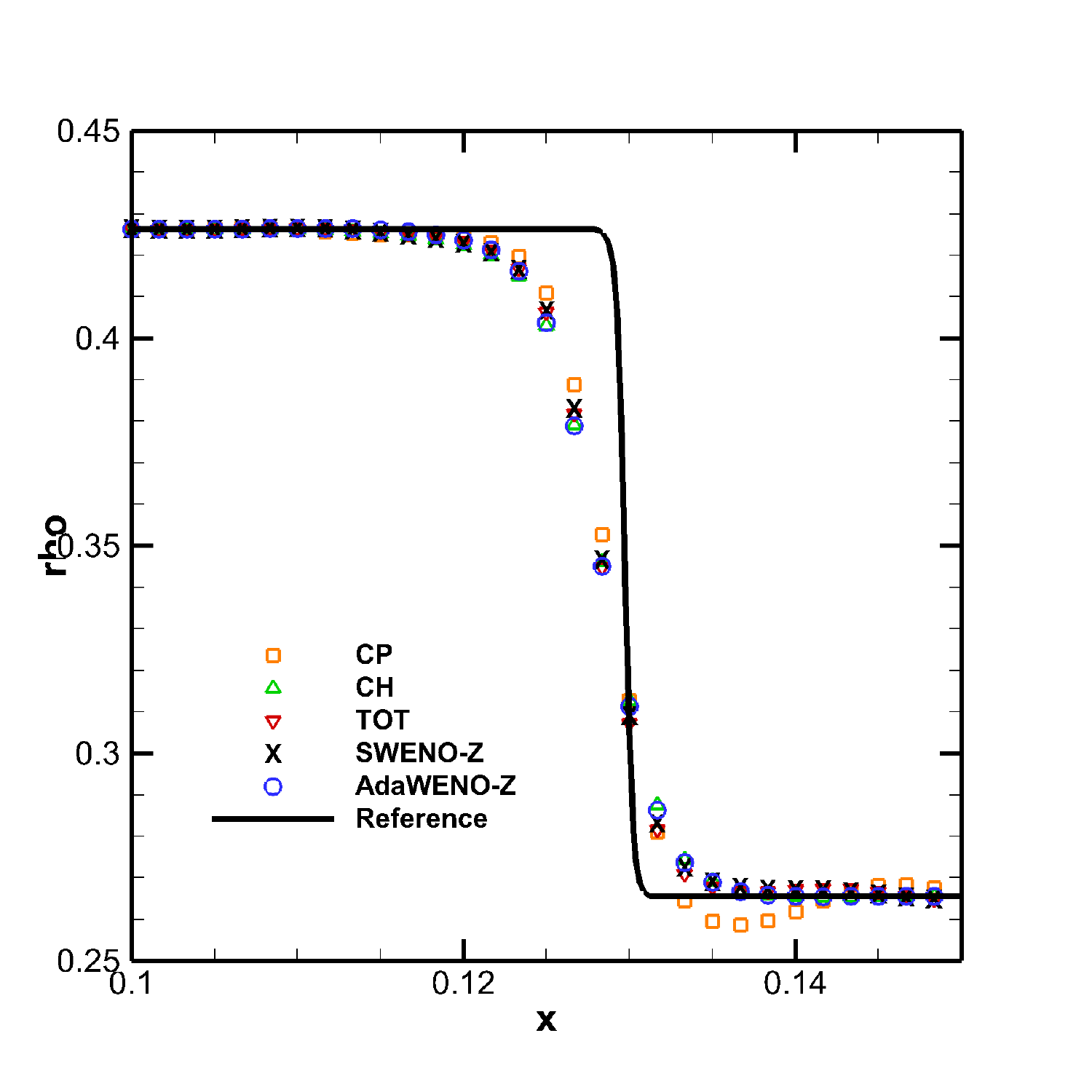}}
    \end{center}
\end{figure}
\begin{figure}[H]
    \begin{center}
    \subfigure[N=800]{
    \includegraphics[width=0.45\textwidth]{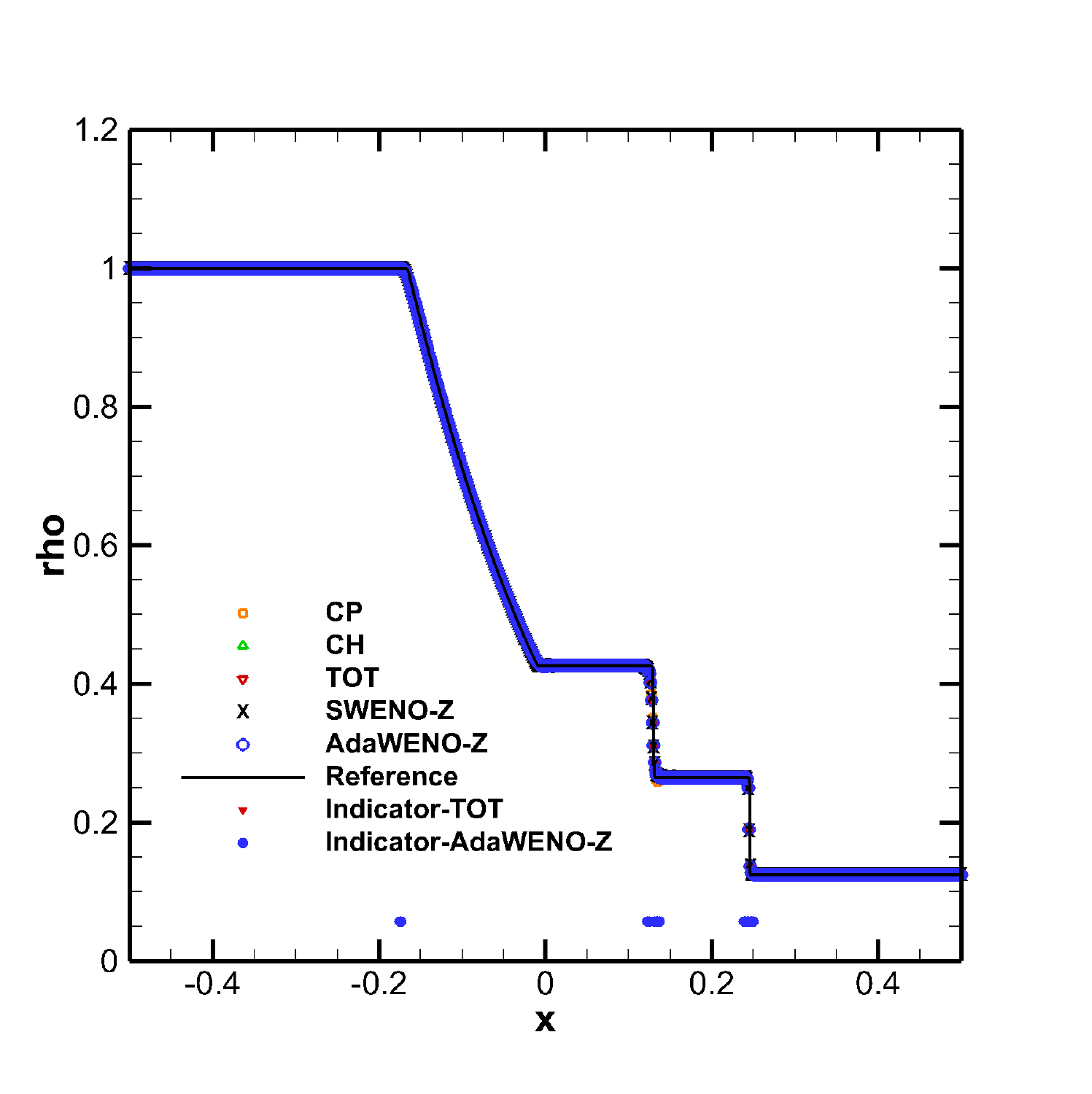}}
    \subfigure[Zoom-in view of (g) near the contact wave]{
    \includegraphics[width=0.45\textwidth]{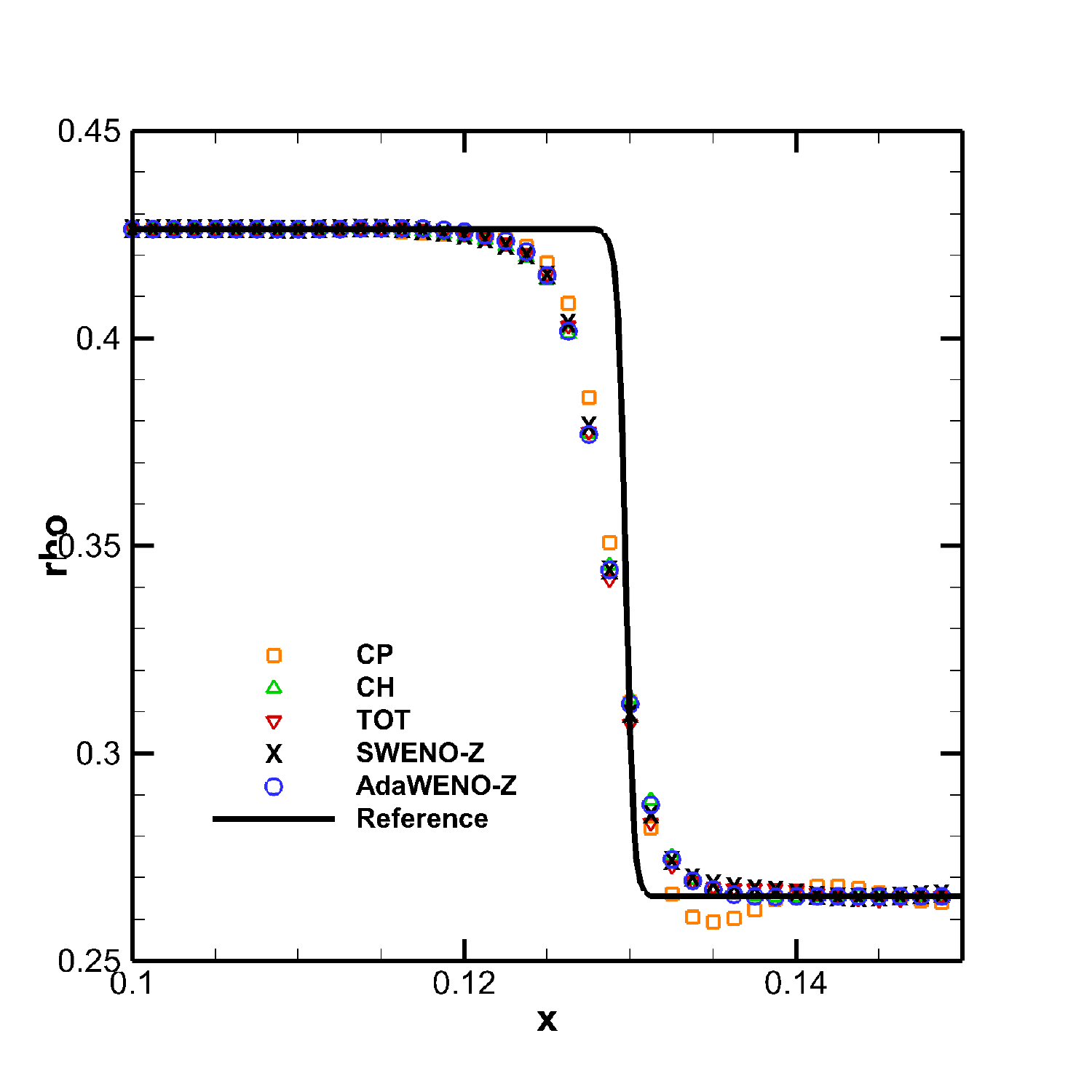}}
    \caption{Distributions of density and indicators where the characteristic-wise reconstruction is applied for the Sod problem at t=0.14.}\label{fig1d.1}
    \end{center}
\end{figure}

\begin{table}[H]
   \caption{The total CPU time (s) of different methods for problem \eqref{eq4.4}}\label{tab:cpu.2}
  \begin{center}\footnotesize
  \begin{tabular}{ccccccccc}
  \toprule
   N  &  CP  &  CH  &  TOT  & SWENO-Z &  AdaWENO-Z      \\
  \midrule
  200  & 0.108 & 0.161 & 0.124 & 7.152E-02 & 8.782E-02  \\
  400  & 0.421 & 0.634 & 0.484 & 0.283     & 0.315      \\
  600  & 0.937 & 1.453 & 1.079 & 0.631     & 0.686      \\
  800  & 1.660 & 2.632 & 1.906 & 1.114     & 1.200      \\
  \bottomrule
  \end{tabular}
  \end{center}
\end{table}

\begin{figure}[H]
\begin{center}
\includegraphics[width=0.8\textwidth]{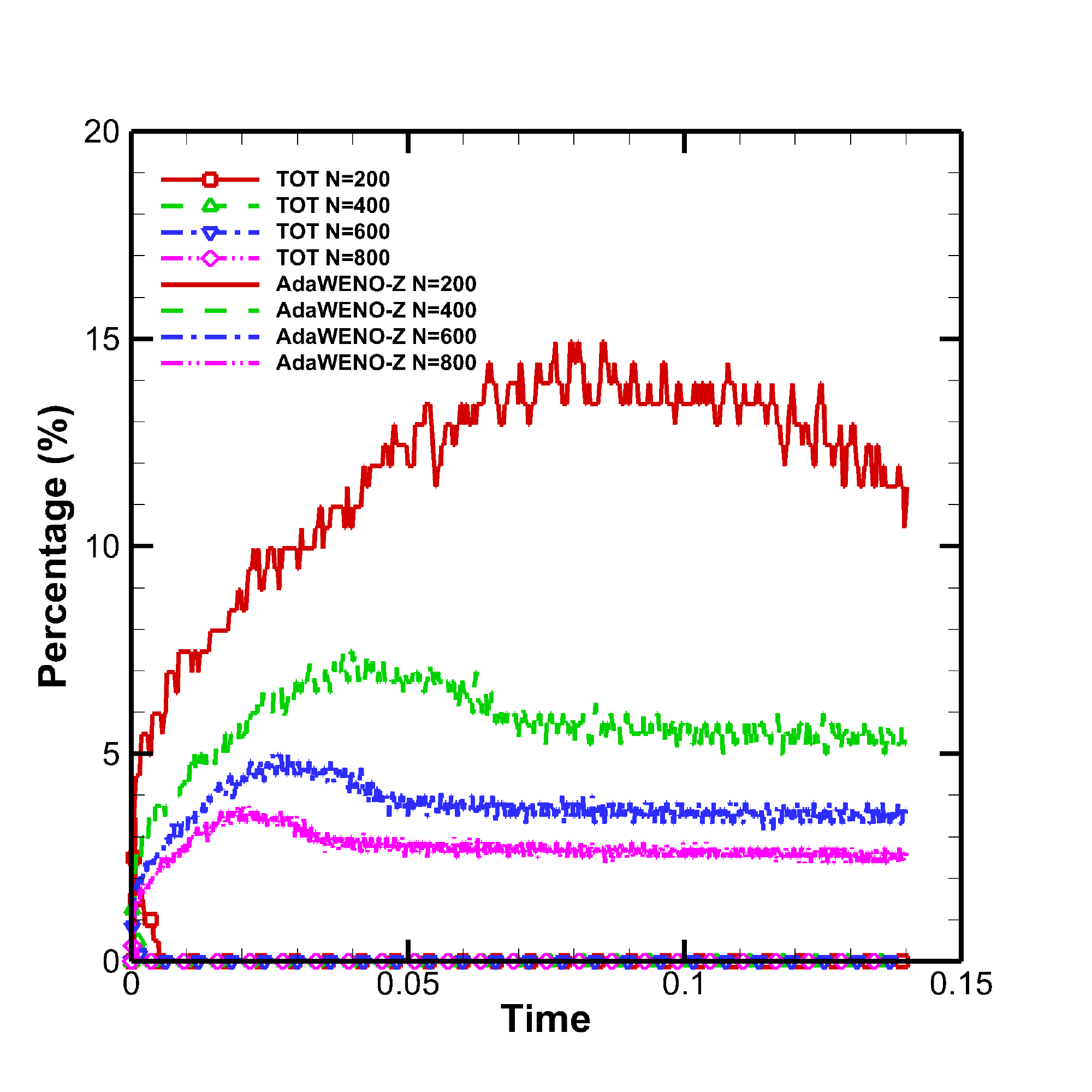}
\caption{Percentage of grids treated by CH vs time of TOT and AdaWENO-Z for the Sod problem.}
\label{fig1d.2}
\end{center}
\end{figure}

Density distributions of each method with different grid numbers are shown in Fig.\ref{fig1d.1}. The reference result is calculated by the CH method with $N=8000$. From the zoom-in figures, it can be seen that the CP method produces obvious spurious oscillations near the contact wave. TOT also shows slightly oscillatory results. AdaWENO-Z shows similar results with those of the CH method that no obvious oscillations can be observed. Markers indicating where characteristic-wise reconstruction is applied are also illustrated. AdaWENO-Z shifts to the characteristic-wise reconstruction near the expansion wave, the contact wave, and the shock wave. TOT does not mark any grid to be treated with CH at $t=0.14$. Ratios of grids that are treated by CH for TOT and AdaWENO-Z at different time are shown in Fig.\ref{fig1d.2}. It reveals that TOT treats a small portion of the grids by CH only at the beginning of the computation. Due to higher resolution of discontinuities at smaller grid size, AdaWENO-Z handles less grids with CH as the grid number increases. The CPU time of different methods are given in Tab.\ref{tab:cpu.2}. TOT is more efficient than CH but is slower than CP. Although performing characteristic-wise reconstruction at more grids, AdaWENO-Z is faster than TOT as the former spends less time on computing the smoothness indicators.

The results and CPU time of SWENO-Z reveal that, when the characteristic-wise part is cut off, AdaWENO-Z leads to faster but more oscillatory results compared to the original one. Nevertheless, the SWENO-Z results are less oscillatory than those of the CP method.
\subsubsection{The Lax problem}\label{subsec4.1.2}
The initial condition of the Lax problem is given by:
\begin{equation}\label{eq4.5}
	(\rho ,u,p)=\left\{ \begin{array}{*{35}{l}}
   (0.445,0.698,3.528) & x\le0  \\
   (0.5,0,0.571) & x>0  \\
\end{array} \right.
\end{equation}
Solutions are integrated to $t=0.13$. The CFL number is set to be 0.1.
\begin{figure}[H]
    \begin{center}
    \subfigure[N=200.]{
    \includegraphics[width=0.45\textwidth]{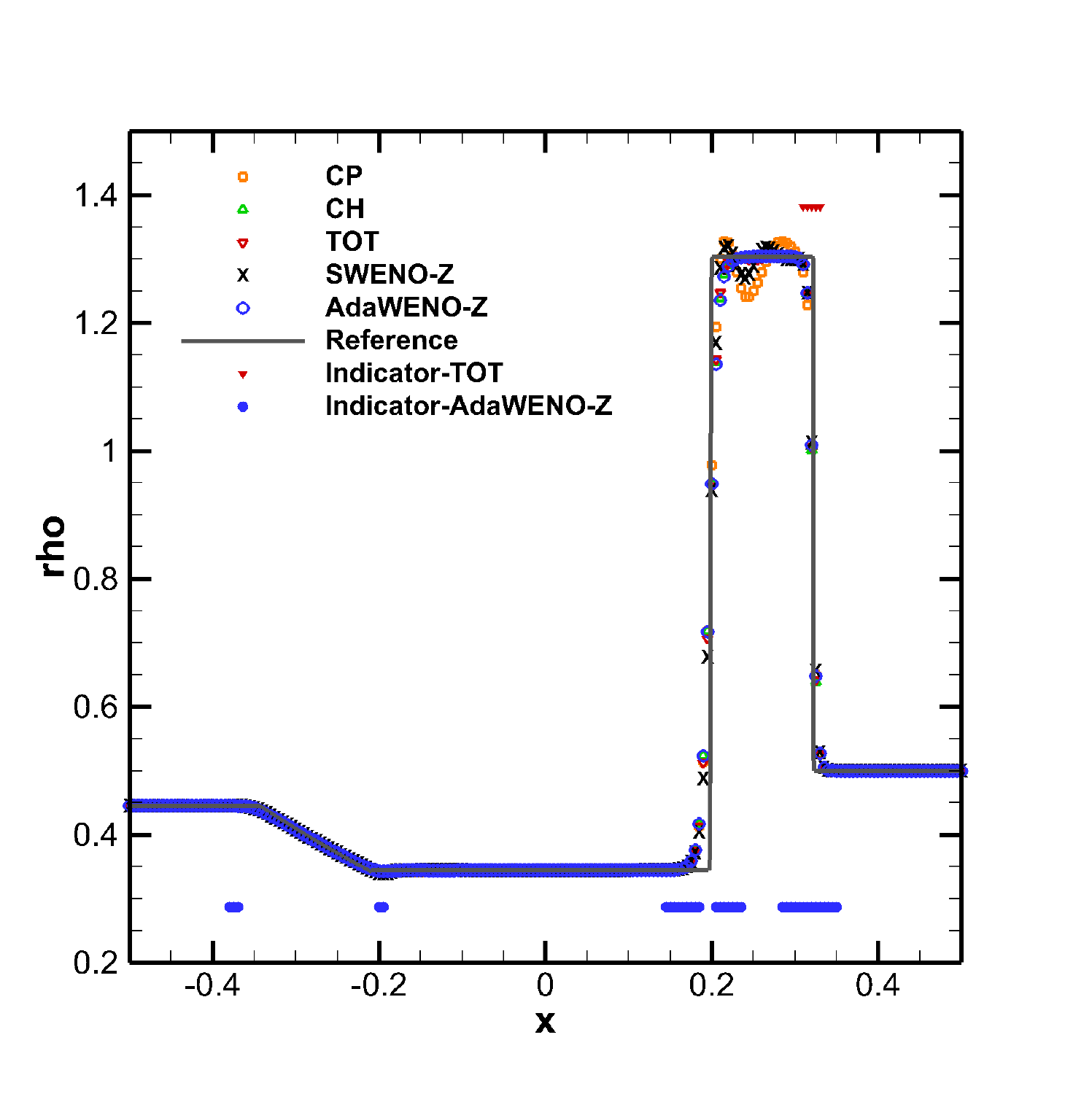}}
    \subfigure[Zoom-in view of (a) near the contact wave and the shock wave.]{
    \includegraphics[width=0.45\textwidth]{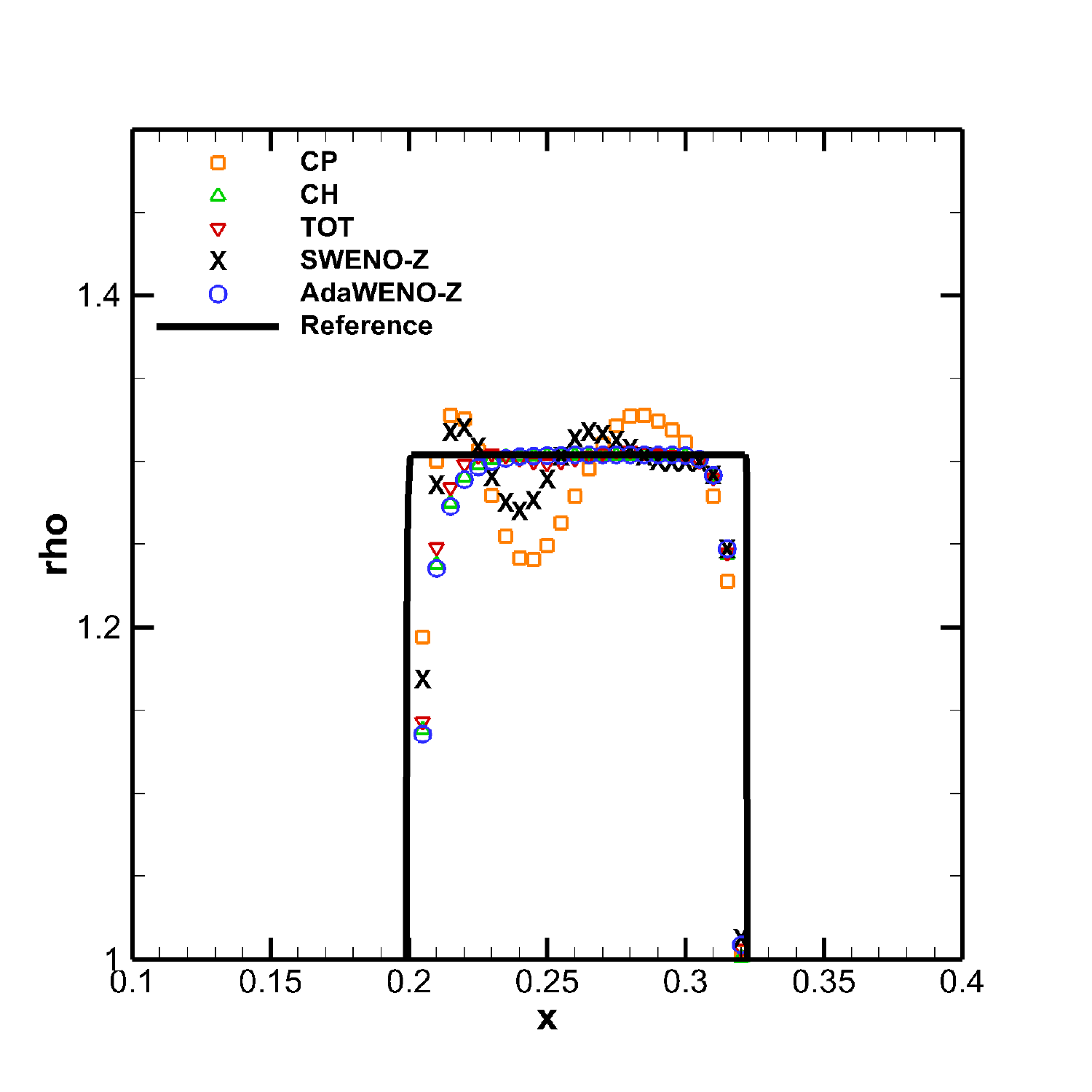}}
    \end{center}
\end{figure}
\begin{figure}[H]
    \begin{center}
    \subfigure[N=400.]{
    \includegraphics[width=0.45\textwidth]{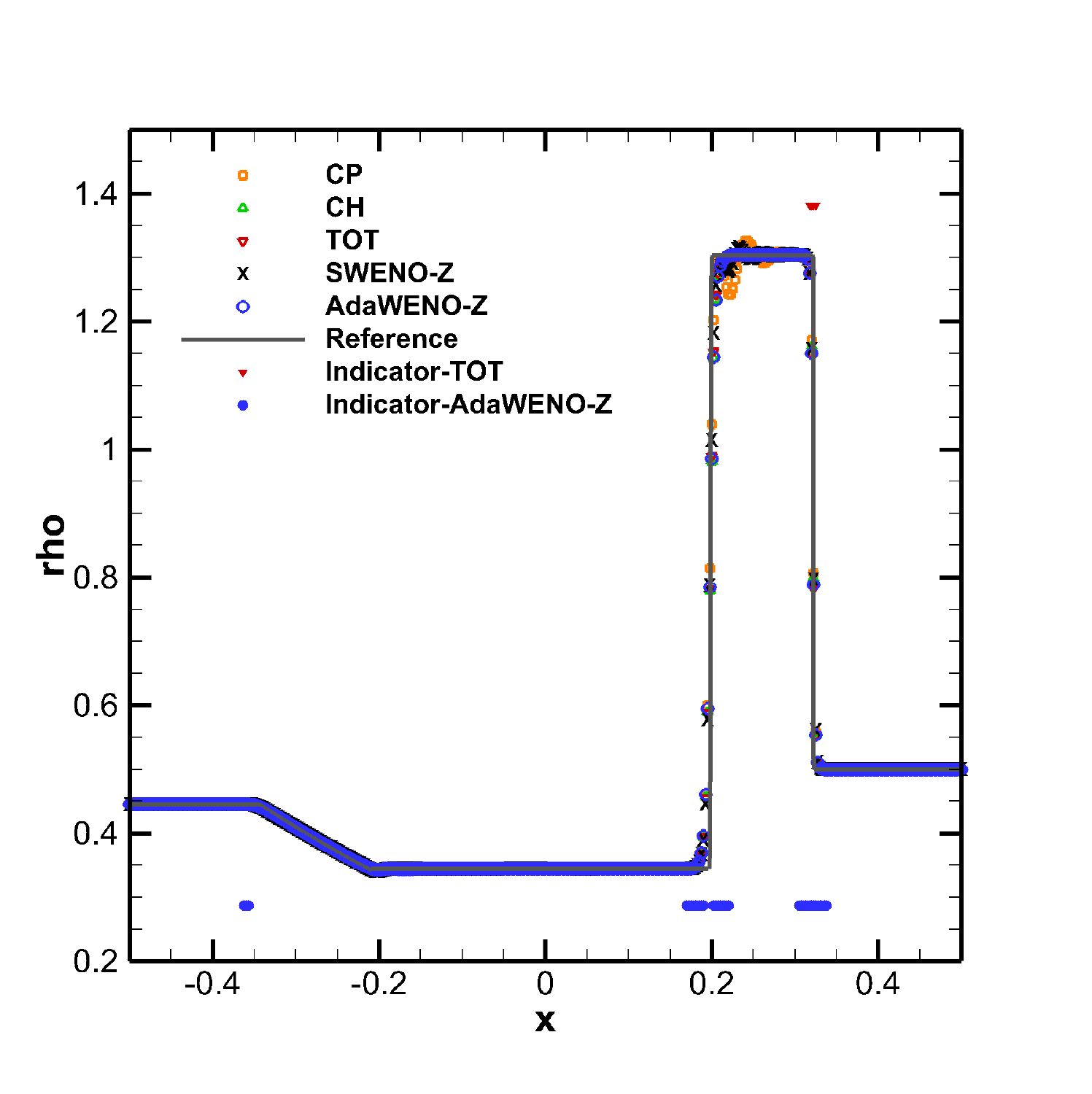}}
    \subfigure[Zoom-in view of (c) near the contact wave and the shock wave.]{
    \includegraphics[width=0.45\textwidth]{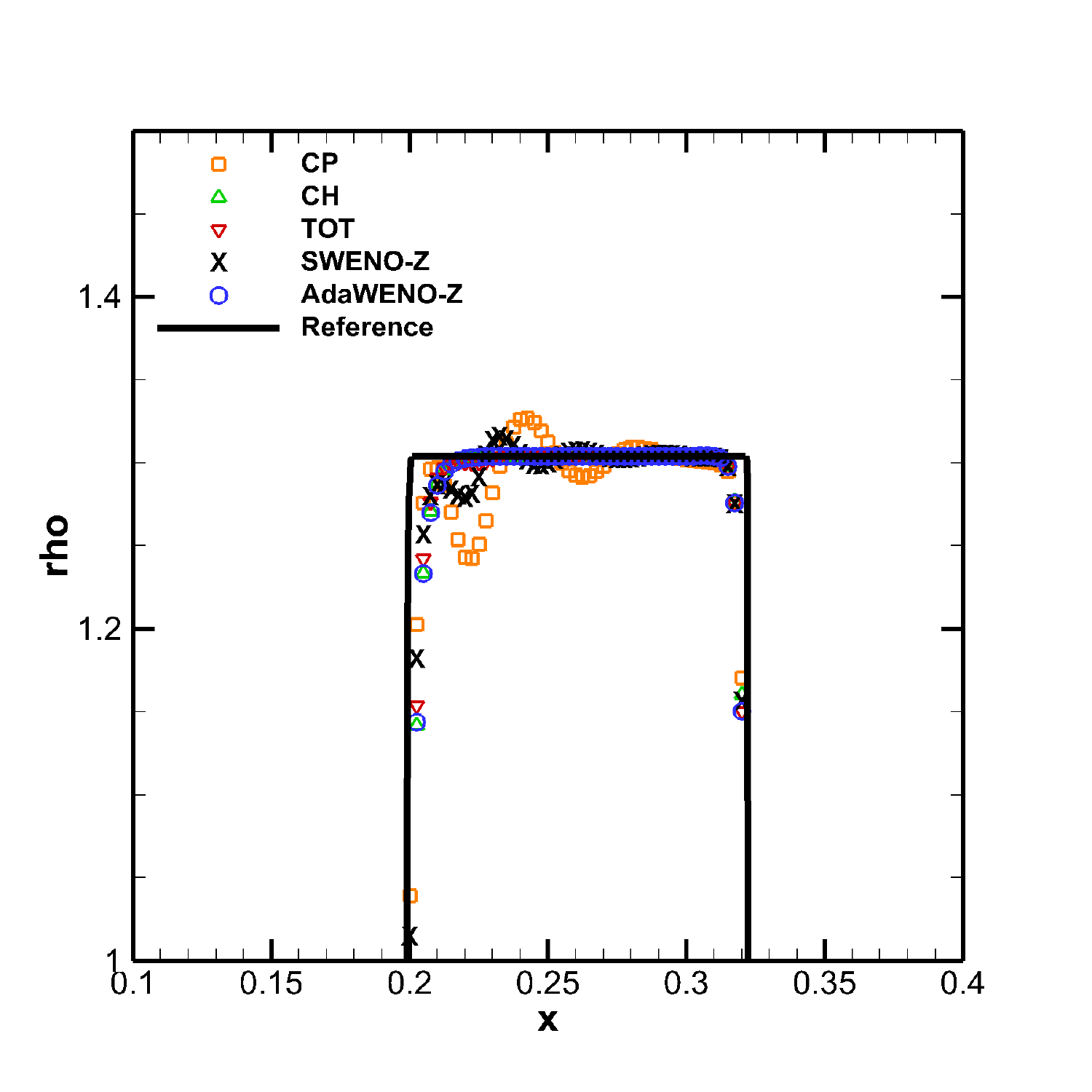}}
    \end{center}
\end{figure}
\begin{figure}[H]
    \begin{center}
    \subfigure[N=600.]{
    \includegraphics[width=0.45\textwidth]{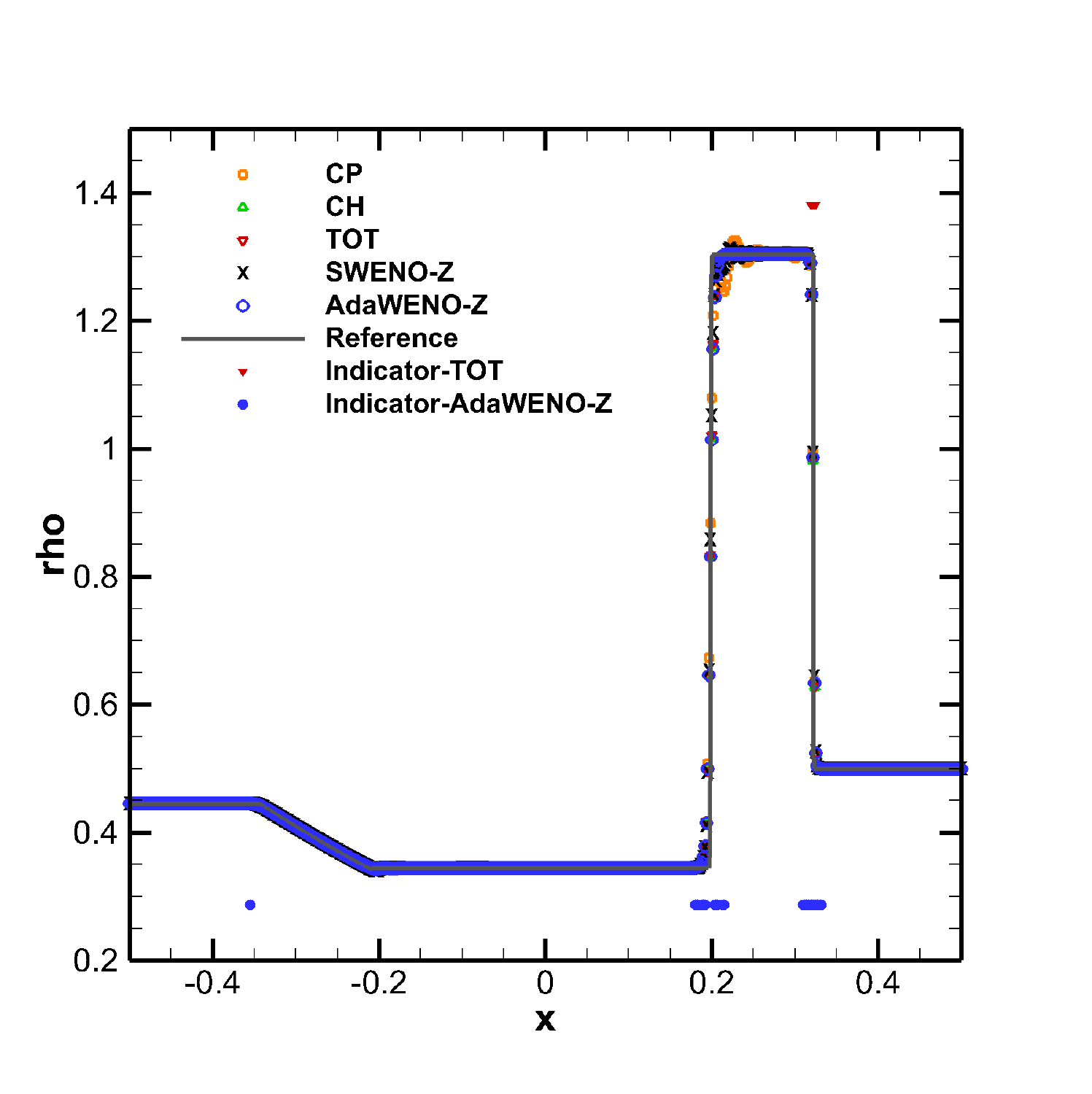}}
    \subfigure[Zoom-in view of (e) near the contact wave and the shock wave.]{
    \includegraphics[width=0.45\textwidth]{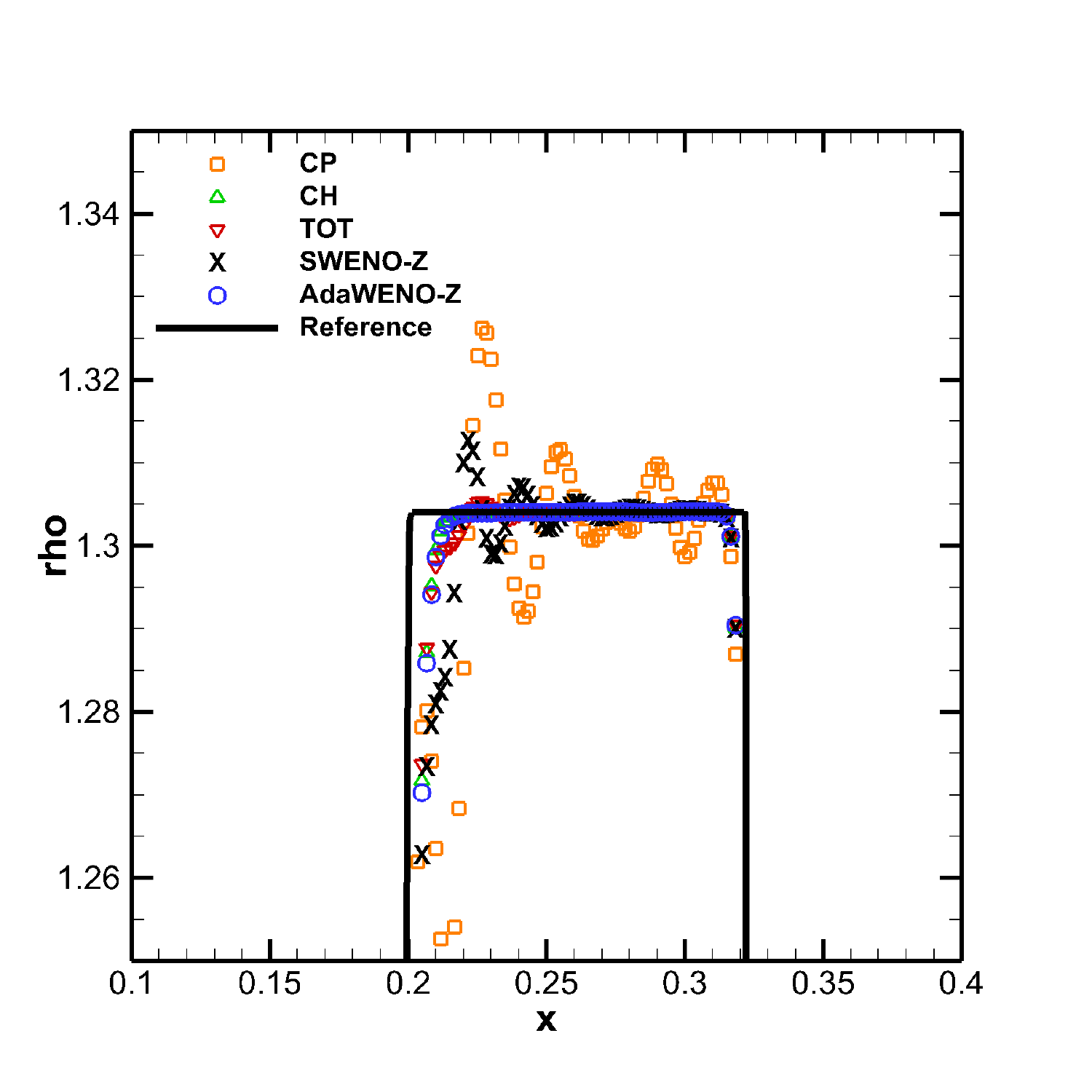}}
    \end{center}
\end{figure}
\begin{figure}[H]
    \begin{center}
    \subfigure[N=800.]{
    \includegraphics[width=0.45\textwidth]{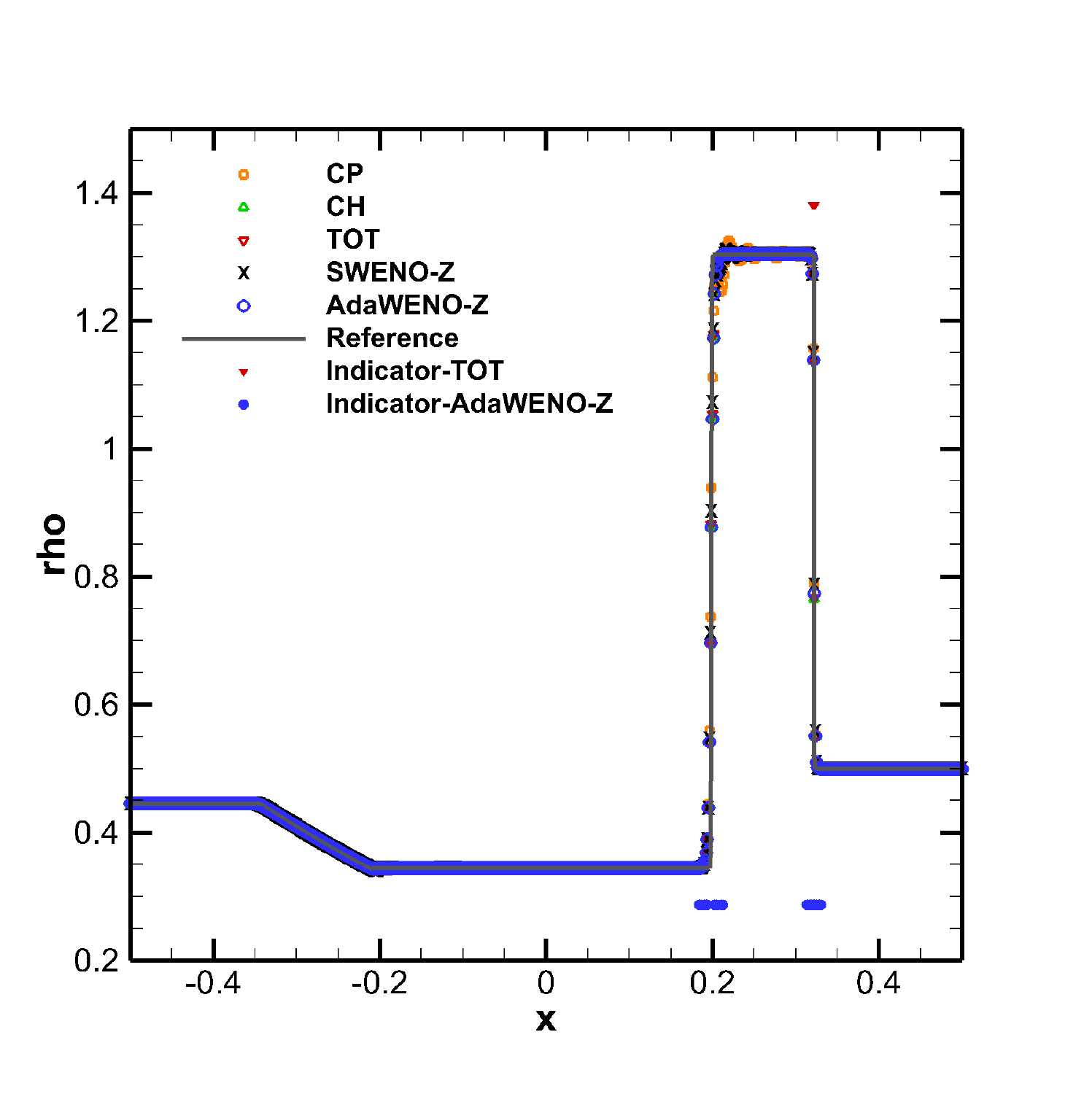}}
    \subfigure[Zoom-in view of (g) near the contact wave and the shock wave.]{
    \includegraphics[width=0.45\textwidth]{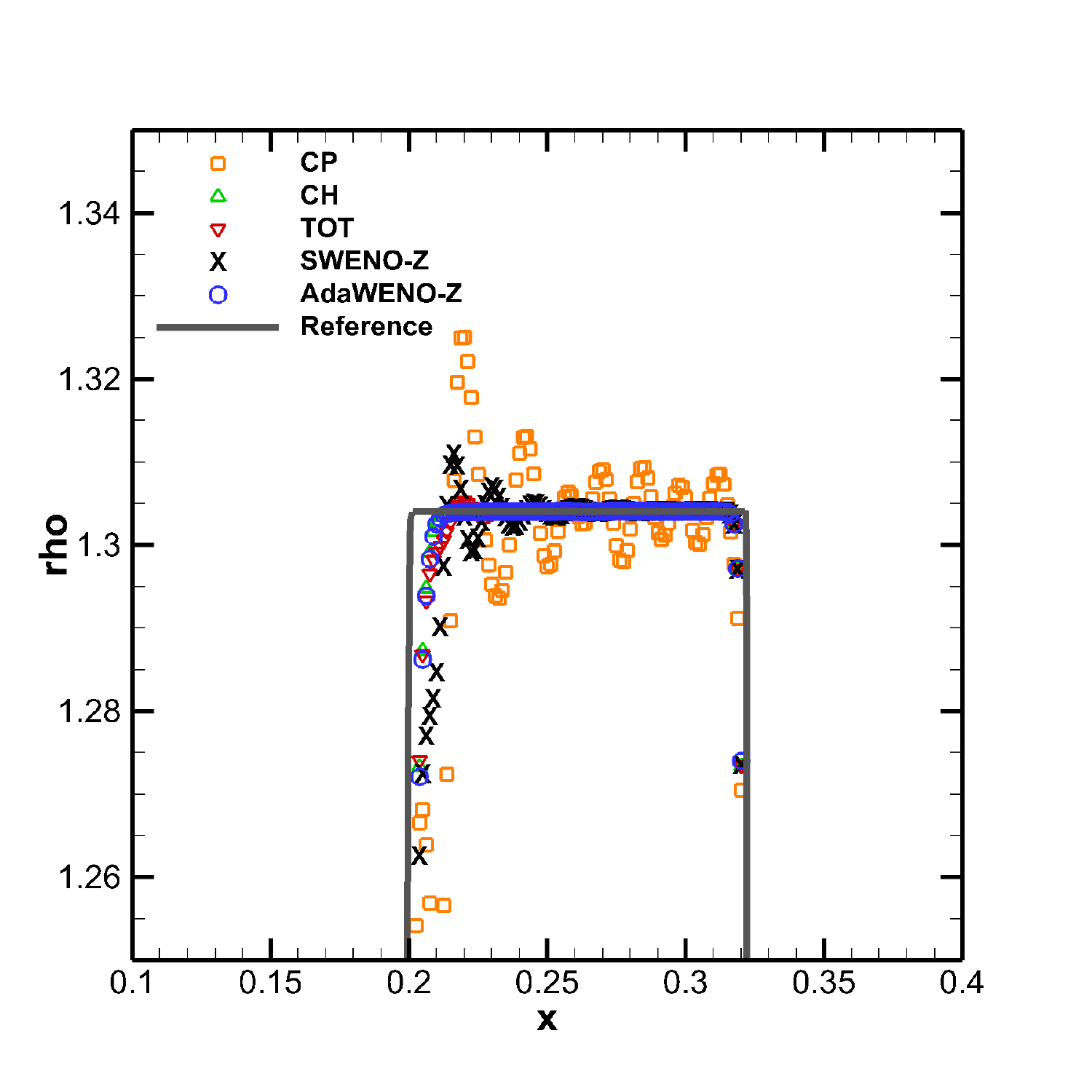}}
    \caption{Distributions of density and indicators where the characteristic-wise reconstruction is applied for the Lax problem at t=0.13.}\label{fig1d.3}
    \end{center}
\end{figure}

\begin{table}[H]
   \caption{The total CPU time (s) of different methods for problem \eqref{eq4.5}.}\label{tab:cpu.3}
  \begin{center}\footnotesize
  \begin{tabular}{ccccccccc}
  \toprule
   N  &  CP  &  CH  &  TOT  & SWENO-Z & AdaWENO-Z \\
  \midrule
  200  & 0.213 & 0.324 & 0.252 & 0.146 & 0.183  \\
  400  & 0.836 & 1.272 & 0.975 & 0.565 & 0.653  \\
  600  & 1.868 & 2.908 & 2.170 & 1.263 & 1.410      \\
  800  & 3.315 & 5.288 & 3.832 & 2.236& 2.448     \\
  \bottomrule
  \end{tabular}
  \end{center}
\end{table}
\begin{figure}[H]
\begin{center}
\includegraphics[width=0.8\textwidth]{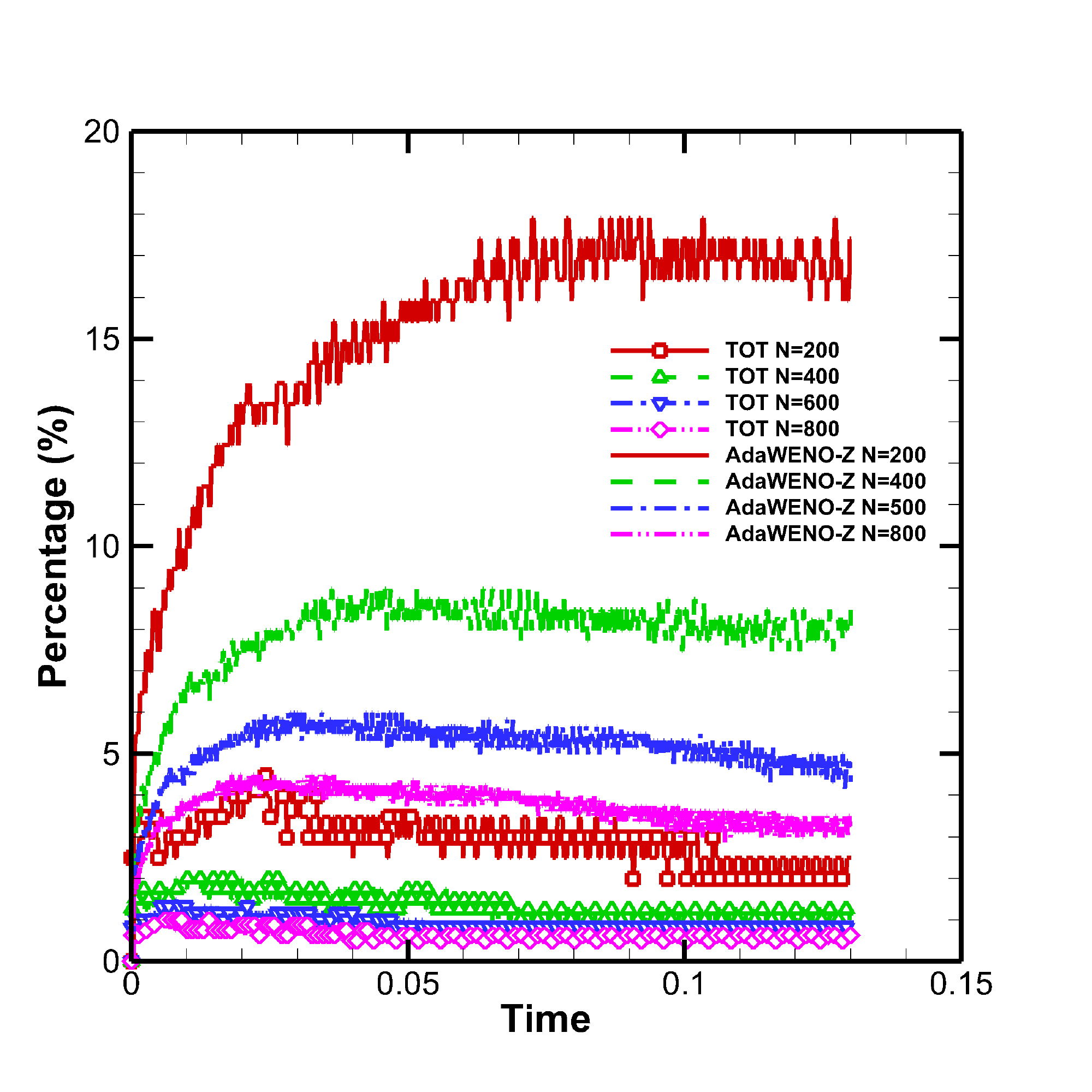}
\caption{Percentage of grids treated by CH vs time of TOT and AdaWENO-Z for the Lax problem.}
\label{fig1d.4}
\end{center}
\end{figure}

Solutions at $t=0.13$ with different grid sizes are illustrated in Fig.\ref{fig1d.3}. The reference result is computed by the CH method with $N=8000$. Similar to the sod problem results, both of the CP method and the TOT method result in spurious oscillations near the contact wave. The solutions of the CH method and AdaWENO-Z almost overlap each other. The indicators shown in Fig.\ref{fig1d.3} imply that, at $t=0.13$, TOT treats the grids near the shock wave with CH while AdaWENO-Z performs characteristic-wise reconstruction at grids in the vicinity of both the contact wave and the shock wave. The percentage of grids solved by the CH part of TOT and AdaWENO-Z at different time are given in Fig.\ref{fig1d.4}. Although more grids are marked, AdaWENO-Z is still about $25\%$ faster than CP and $35\%$ faster than TOT according to the CPU time given in Tab.\ref{tab:cpu.3}.

As with being observed from the Sod problem, without the adaptive characteristic-wise reconstruction part, although faster, SWENO-Z results to more numerical oscillations than AdaWENO-Z.
\subsubsection{Shu-Osher problem}\label{subsec4.1.3}
The initial condition of the Shu-Osher problem is given by:
\begin{equation}\label{eq4.6}
	(\rho ,u,p)=\left\{ \begin{array}{*{35}{l}}
   (\frac{27}{7},\frac{4\sqrt{35}}{9},\frac{31}{3}) & x<-4  \\
   (1+\frac{1}{5}sin5x,0,1) & x\ge-4  \\
\end{array} \right.
\end{equation}
Solutions are integrated to $t=1.8$. The CFL number is set to be 0.1.
\begin{figure}[H]
    \begin{center}
    \subfigure[N=200]{
    \includegraphics[width=0.45\textwidth]{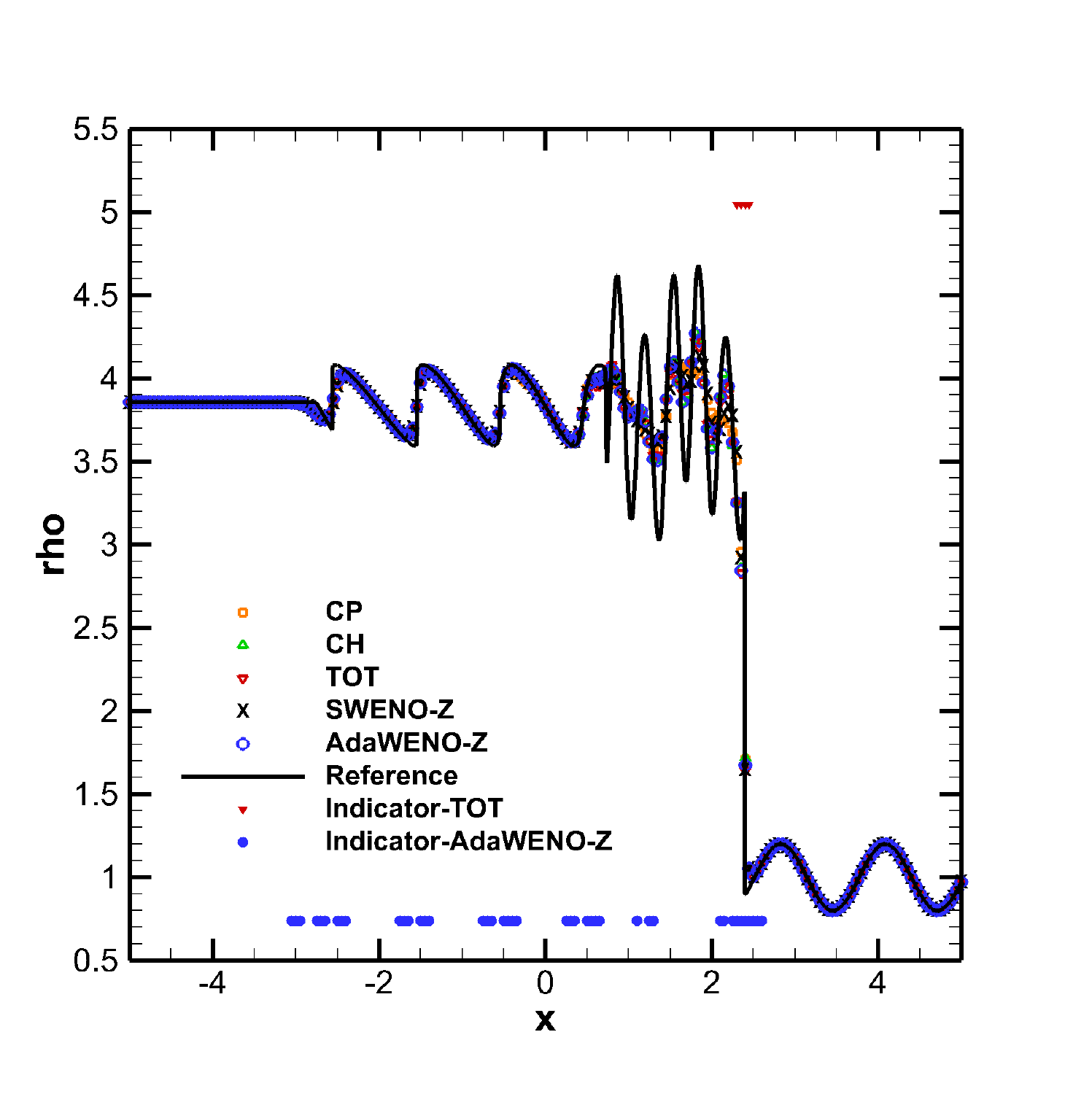}}
    \subfigure[Zoom-in view of (a) near the post-shock waves.]{
    \includegraphics[width=0.45\textwidth]{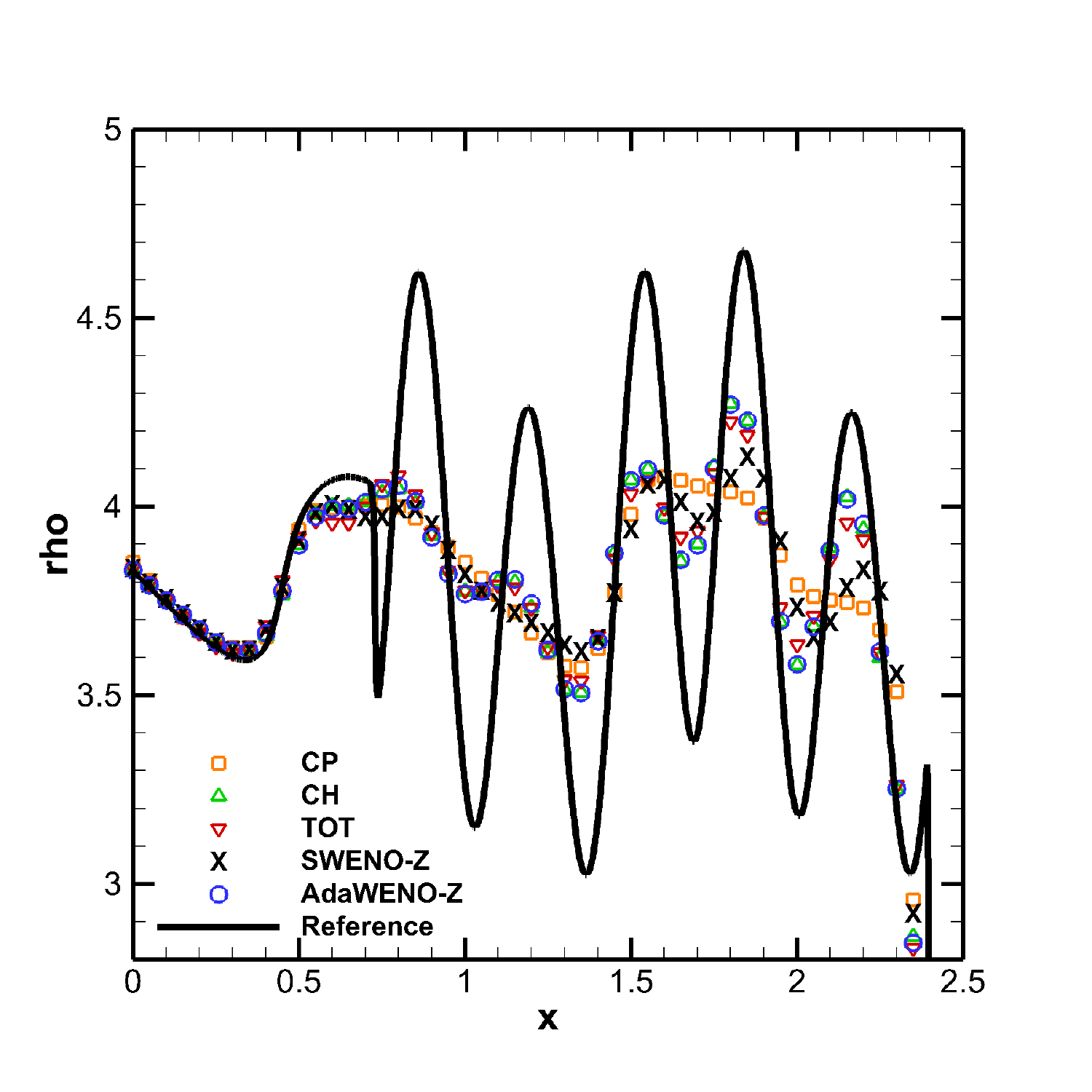}}
    \end{center}
\end{figure}
\begin{figure}[H]
    \begin{center}
    \subfigure[N=400]{
    \includegraphics[width=0.45\textwidth]{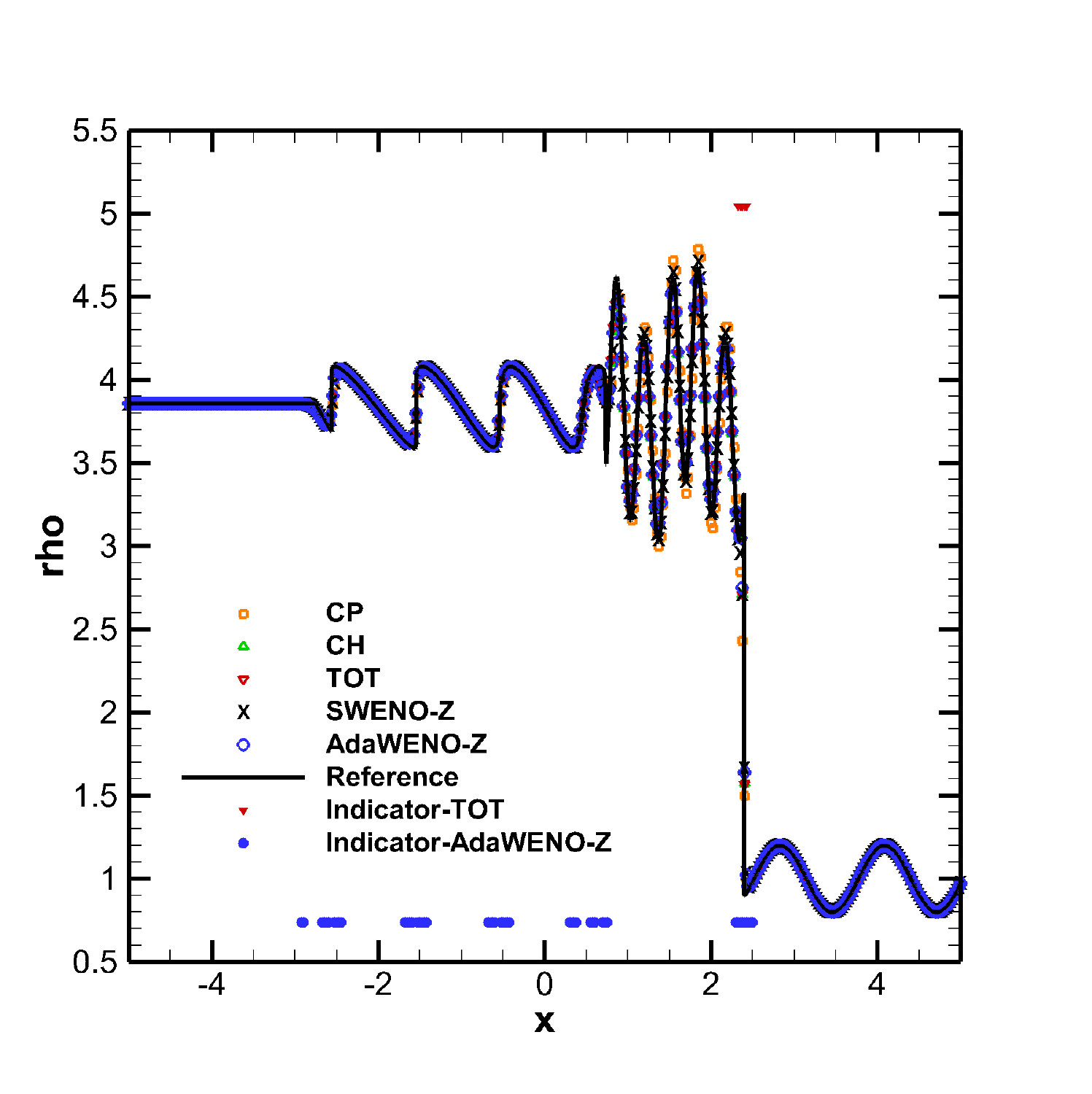}}
    \subfigure[Zoom-in view of (c) near the post-shock waves.]{
    \includegraphics[width=0.45\textwidth]{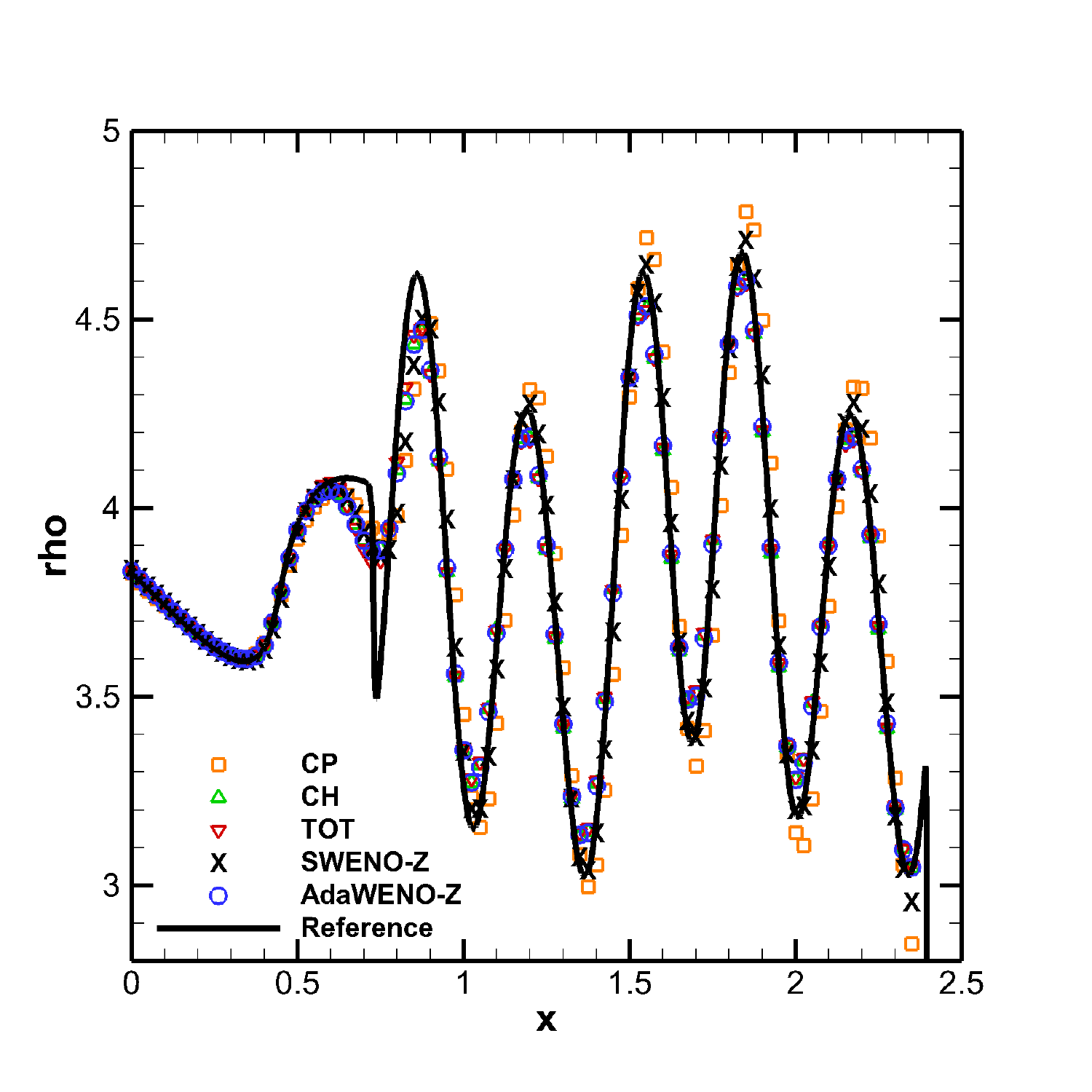}}
    \end{center}
\end{figure}
\begin{figure}[H]
    \begin{center}
    \subfigure[N=600]{
    \includegraphics[width=0.45\textwidth]{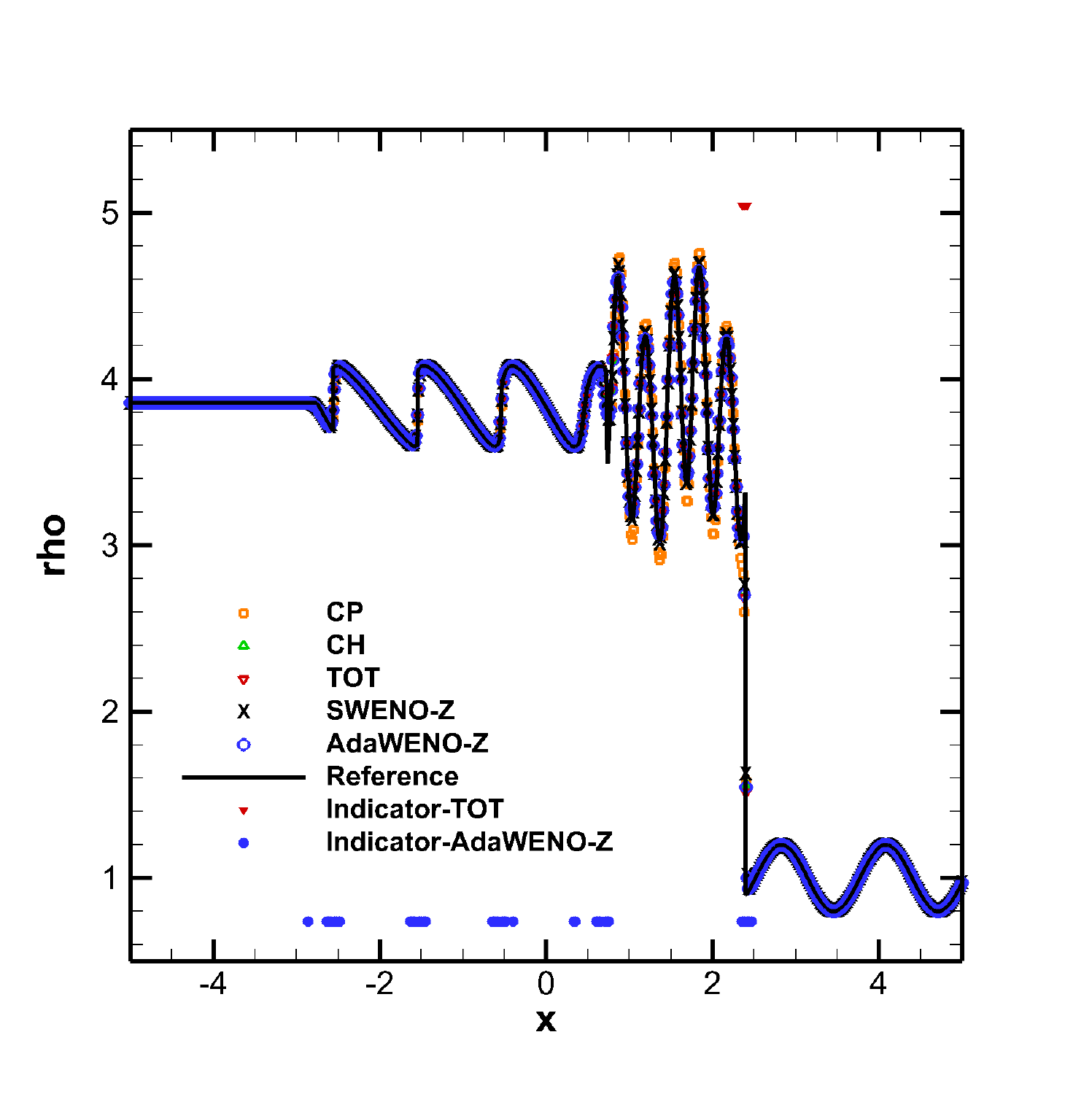}}
    \subfigure[Zoom-in view of (e) near the post-shock waves.]{
    \includegraphics[width=0.45\textwidth]{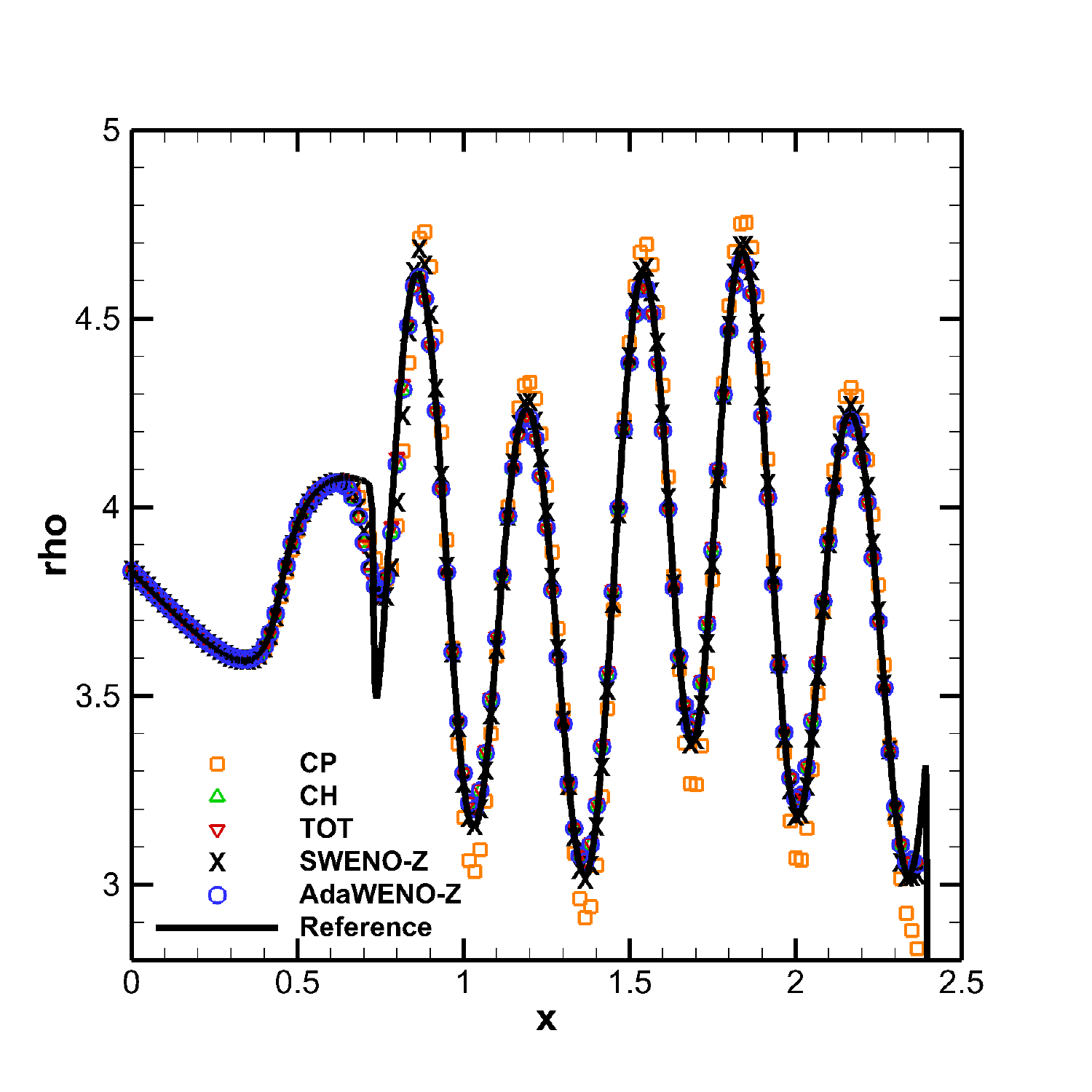}}
    \end{center}
\end{figure}
\begin{figure}[H]
    \begin{center}
    \subfigure[N=800]{
    \includegraphics[width=0.45\textwidth]{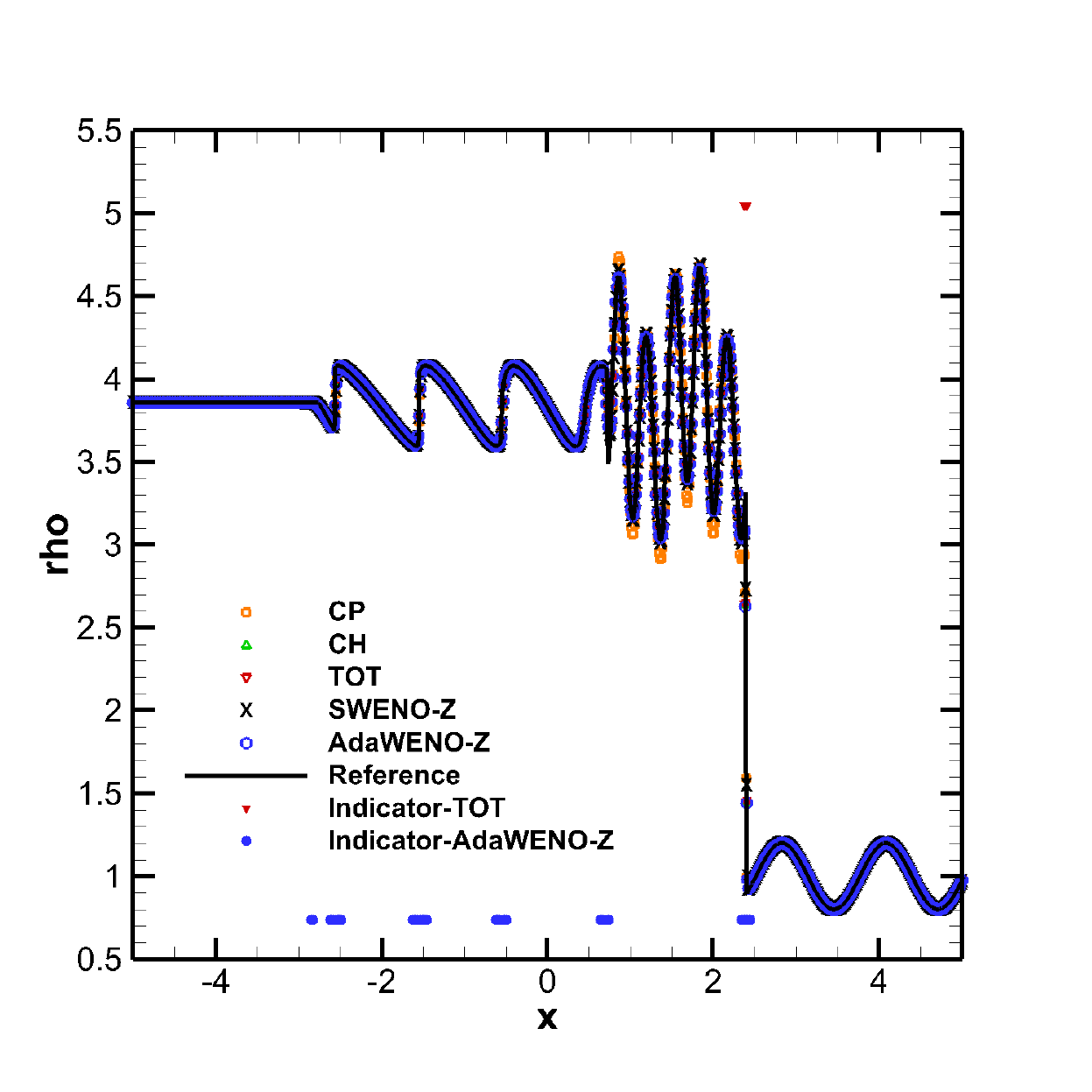}}
    \subfigure[Zoom-in view of (g) near the post-shock waves.]{
    \includegraphics[width=0.45\textwidth]{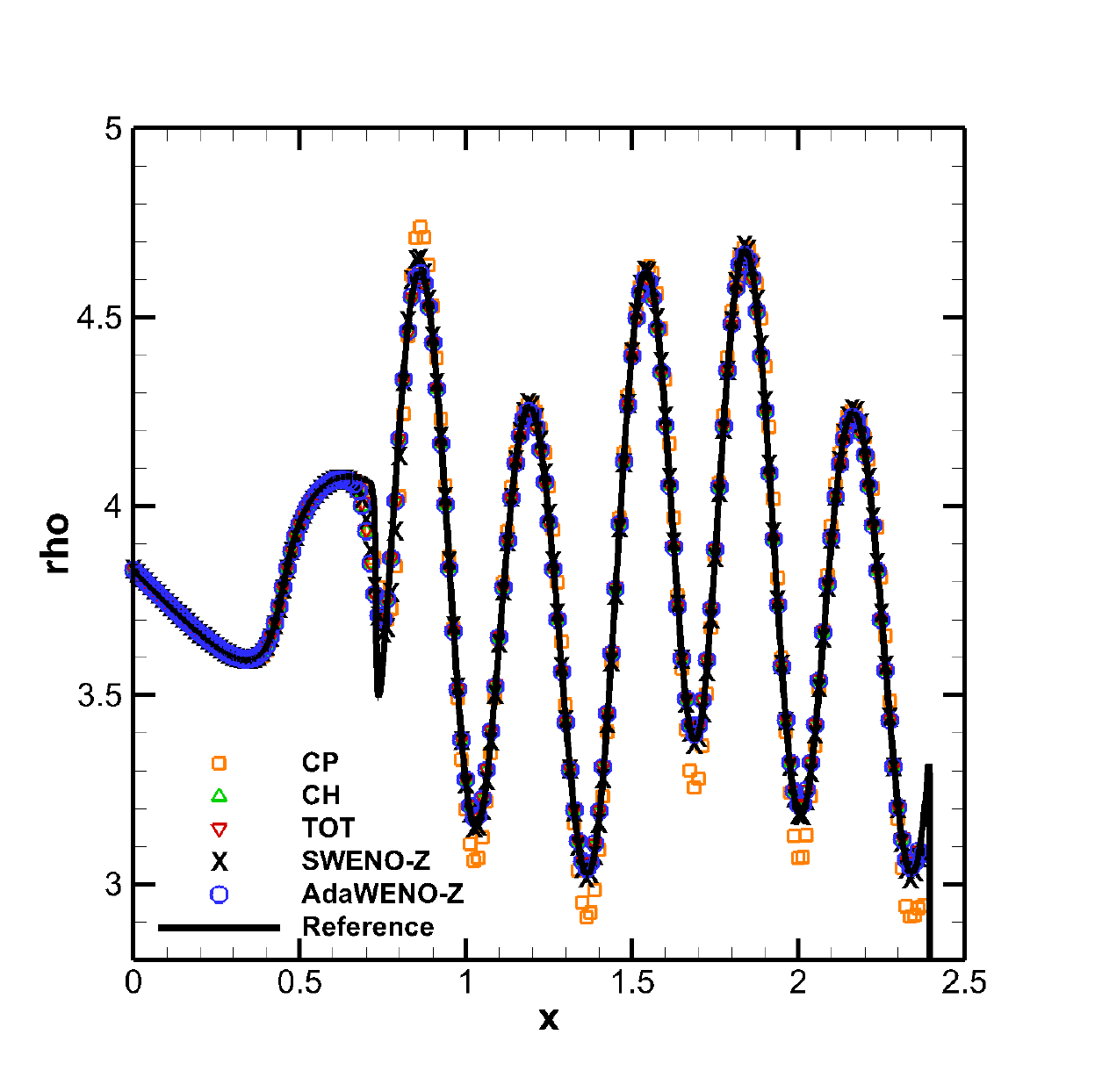}}
    \caption{Distributions of density and indicators where the characteristic-wise reconstruction is applied for the Shu-Osher problem at t=1.8.}\label{fig1d.5}
    \end{center}
\end{figure}

\begin{table}[H]
   \caption{The total CPU time (s) of different methods for problem \eqref{eq4.6}}\label{tab:cpu.4}
  \begin{center}\footnotesize
  \begin{tabular}{ccccccccc}
  \toprule
   N  &  CP  &  CH  &  TOT  &  SWENO-Z  &  AdaWENO-Z \\
  \midrule
  200  & 0.294 & 0.447 & 0.350 & 0.200 & 0.249  \\
  400  & 1.179 & 1.785 & 1.364 & 0.793 & 0.943  \\
  600  & 2.656 & 4.117 & 3.055 & 1.791 & 1.966  \\
  800  & 4.703 & 7.470 & 5.409 & 3.142 & 3.420  \\
  \bottomrule
  \end{tabular}
  \end{center}
\end{table}
\begin{figure}[H]
\begin{center}
\includegraphics[width=0.8\textwidth]{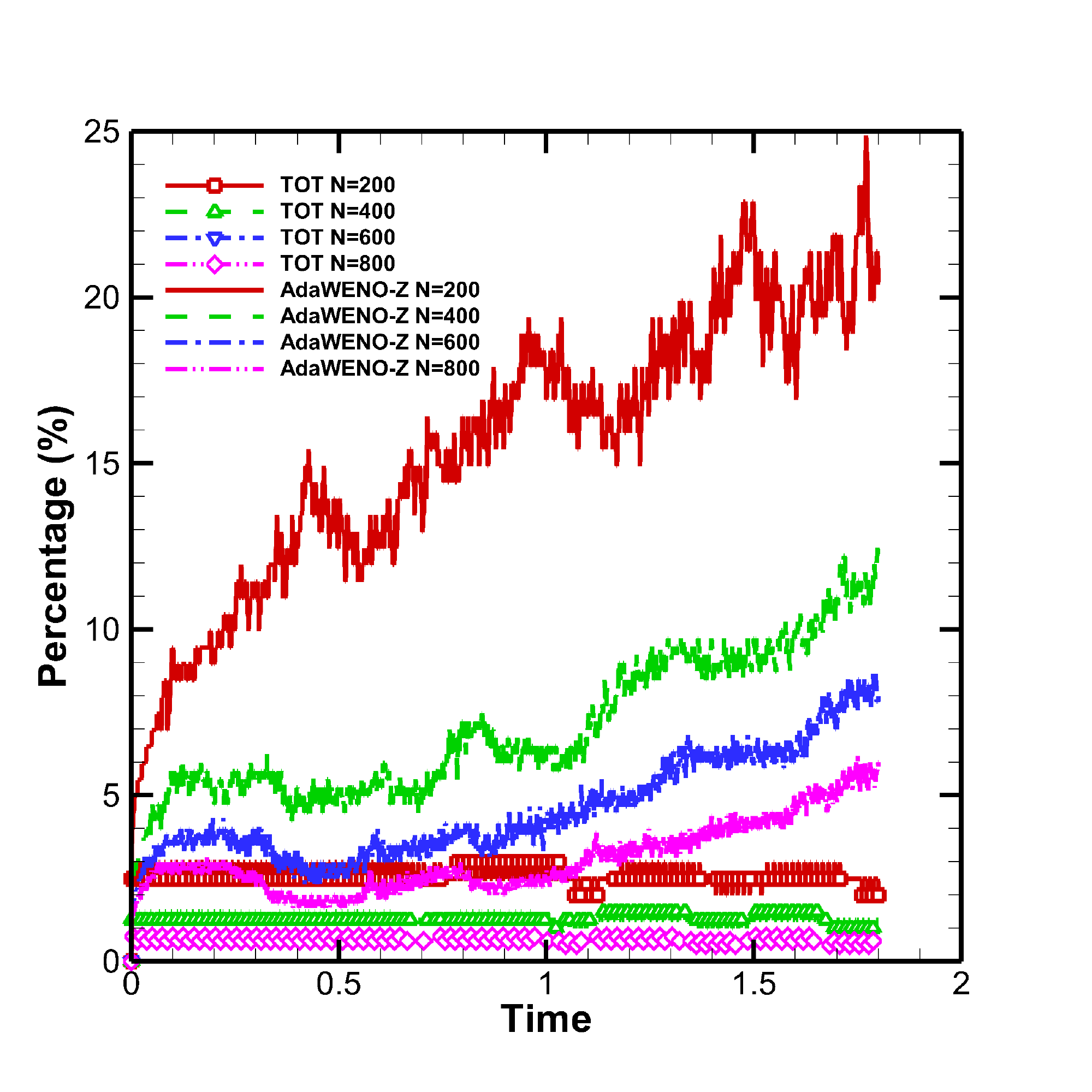}
\caption{Percentage of grids treated by CH vs time of TOT and AdaWENO-Z for the Shu-Osher problem.}
\label{fig1d.6}
\end{center}
\end{figure}

The CH method is used to give the reference result with grid number $N=8000$. Solutions of different methods at $t=1.8$ are shown in Fig.\ref{fig1d.5}. When the grid number N is 200, the CP method is unable to resolve the short waves while the other three methods give better resolutions. The short waves are well resolved when the grid number is more than 200 for these fifth order schemes. However, the CP method leads to overshoot of the post-shock waves as the grid size decreases. The CH method, the TOT method, and AdaWENO-Z show almost coinciding solutions. The indicators drawn in Fig.\ref{fig1d.5} reveal that the TOT method discerns the main shock wave and shifts to the characteristic-wise reconstruction at $t=1.8$. AdaWENO-Z marks grids near not only the main shock wave but also those weaker shock waves behind the short-wave region to be treated with CH. Tab.\ref{tab:cpu.4} shows the CPU time of different methods. AdaWENO-Z is more efficient than the other three methods. The SWENO-Z results implies that although no characteristic-wise reconstruction is performed, sharing the WENO weights among each component reduce the overshoot.

\subsection{Two dimensional cases}\label{subsec4.2}
\subsubsection{2D Density perturbation advection}\label{subsec4.2.1}
The initial condition of this problem is given by:
\begin{equation}\label{eqDensity2d}
\rho(x,y,0) = 1+0.2sin(\pi x), \quad u(x,y,0)=1, \quad v(x,y,0)=0, \quad p(x,y,0)=1.
\end{equation}
It is a simple extension of Eq.\eqref{eqDensity} to two dimensional. The exact solution is:
\begin{equation}
\rho(x,y,t) =1+0.2sin(\pi (x-t)),\ u(x,y,t) =1,\ v(x,y,t) =0,\ p(x,y,t)  =1.
\end{equation}
The computational domain is $[0,2]\times[0,2]$. Solution is integrated to $t=2.0$. To rule out the effect of the time integration method on the order of accuracy, the time step is set to be $\Delta t = 0.05 h^{5/3}$ where $h=\Delta x = \Delta y$.

The $L_2$ errors:
\begin{equation}
L_2 = \sqrt{\frac{\sum_{i}((\rho_i-\rho_{i,exact})^2+(u_i-u_{i,exact})^2+(v_i-v_{i,exact})^2+(p_i-p_{i,exact})^2)}{N_xN_y}}
\end{equation}
 and orders of accuracy of different methods at $t=2.0$ are shown in Tab.\ref{tab:l2.2d}. As with the one dimensional results, for smooth problem, the four tested methods obtain approximately the same $L_2$ error and order of accuracy. The total CPU time of different methods are given in Tab.\ref{tab:cpu.2d}. AdaWENO-Z requires the least CPU time among all the methods.
\begin{table}[H]
   \caption{The $L_2$ errors and orders of accuracy of different methods at $t=2.0$ for problem \eqref{eqDensity2d}.}\label{tab:l2.2d}
  \begin{center}\footnotesize
  \begin{tabular}{ccccccccc}
  \toprule
  \multirow{2}{*}{$(N_x,N_y)$} &
  \multicolumn{2}{c}{CP} &
  \multicolumn{2}{c}{CH} &
  \multicolumn{2}{c}{TOT} &
  \multicolumn{2}{c}{AdaWENO-Z} \\
  & $L_2$ & order & $L_2$ & order & $L_2$ & order & $L_2$ & order\\
  \midrule
  (32,32)    & 1.11E-05 & -    & 1.11E-05 & -     & 1.11E-05 & -    & 1.11E-05  & -    \\
  (64,64)    & 3.48E-07 & 4.99 & 3.47E-07 & 4.99  & 3.47E-07 & 4.99 & 3.47E-04 & 4.99 \\
  (128,128)  & 1.09E-08 & 5.00 & 1.08E-08 & 5.00  & 1.08E-08 & 5.00 & 1.08E-08 & 5.00 \\
 (256,256)   & 3.39E-10 & 5.00 & 3.39E-10 & 5.00  & 3.39E-10 & 5.00 & 3.39E-10 & 5.00 \\
  \bottomrule
  \end{tabular}
  \end{center}
\end{table}
\begin{table}[H]
   \caption{The total CPU time (s) of different methods for problem \eqref{eqDensity2d}.}\label{tab:cpu.2d}
  \begin{center}\footnotesize
  \begin{tabular}{ccccccccc}
  \toprule
   $(N_x,N_y)$  &  CP  &  CH  &  TOT  &  AdaWENO-Z \\
  \midrule
  (32,32)  & 0.72     & 1.02     & 0.78     & 0.40      \\
  (64,64)  & 8.63     & 12.42     & 9.56    & 4.12      \\
  (128,128) & 107.87    & 156.56    & 117.63    & 51.34     \\
  (256,256) & 1368.26    & 1999.39   & 1487.42   &  667.04    \\
  \bottomrule
  \end{tabular}
  \end{center}
\end{table}

\subsubsection{Double Mach reflection}\label{subsec4.2.2}
The double mach reflection test is a mimic of the planar shock reflection in the air from wedges. It is a widely used benchmark to test the ability of shock capturing and the small scale structure resolution of a certain scheme. In the present simulation, the computation domain is taken as $[0,4]\times[0,1]$. The lower boundary is set to be a reflecting wall starting from $x=\frac{1}{6}$. At $t=0$, a right-moving $60^\circ$ inclined Mach 10 shock is positioned at $(\frac{1}{6} ,0)$. The upper boundary is set to describe the exact motion of the Mach $10$ shock. The left boundary at $x=0$ is assigned with post-shock values. Zero gradient outflow condition is set at $x=4$. Readers may refer to \cite{woodward1984numerical,kemm2016proper} for detailed descriptions of the double Mach reflection problem. An uniform grid is used with $\Delta x=\Delta y=\frac{1}{240}$. The reference result is given by the CH method with $\Delta x=\Delta y=\frac{1}{480}$.
\begin{figure}[H]
    \begin{center}
    \subfigure[Component-wise reconstruction]{
    \includegraphics[width=0.8\textwidth]{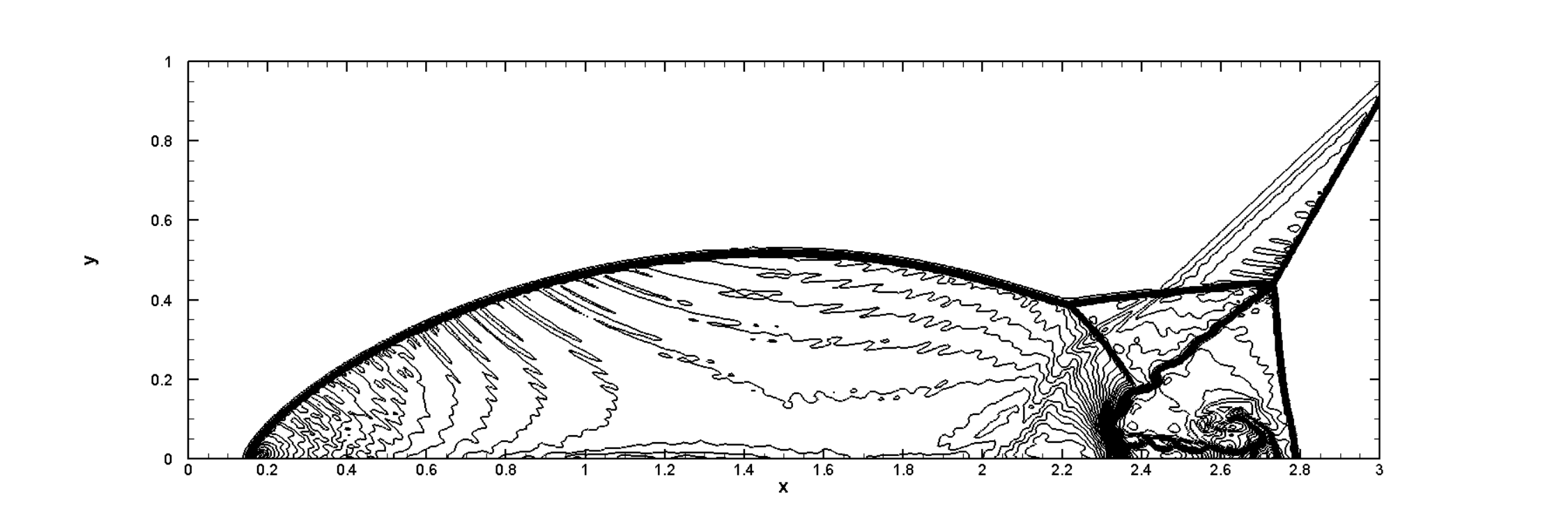}}
    \end{center}
\end{figure}
\begin{figure}[H]
    \begin{center}
    \subfigure[Characteristic-wise reconstruction]{
    \includegraphics[width=0.8\textwidth]{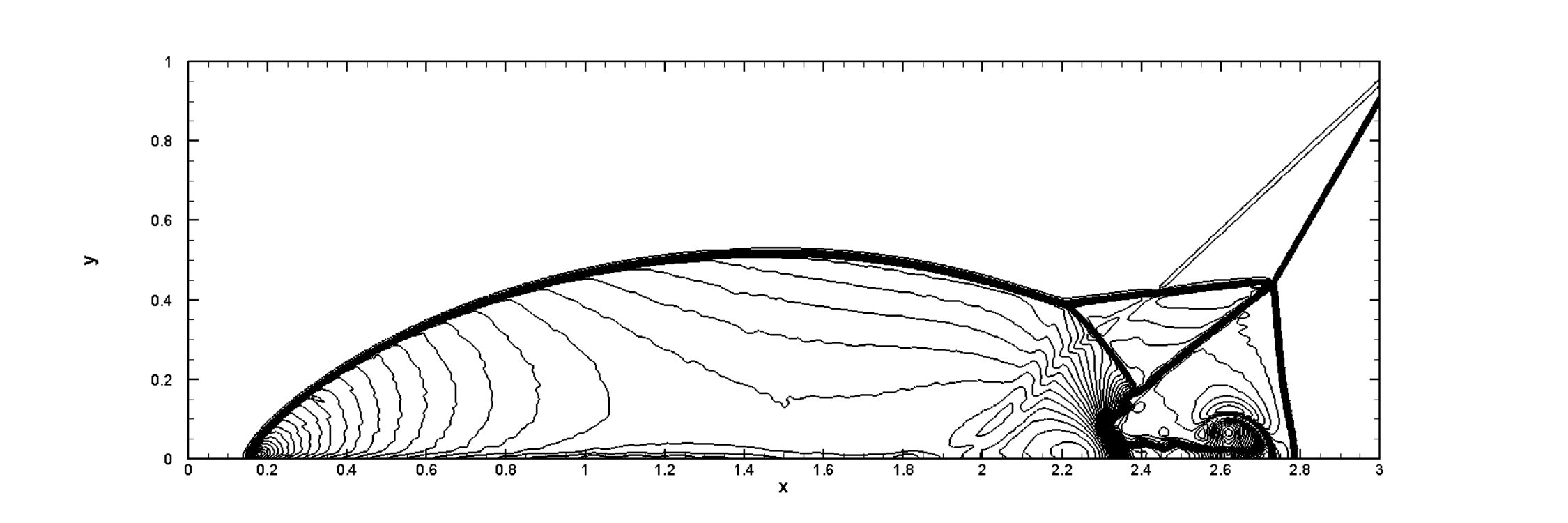}}
    \end{center}
\end{figure}
\begin{figure}[H]
    \begin{center}
    \subfigure[TOT]{
    \includegraphics[width=0.8\textwidth]{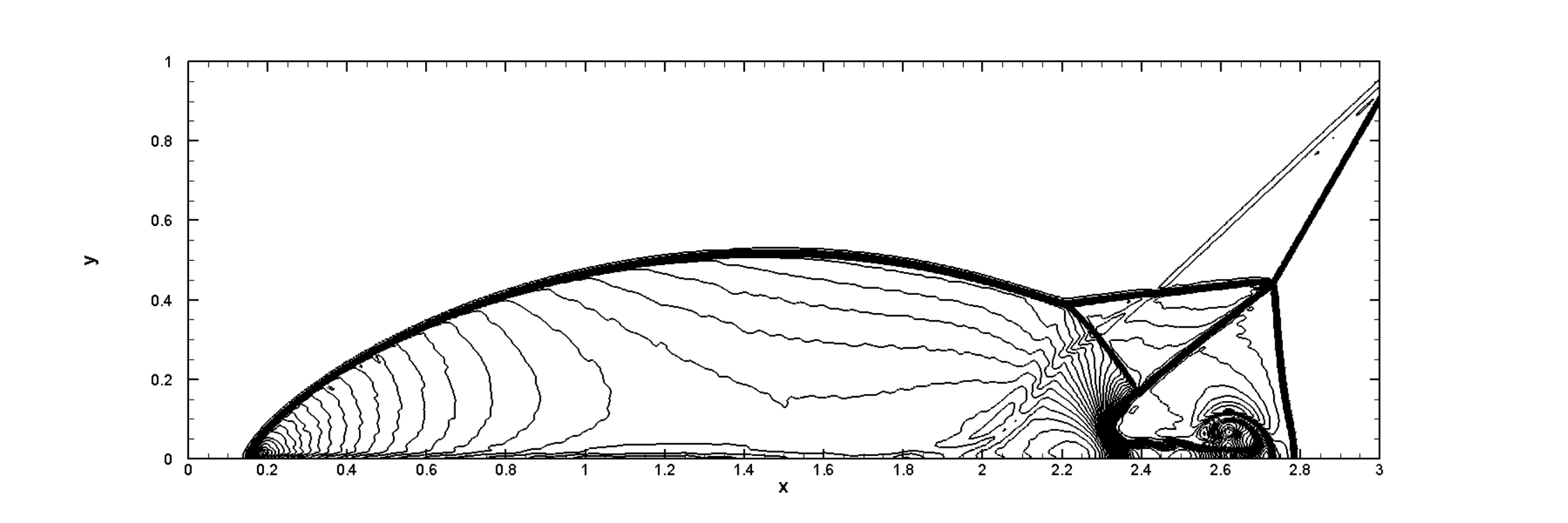}}
    \end{center}
\end{figure}
\begin{figure}[H]
    \begin{center}
    \subfigure[Present]{
    \includegraphics[width=0.8\textwidth]{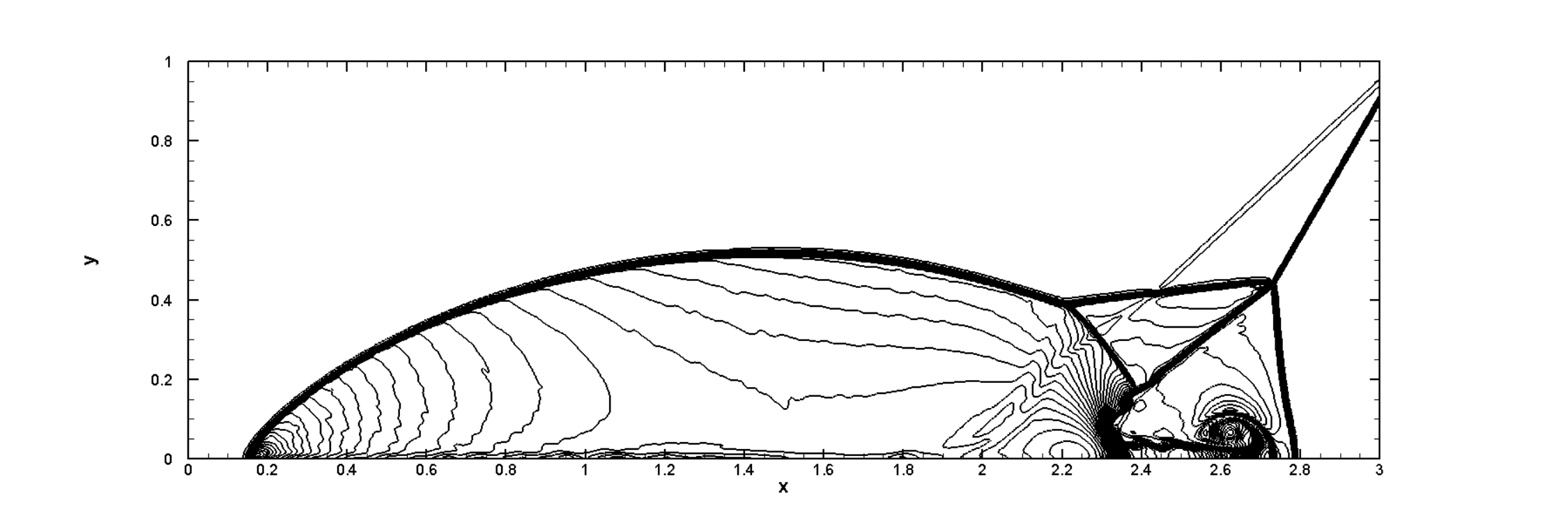}}
    \end{center}
\end{figure}
\begin{figure}[H]
    \begin{center}
    \subfigure[Reference]{
    \includegraphics[width=0.8\textwidth]{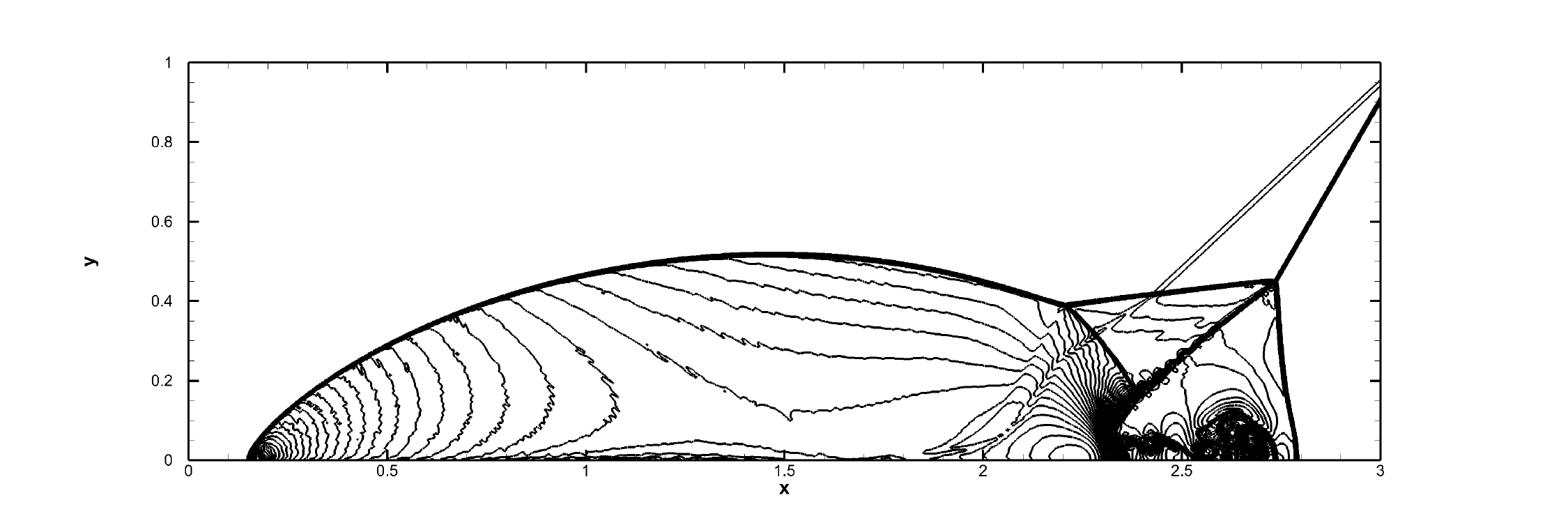}}
    \caption{Density contours of the double Mach reflection problem at t=0.2, ranging from $\rho=2.1379$ to 24 with 90 equally separated levels}\label{figdm1}
    \end{center}
\end{figure}

\begin{figure}[H]
    \begin{center}
    \subfigure[Component-wise reconstruction]{
    \includegraphics[width=0.5\textwidth]{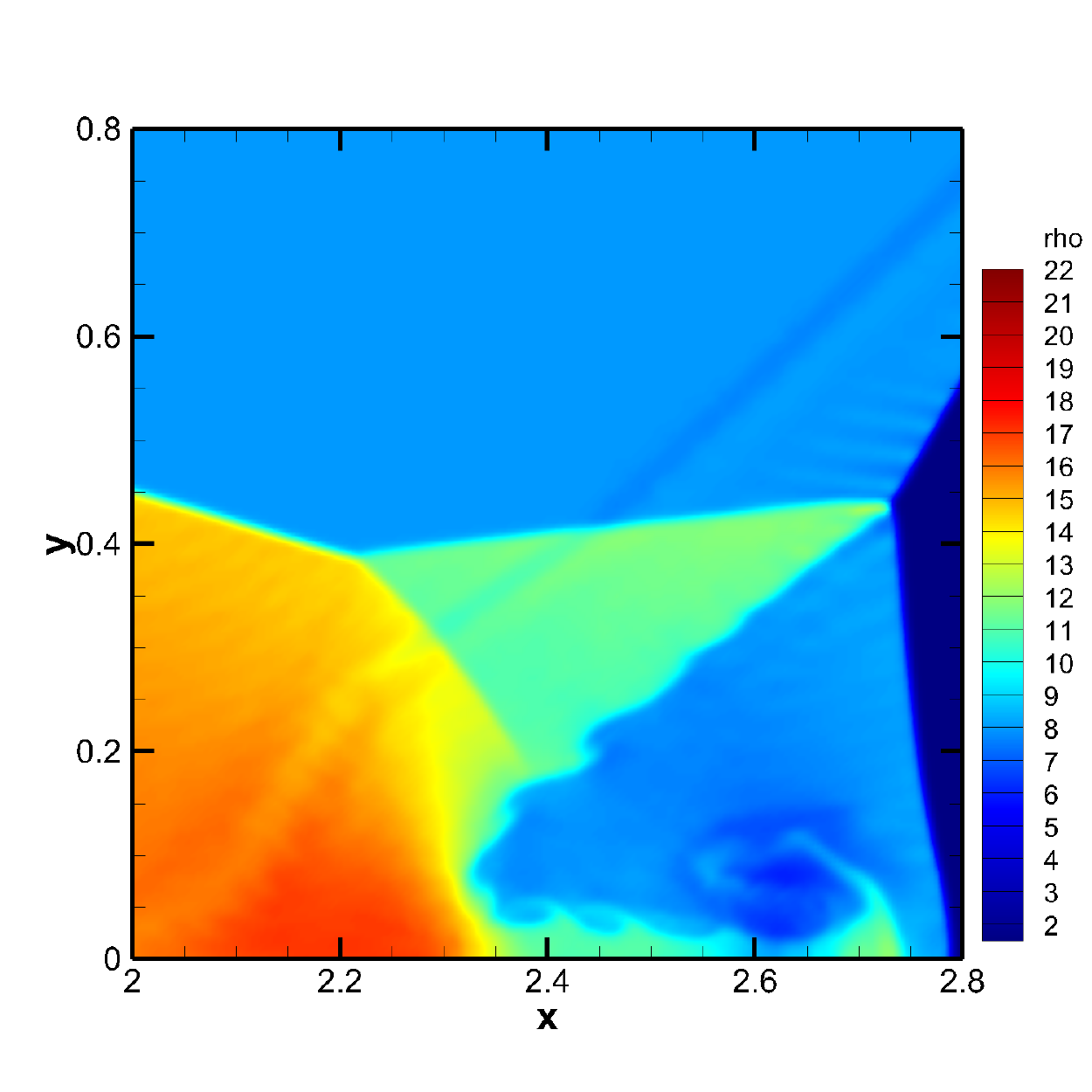}}
    \end{center}
\end{figure}
\begin{figure}[H]
    \begin{center}
    \subfigure[Characteristic-wise reconstruction]{
    \includegraphics[width=0.5\textwidth]{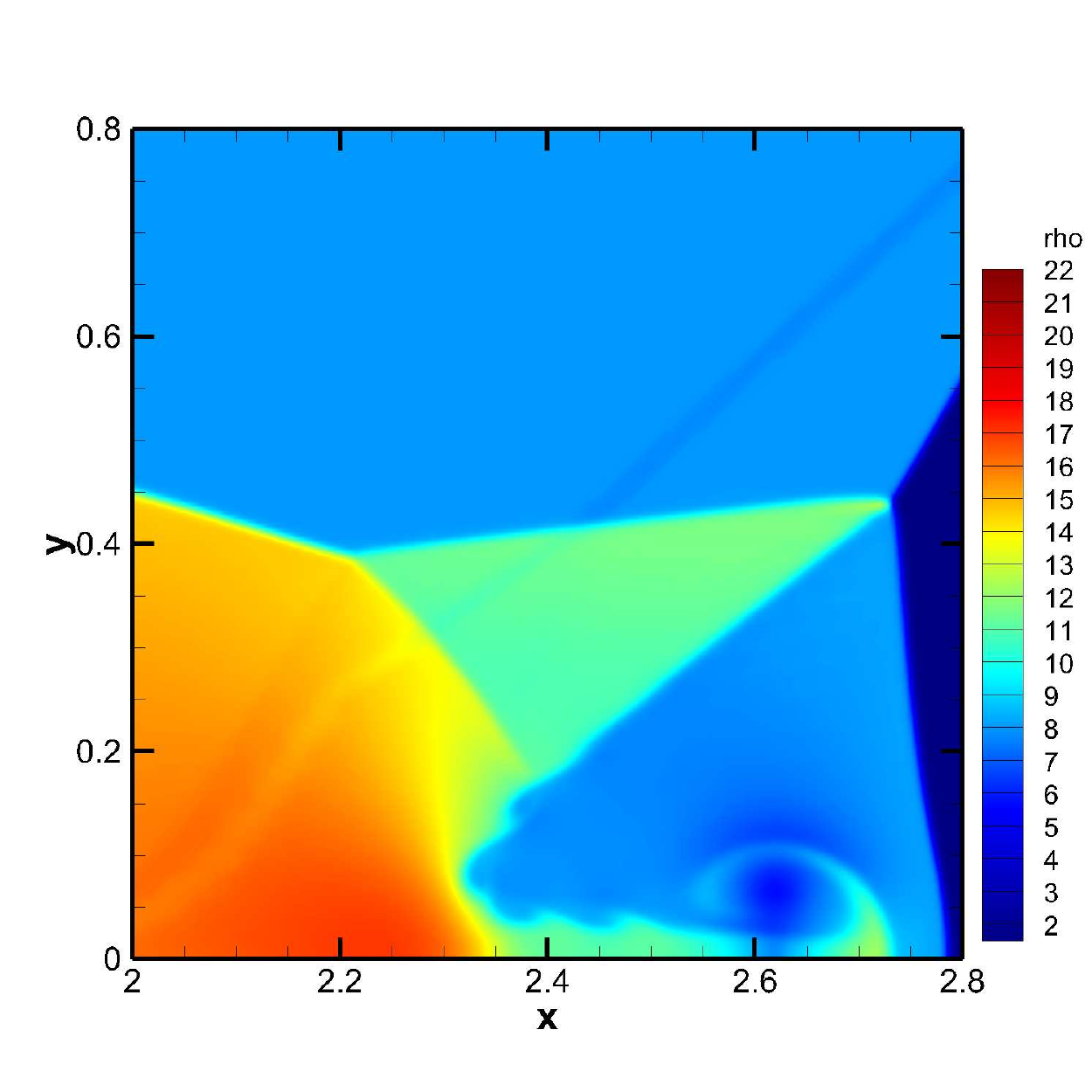}}
    \end{center}
\end{figure}
\begin{figure}[H]
    \begin{center}
    \subfigure[TOT]{
    \includegraphics[width=0.5\textwidth]{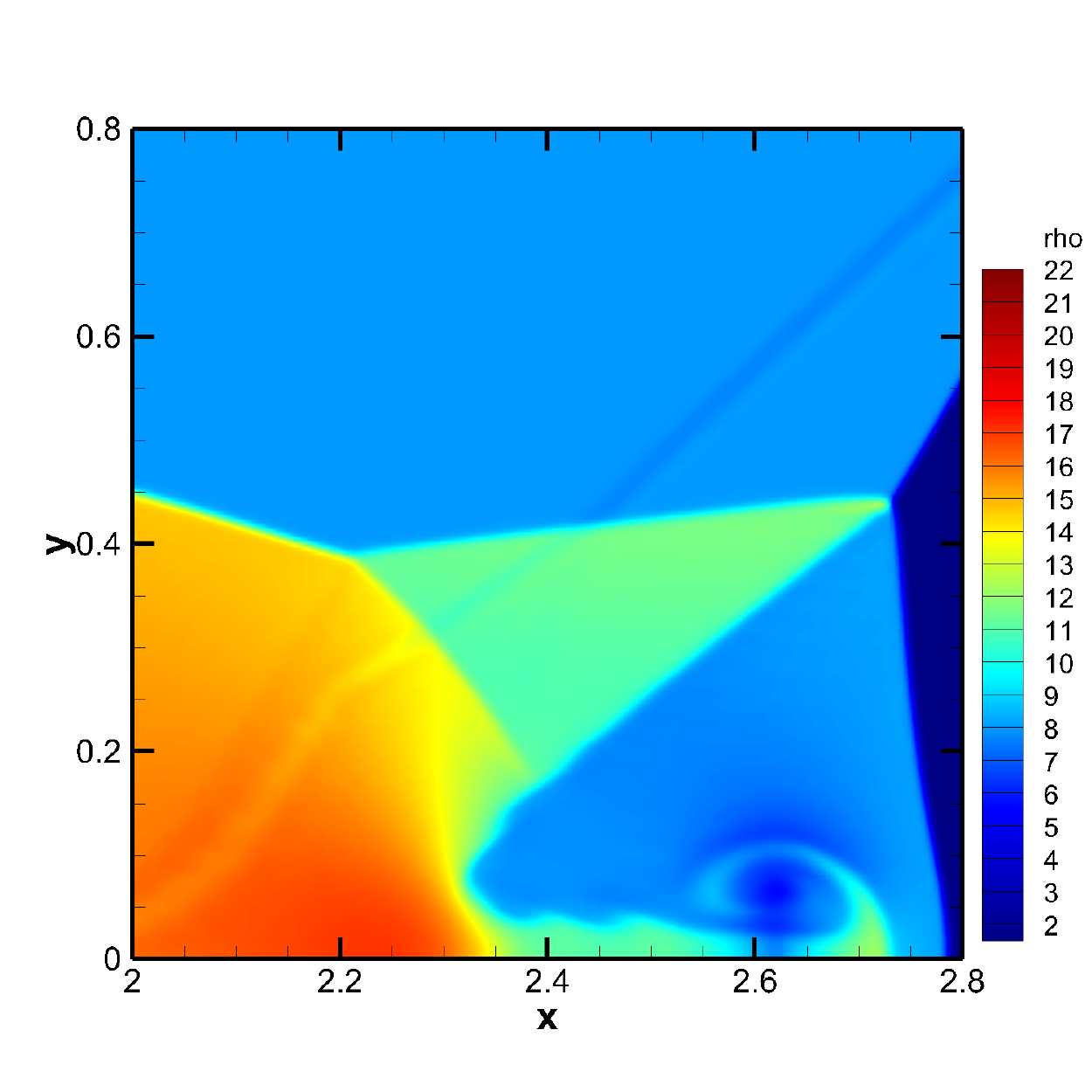}}
    \end{center}
\end{figure}
\begin{figure}[H]
    \begin{center}
    \subfigure[Present]{
    \includegraphics[width=0.5\textwidth]{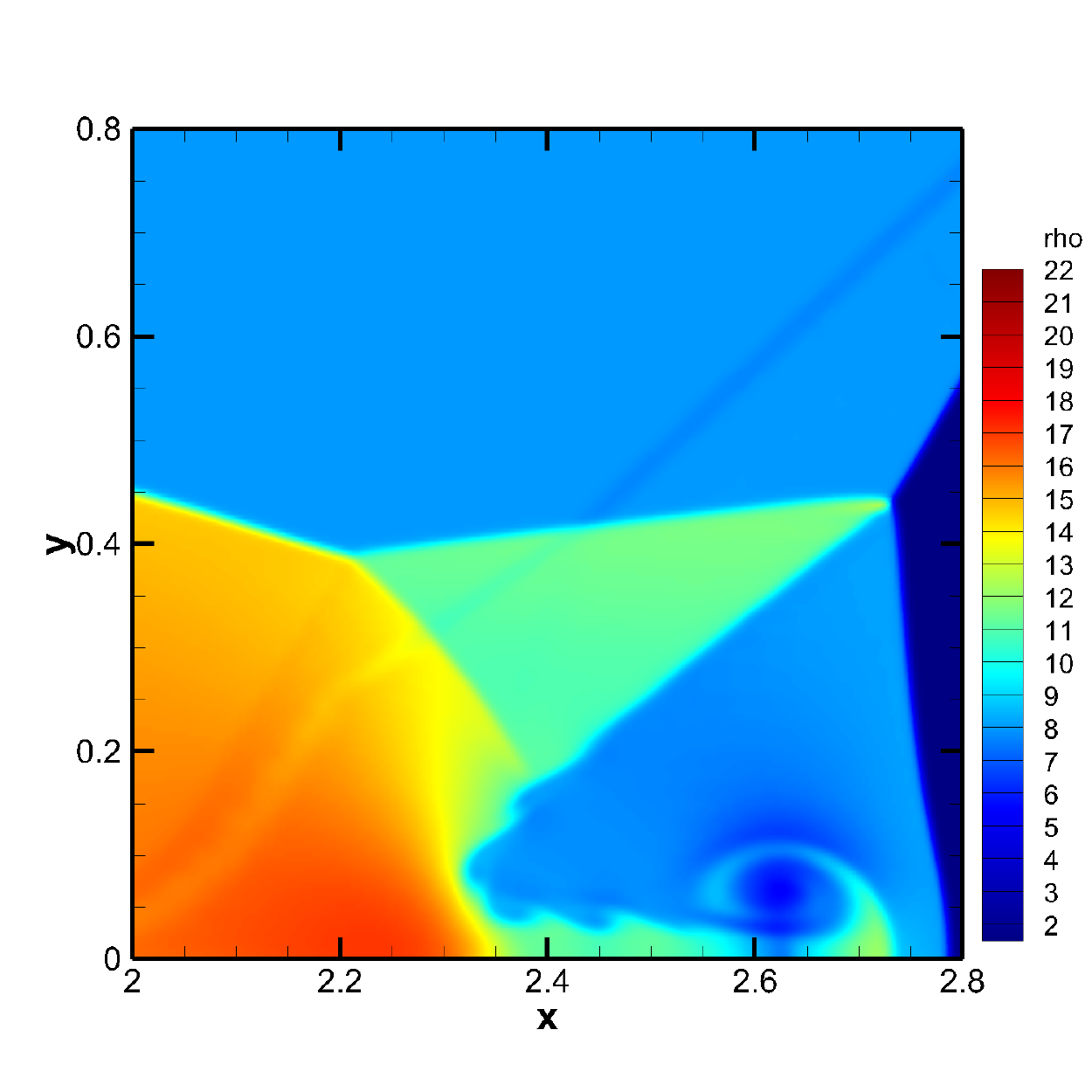}}
    \end{center}
\end{figure}
\begin{figure}[H]
    \begin{center}
    \subfigure[Reference]{
    \includegraphics[width=0.5\textwidth]{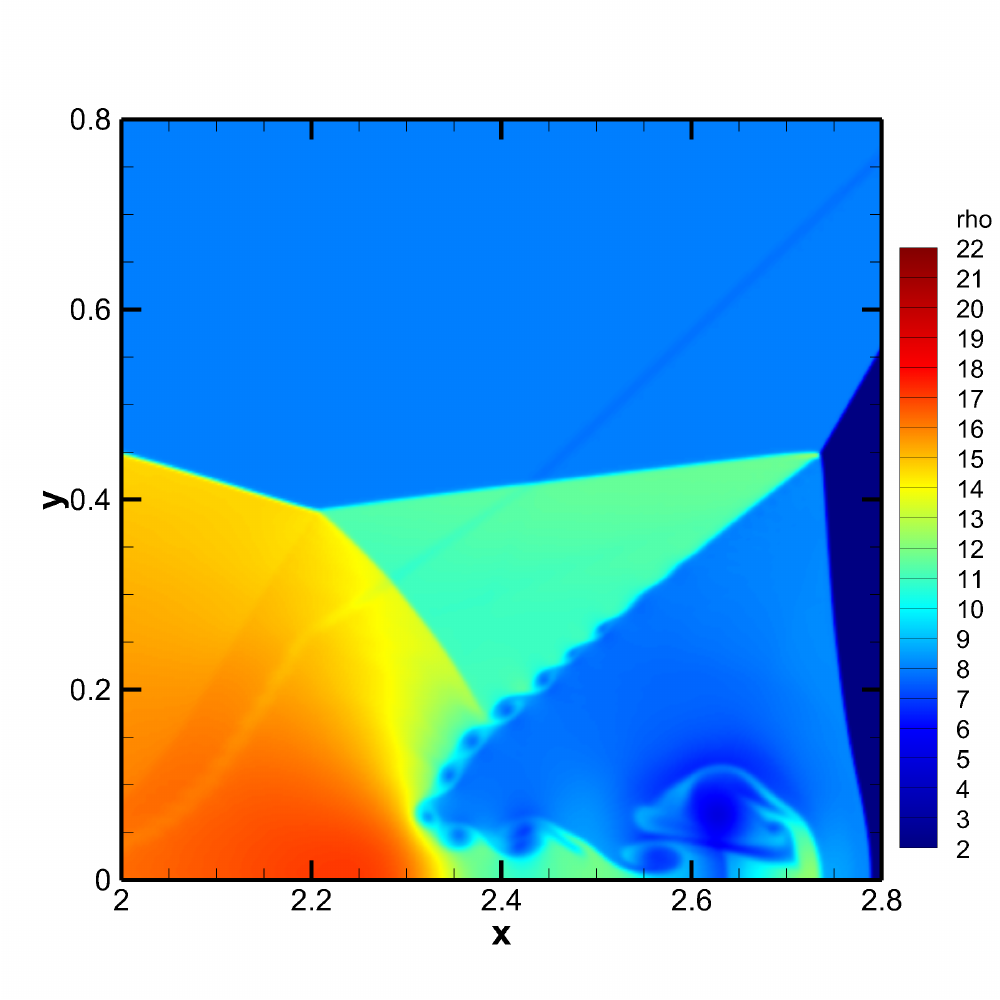}}
    \caption{Density contours of the up-rolling region at t=0.2}\label{figdmrmg}
    \end{center}
\end{figure}

\begin{figure}[H]
    \begin{center}
    \subfigure[TOT]{
    \includegraphics[width=0.8\textwidth]{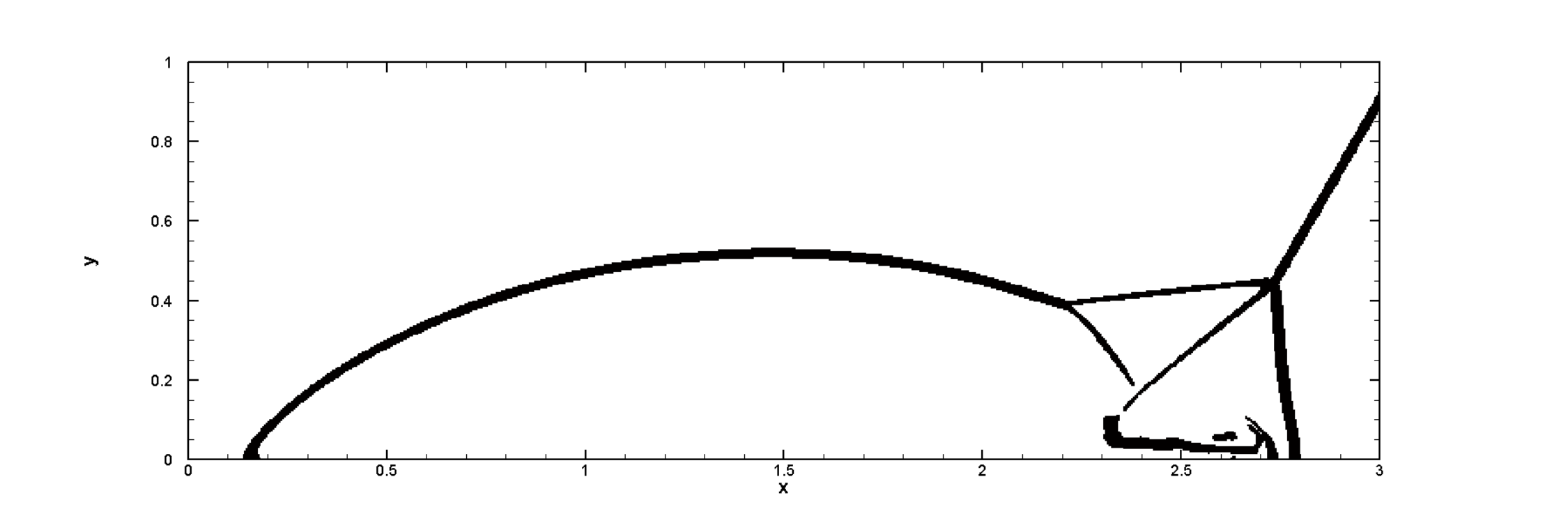}}
    \end{center}
\end{figure}
\begin{figure}[H]
    \begin{center}
    \subfigure[Present]{
    \includegraphics[width=0.8\textwidth]{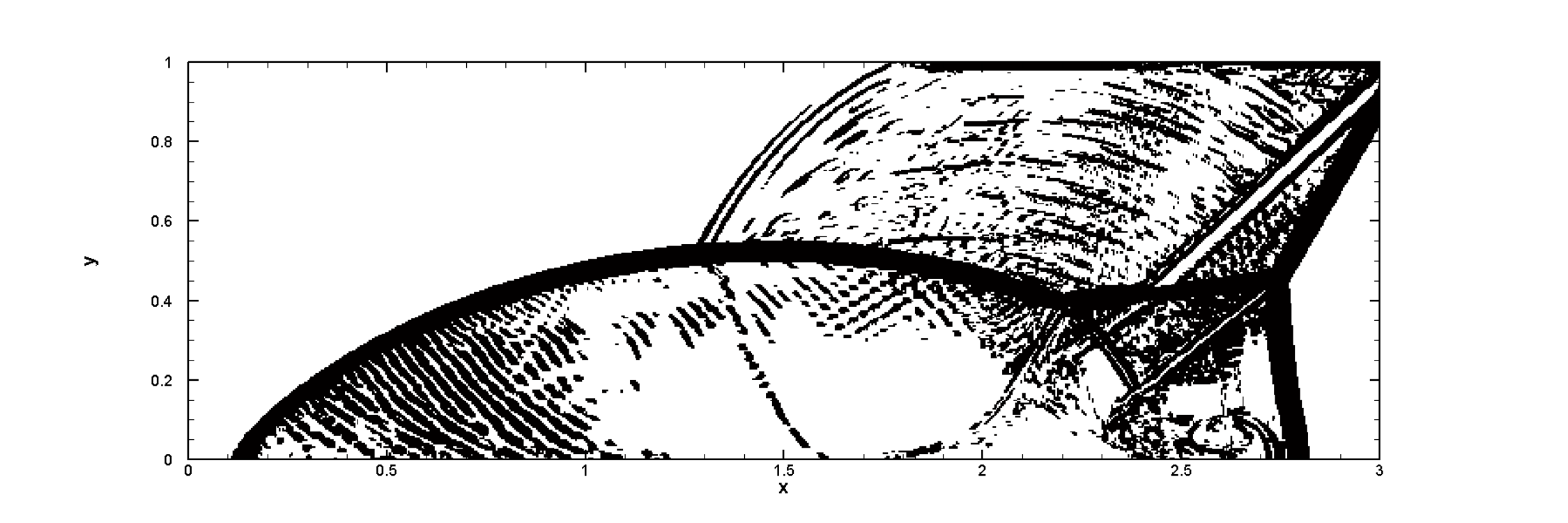}}
    \caption{Grids treated with the characteristic-wise reconstruction at t=0.2}\label{figdm2}
    \end{center}
\end{figure}

\begin{table}[h]
   \caption{The total CPU time  for the double Mach reflection problem}\label{tab:timeDM}
  \begin{center}\footnotesize
  \begin{tabular}{cc}
  \toprule
  Method & CPU time(s)\\
  \midrule
   CP & 10921.189  \\
   CH & 16646.573 \\
   TOT & 12107.236\\
   AdaWENO-Z & 5858.64\\
  \bottomrule
  \end{tabular}
  \end{center}
\end{table}
Fig.\ref{figdm1} shows the density contours of different methods at $t=0.2$. All methods capture discontinuities. The result of the CP method shows more numerical noises behind the reflected shock than those of the other three. Density contours of the up-rolling region of different methods are shown in Fig.\ref{figdmrmg}. The CP method result shows obvious oscillations near the triple-wave interaction point while the CH method, the TOT method, and the present method give similar and clean structures. Compared to TOT, AdaWENO-Z resolves the K-H instability structures better. The CPU time of different methods are given in Tab.\ref{tab:timeDM}, AdaWENO-Z is about 3 times faster than the CH method and is about 40\% faster than the CP method. Although marking more grids to be treated with CH as shown in Fig.\ref{figdm2}, AdaWENO-Z is more efficient than TOT due to the performance gain brought by the new shared smoothness indicators.

\subsubsection{Shock/shear layer interaction}\label{subsec4.2.3}
The shock wave impingement problem is designed to measure the resolution of schemes when shock waves interact with vortices \cite{yee1999low}. A Mach $0.6$ shear layer evolves and impacts on an oblique shock. The vortices produced by the shear layer instability pass, firstly, through the oblique shock and then a second shock reflected from the slip wall at the lower boundary. The computation domain is $[0,200]\times[-20,20]$. At $x=0$, the inlet condition is specified as:
\begin{equation}
 u=2.5+0.5tanh(2y).
\end{equation}
For the upper stream ($y>0$), $\rho=1.6374$, $p=0.3327$ and for lower stream ($y<0$), $\rho=0.3626$, $p=0.3327$. Post shock condition $(\rho,u,v,p)=(2.1101,2.9709,-0.1367,0.4754)$ is set at the upper boundary, and slip wall condition is applied at the lower boundary. Besides, fluctuations are added to the vertical velocity component at the inlet:
\begin{align}
 v'=\sum_{k=1}^{2}a_kcos(2\pi kt/T+\phi_k)exp(-y^2/b) \\
 b=10, a_1=a_2=0.05, \phi_1=0, \phi_2=\pi/2
\end{align}
in which $T=\lambda /u_c$ is the period, $\lambda=30$ is the wavelength, $u_c=2.68$ is the convective velocity. To illustrate the performance of each method, the two dimensional Euler equations instead of the Navier-Stokes equations are solved. A equally spaced grid with grid number $(N_x,N_y)=(500,100)$ is used. The reference result is calculated on a refined gird with $(N_x,N_y)=(2000,400)$ by the CH method.
\begin{figure}[H]
    \begin{center}
    \subfigure[Component-wise reconstruction]{
    \includegraphics[width=0.9\textwidth]{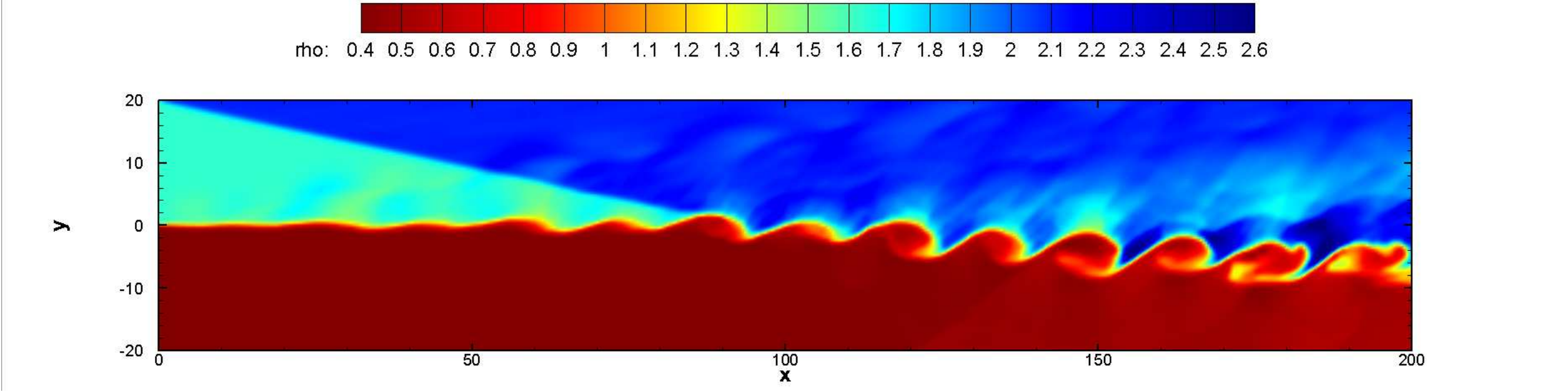}}
    \end{center}
\end{figure}
\begin{figure}[H]
    \begin{center}
    \subfigure[Characteristic-wise reconstruction]{
    \includegraphics[width=0.9\textwidth]{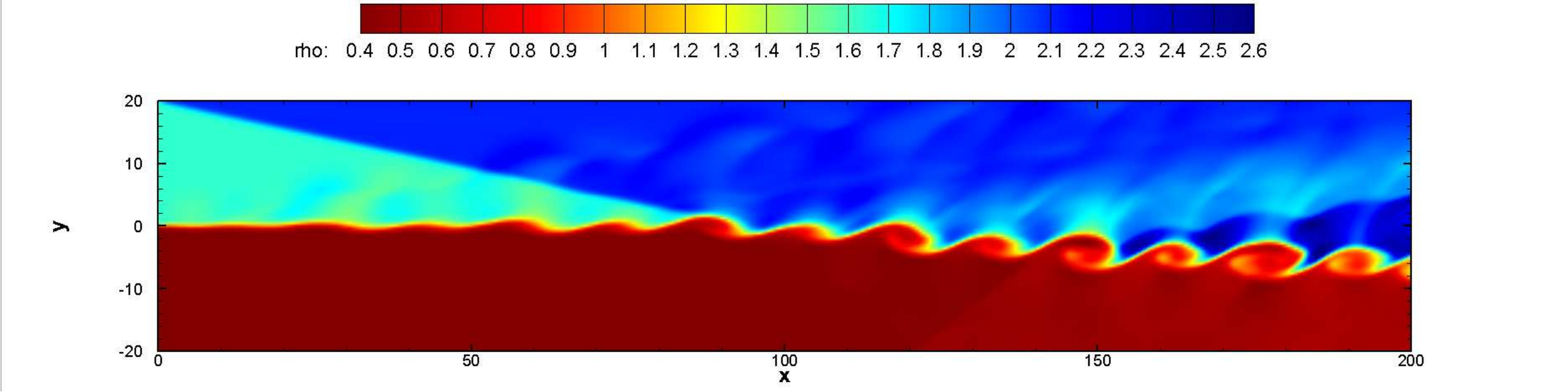}}
    \end{center}
\end{figure}
\begin{figure}[H]
    \begin{center}
    \subfigure[TOT]{
    \includegraphics[width=0.9\textwidth]{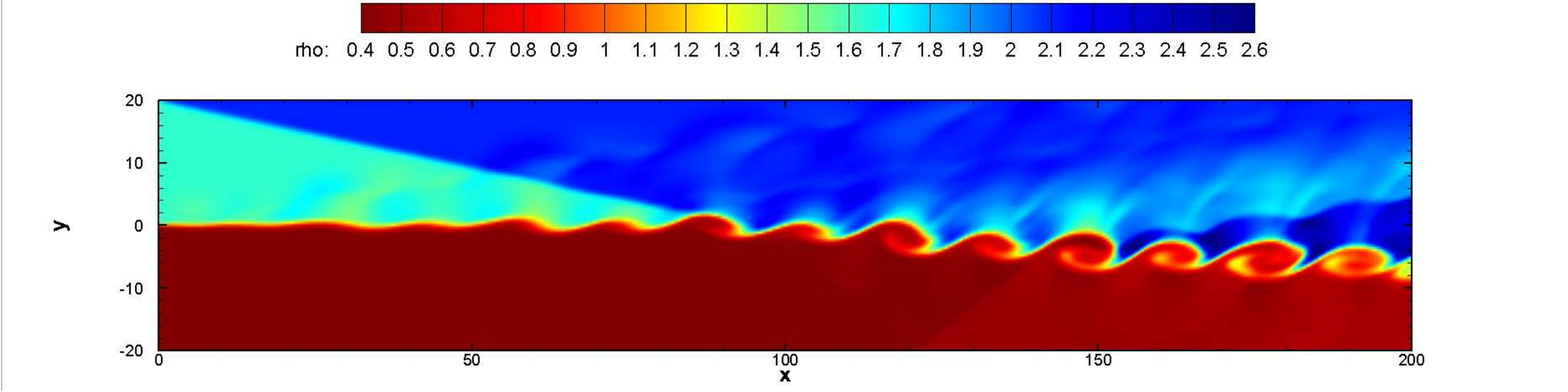}}
    \end{center}
\end{figure}
\begin{figure}[H]
    \begin{center}
    \subfigure[Present]{
    \includegraphics[width=0.9\textwidth]{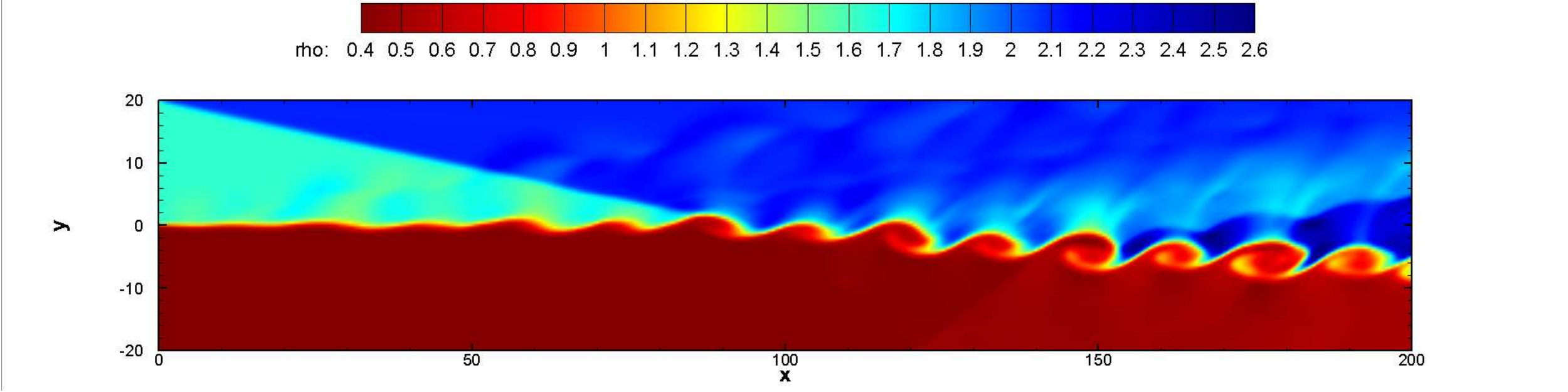}}
    \end{center}
\end{figure}
\begin{figure}[H]
    \begin{center}
    \subfigure[Reference]{
    \includegraphics[width=0.9\textwidth]{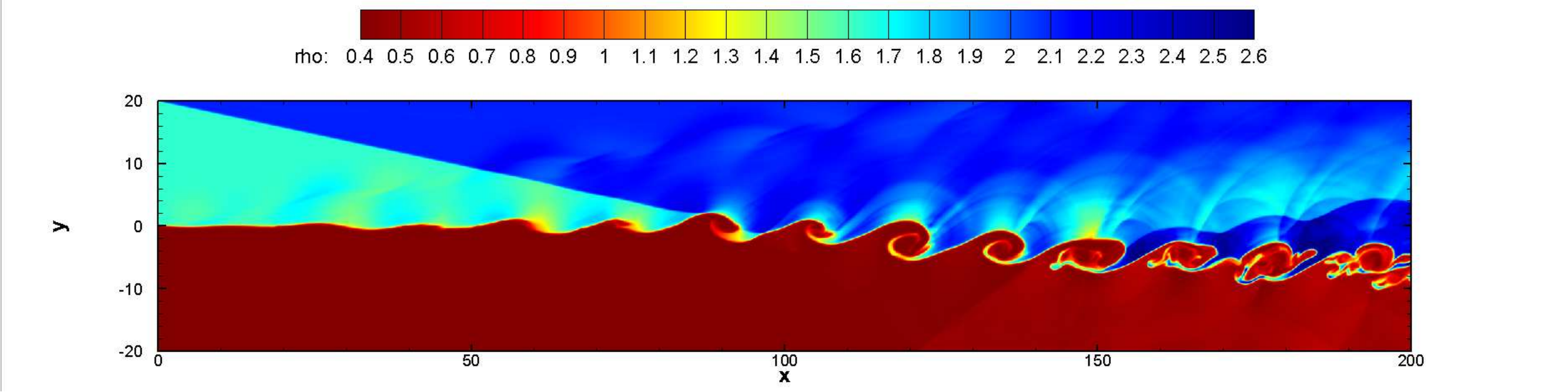}}
    \caption{Density contours of the shock/shear layer interaction problem at t=120}\label{figssl1}
    \end{center}
\end{figure}

\begin{figure}[H]
    \begin{center}
    \includegraphics[width=0.9\textwidth]{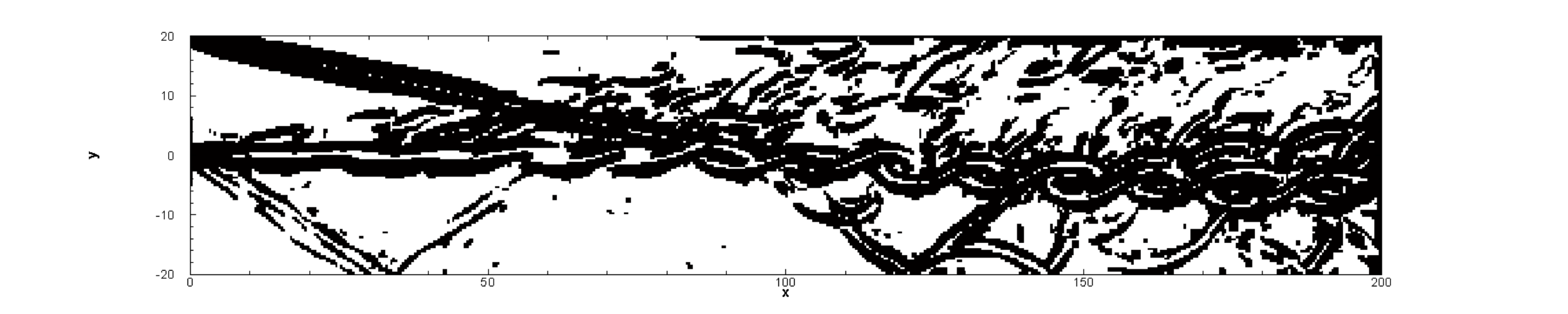}
    \caption{Grids treated with the characteristic-wise reconstruction of AdaWENO-Z at t=120}\label{figssl2}
    \end{center}
\end{figure}

\begin{table}[h]
   \caption{The total CPU time for the shock/shear layer interaction problem}\label{tab:timeSSL}
  \begin{center}\footnotesize
  \begin{tabular}{cc}
  \toprule
  Method & CPU time(s)\\
  \midrule
   CP & 2742.574  \\
   CH & 3949.063 \\
   TOT & 2968.359\\
   AdaWENO-Z & 1757.452\\
  \bottomrule
  \end{tabular}
  \end{center}
\end{table}
Density contours are shown in Fig.\ref{figssl1} at $t=120$. It can be observed that the CH method, the TOT method, and AdaWENO-Z obtain similar results which are very comparable to the reference result, while the self-similar structures of the downstream vortices are twisted in the CP method result. While the TOT method does not treat any grid with CH for this problem, the grids treated by CH of the AdaWENO-Z result at $t=120$ are shown in Fig.\ref{figssl2}. It indicates that the switch method of AdaWENO-Z is able to resolve both strong and weak discontinuities.

\section{Conclusion}\label{sec5}
In this paper, we present an adaptive characteristic-wise reconstruction WENO scheme (AdaWENO-Z) for the gas dynamic Euler equations. By defining shared smoothness functions, shared smoothness indicators are introduced to reduce the computational cost of the component-wise reconstruction procedure and to develop a global switch function. The switch function is based on the WENO-Z non-linear weights calculated from the shared smoothness indicators and is capable of detecting discontinuities. With the help of the switch function, the new method performs the component-wise reconstruction in smooth region and shifts to the characteristic-wise reconstruction near discontinuities. Numerical tests show that AdaWENO-Z achieves high efficiency without producing obvious numerical oscillations. AdaWENO-Z is about 2 to 3 times faster than the CH method and about 20\% to 40\% faster than the CP method. With good performance and high efficiency, AdaWENO-Z is suitable for large scale compressible flow simulation.

\section*{Acknowledgement}
This research work was supported by SCP (No. TZ2016002) and NSAF (No.U1530145) of Yiqing Shen and the Postdoctoral Science Foundation of China under Grand No.2017M610822 of Jun Peng. The authors acknowledge the anonymous reviewers for their valuable comments and suggestions.
\section*{References}
\bibliography{mybibfile}
\end{document}